\documentclass[12pt,preprint]{aastex}
\usepackage{amsmath}
\shorttitle{curvature of blazar SED}
\shortauthors{Liang Chen}

\slugcomment{accepted for publication in the ApJ}

\begin{document}
\title{The curvature of spectral energy distribution of blazars}

\author{Liang Chen{$^{1,\,2}$}}

\altaffiltext{1}{Key Laboratory for Research in Galaxies and
Cosmology, Shanghai Astronomical Observatory, Chinese Academy of
Sciences, 80 Nandan Road, Shanghai 200030, China; chenliang@shao.ac.cn}
\altaffiltext{2}{Key Laboratory of Modern Astronomy and Astrophysics (Nanjing University)£¬Ministry of Education, Nanjing 210093, China}


\begin{abstract}

The broadband spectral energy distribution (SED) of blazars show significant curvature. In this paper, we study the curvature properties for a large sample of \emph{Fermi}/LAT bright blazars based on quasi-simultaneous SED. Both SEDs of synchrotron and inverse Compton (IC) components are fitted by a log-parabolic law in $\log\nu$-$\log\nu f_{\nu}$ diagram. The second-degree term of log-parabola measures the curvature of SED. We find a statistically significant correlation between synchrotron peak frequency and its curvature. This result is in agreement with the theoretical prediction, and confirms previous studies, which dealt with single source with various epoch observations or a small sample. If a broken power-law is employed to fit the SED (spectral indexes $\alpha_{1}$ and $\alpha_{2}$, before and after the peak frequency, respectively), the difference between the two spectral indexes (i.e., $|\alpha_{2}-\alpha_{1}|$) can be considered as a ``surrogate" of the SED curvature. We collect spectral parameters of a sample blazars from literature, and find a correlation between the synchrotron peak frequency and the spectral difference. We do not find a significant correlation between the IC peak frequency and its curvature, which may be caused by complicated seed photon field. It is also found that the synchrotron curvatures are on average larger than that of IC curvatures, and there is no correlation between these two parameters. As suggested by previous works in literature, both the log-parabolic law of SED and above correlation can be explained by statistical and/or stochastic particle accelerations. Stochastic particle acceleration predicts a different slope of the correlation from that of statistical one, and our result seems favor stochastic acceleration mechanisms and emission processes. Some of other evidences also seem to support that the electron energy distribution (and/or synchrotron SED) may be log-parabolic, which include SED modeling, particle acceleration simulation, and comparisons between some predictions and empirical relations/correlations.

\end{abstract}


\section{Introduction}
\label{Introduction}

As early as 1970s, people noted that the non-thermal spectra of blazar show significant curvature. \citet{1974ApJ...192L.115R} found the infrared (IR) spectrum is significantly flatter than that of optical for OJ 287. Many other blazars, like BL Lacertae, 0735+178, ON 231, B2 0912+29, B2 1215+30, and AO 0253+164 also present similar properties \citep{1974ApJ...192L.115R, 1977ApJ...214L.105O, 1983PASP...95..724S}. \citet{1986ApJ...310..317G} claimed that there is apparent difference between slopes of IR and ultraviolet (UV), with a mean value of $\Delta\alpha=0.49\pm0.14$ \citep[see also,][]{1989ApJ...340..129B}. Blazar SED shows more clear curvature and some present significant breaks, when combine broadband emissions \citep[radio through X-ray bands,][]{1981ApJ...243...47L, 1985ApJ...298..630L, 1987ApJ...318..175B}. Further researches revealed a component on blazar SED from radio to UV/X-ray bands in $\log\nu$-$\log\nu f_{\nu}$ diagram, which is widely interpreted as synchrotron emissions of high energy electrons in a relativistic jet closely aligned to our line of sight \citep[see e.g.,][]{1978bllo.conf..328B, 1986ApJ...308...78L, 1995PASP..107..803U, 1996ApJ...463..444S, 1989MNRAS.241P..43G, 1998MNRAS.301..451G}. Some high energy detectors have broad energy coverage, e.g, \emph{Beppo}SAX covering 0.1-300 keV (LECS: 0.1-10 keV, MECS: 1.3-10 keV, and PDS: 13-300 keV). And hence, with observations of some of the telescopes alone, blazar SEDs are found to be curved, some of which can be fitted by a log-parabolic law. For instance, X-ray observations of Mrk 501 and Mrk 421 \citep[see,][and references and citations therein]{2004A&A...413..489M, 2004A&A...422..103M, 2008A&A...478..395M}, together with several other BL Lacs \citep[PKS 0548-322, 1H 1426+418, 1ES 1959+650, and PKS 2155-304; see,][]{2008A&A...478..395M} show that X-ray spectra can be fitted by log-parabolic law, i.e., $\log\nu f_{\nu}=-b\left(\log\nu-\log\nu_{p}\right)^{2}+\log\nu_{p}f_{\nu_{p}}$. \citet{2011ApJ...739...73M} studied a sample of high-frequency-peaked BL Lac objects (HBLs), and found that X-ray spectral curvature (measured by the second-term $b$) of TeV HBLs are systematically larger than that of those HBLs non-detected at TeV energies (NBL), implying that the NBL X-ray spectra are systematically narrower \citep[see also,][]{2013ApJS..207...16M}.

For the first time, BL Lac object Mrk 421 was detected by Whipple emitting $\gamma$-ray photons at TeV band \citep{1992Natur.358..477P}. Thanks to development of many ground based Cherenkov telescopes (e.g., Whipple, MAGIC, VERITAS, H.E.S.S., CANGAROO), more than 50 blazars have now been discovered bearing TeV $\gamma$-ray emissions (see, http://tevcat.uchicago.edu). At GeV $\gamma$-ray band, \emph{CGRO}/EGRET successfully detected 68 AGNs with significance $\sigma>5$ \citep[working at 20 MeV - 30 GeV, most of them are blazars,][]{1999ApJS..123...79H}. As a successor, \emph{Fermi}/LAT works between 20 MeV - 300 GeV, which was launched at June 2008. During the first 11 months survey of \emph{Fermi}/LAT, 671 AGNs were detected with high confidence level at high Galactic latitude \citep[$|b|>10^{\circ}$, $\sigma>5$, and most of them are blazars, see][]{2010ApJ...715..429A}. The number increases to 1017, through first two years survey of \emph{Fermi}/LAT \citep{2011ApJ...743..171A}. Based on these $\gamma$-ray observations, another component on blazar SED was discovered, peaking at $\gamma$-ray band. These $\gamma$-ray emissions are usually explained as inverse Compton emissions (IC) of the same electron population accounting for the synchrotron component emissions \citep[see,][and references therein]{1998MNRAS.301..451G, 2010MNRAS.402..497G, 2010ApJ...716...30A}. Many blazars observed by \emph{Fermi}/LAT alone present curved $\gamma$-ray spectra and some can be well fitted by a log-parabolic law \citep[see, e.g.,][]{2011ApJ...743..171A}. Higher energy observation of some balzars \citep[e.g., PKS 2155-304, Mrk 501, Mrk 421, see,][]{2007ApJ...664L..71A, 2008MNRAS.386L..28G, 1998ApJ...501L..17S, 1999ApJ...511..149K, 2012A&A...542A.100A} by Imaging Atmospheric Cherenkov Telescopes (IACT) show substantial curvature at TeV band, and some can also be fitted by log-parabolic law. If combine the GeV data with those of TeV, the spectra present more significant curvature \citep[see,][]{2009ApJ...707.1310A, 2012ApJ...752..157Z, 2013ApJ...764..119S}.

To characterize a broadband peaked component, one needs at least two parameters: peak frequency and peak flux/luminosity. These two parameters are extensively used in blazar studies, especially in the so called balzar sequence \citep[see e.g.,][]{1998MNRAS.299..433F, 1998MNRAS.301..451G, 2011ApJ...735..108C, 2012MNRAS.420.2899G}. If log-parabolic law is adopted to fit the peaked component, i.e., $\log\nu f_{\nu}=-b\left(\log\nu-\log\nu_{p}\right)^{2}+\log\nu_{p}f_{\nu_{p}}$, the second-term $b$ measures the curvature around the SED peak, which is the third important parameter to characterize the component. Using multi-epoch observations of Swift, XMM-Newton, and \emph{Beppo}SAX, it was found that each observational X-ray spectra of Mrk 421 can be well fitted by a log-parabolic law, and an anti-correlation between the peak frequency $\nu_{p}$ and the curvature $b$ is discovered \citep[see,][]{2004A&A...413..489M, 2007A&A...466..521T, 2009A&A...501..879T}. Further, \citet{2008A&A...478..395M} extend the study to several other BL Lacs, e.g., Mrk 501, PKS 2155-304, PKS 0548-322, and found that all these BL Lacs present similar behavior: an anti-correlation between the peak frequency $\nu_{p}$ and the curvature $b$. As expected by previous theoretical investigations, the anti-correlation between peak frequency and curvature can be explained in the framework of acceleration processes of emitting electrons \citep[see,][and the Section \ref{TheoryoftheCorrelation} in this work]{2007A&A...466..521T, 2009A&A...508L..31P, 2009A&A...504..821P}. More detailed study on this third parameter, the curvature, is necessary to understand the particle acceleration, and energy dissipation mechanism in blazars.

Besides the high energy study, low energy observations from radio to optical also illustrate a similar feature. For example, \citet{1986ApJ...308...78L} fitted the SED from radio to optical with a log-parabolic law of a sample of 18 balzars. Except for 3 bad fitted or steep radio spectral blazars, they found an anti-correlation between peak frequency and curvature for the remaining 15 blazars \citep[see the top panel of Fig. 3 in][]{1986ApJ...308...78L}. In the past nearly three decades, many large samples of blazars were used to study the properties of the peak frequency, luminosity, and jet emission parameters \citep[see,][]{1996ApJ...463..444S, 1998MNRAS.299..433F, 1998MNRAS.301..451G, 2006A&A...445..441N, 2009MNRAS.397.1713C, 2009RAA.....9..168W, 2010MNRAS.402..497G, 2010ApJ...716...30A, 2011ApJ...735..108C}. However, few works focused on the curvature properties by using broadband SEDs. Until recently, \citet{2011MNRAS.417.1881R} presented an anti-correlation between synchrotron peak frequency and curvature by fitting the SEDs from radio to optical of 10 BL Lacs in both high and low states.

In this paper, we collect quasi-simultaneous broadband SEDs, from radio to $\gamma$-ray, of a large sample of blazars. Both SEDs of synchrotron and IC components are fitted by log-parabolic law in $\log\nu$-$\log\nu f_{\nu}$ diagram, respectively, and we obtain the synchrotron and IC curvatures. We then present detailed studies on the correlations between the peak frequency and curvature, and implications of the results. This paper is organized as follows: Section \ref{TheoryoftheCorrelation} shows the theoretical interpretation of the correlation between the peak frequency and curvature. Section \ref{TheSample} describes the sample, and Section \ref{TheFittingProcedure} shows the fitting procedures. After providing the results in Section \ref{TheResults}, the detailed discussions and implications of the results are presented in Section \ref{DiscussionandConclusions}. We end with a summary of the findings in Section \ref{Summary}.
Throughout the paper, a $\Lambda$CDM cosmology with values within 1$\sigma$ of the \emph{Wilkinson Microwave Anisotropy Probe} (\emph{WMAP}) results \citep{2011ApJS..192...18K} is used; in particular, $H_{0}=70$ km s$^{-1}$ Mpc$^{-1}$, $\Omega_{\Lambda}=0.73$, and $\Omega_{\rm M}=0.27$.

\section{Theory of the Correlation}
\label{TheoryoftheCorrelation}

Two different scenarios can explain the correlation between peak frequency and curvature, i.e., statistical and the stochastical acceleration mechanisms, both of which can reproduce the electron energy distribution as a log-parabolic law. A log-parabolic distribution of electron energy also emits a log-parabolic SED approximately, which consists with the fitting methods \citep[see,][and references therein]{2004A&A...413..489M, 2004A&A...422..103M, 2006A&A...448..861M, 2007A&A...466..521T, 2008A&A...478..395M, 2011ApJ...739...66T}. In the following, we will explain the difference between these two acceleration mechanisms, that will be tested with our results in this work.

The first scenario is in the framework of statistical acceleration, which needs either an energy-dependent acceleration probability ($p_{a}$) or a fluctuation of fractional acceleration gain ($\varepsilon$). For the case of energy-dependent acceleration probability, \citet{2004A&A...413..489M} showed that when acceleration efficiency is inversely proportional to the energy itself ($p_{a}=g/\gamma^{q}$ and $\varepsilon=cons.$ in this case), the log-parabolic law would be a very good approximation of the electron energy distribution,
\begin{equation}
N(\gamma)\approx cons.\left(\frac{\gamma}{\gamma_{0}}\right)^{-s-r\log(\gamma/\gamma_{0})}.
\end{equation}
Where $\gamma_{0}$ is initial electron energy, $r=q/(2\log\varepsilon)$ is curvature of electron energy distribution, and $s=-2r/q\log(g/\gamma_{0})-(q-2)/2$. The SED of synchrotron emissions by these electrons is also approximately log-parabolic, i.e., $\log\nu L_{\nu}=-b_{sy}\left(\log\nu-\log\nu_{p}\right)^{2}+\log\nu_{p}L_{\nu_{p}}$. With monochromatic emission assumption (i.e., $\delta$-approximation), one can derive the synchrotron peak frequency $\nu_{p}\propto\gamma_{p}^{2}$ and curvature $b_{sy}=r/4$, where $\log\gamma_{p}=\log\gamma_{0}+(3-s)/2r$ is the peak energy of electron in diagram $\log\gamma$-$\log\gamma^{3}N(\gamma)$. As suggested by \citet{2006A&A...448..861M}, one obtains $b_{sy}\approx r/5$ instead of $b_{sy}=r/4$, when release the assumption of $\delta$-approximation. After substituting the $\gamma_{p}$, $s$ and $r$, we obtain $\log\nu_{p}=C+2\log\gamma_{0}+(3-s)/r
=C+2\log\gamma_{0}+2/q\log(g/\gamma_{0})+\log\varepsilon+2/r$. Assuming that $\gamma_{0}$, $q$, $g$, and $\varepsilon$ are independent variables and substituting the relation $b_{sy}\approx r/5$, we get $\log\nu_{p}\approx C_{1}+2/(5b_{sy})$. For the case of fluctuations of fractional acceleration gain, \citet{2011ApJ...739...66T} showed that when $\varepsilon$ is a random variable around a systematic energy gain $\overline{\varepsilon}$ ($p_{a}=1$ and $\varepsilon=\overline{\varepsilon}+\chi$ in this case, where random variable $\chi$ has a probability density function with zero mean value and variance $\sigma_{\varepsilon}^{2}$), applying the multiplicative case of the central limit theorem, one derives the electron energy distribution at acceleration step $n_{s}$,
\begin{equation}
N(\gamma)=\frac{N_{0}}{\gamma\sigma_{\gamma}\sqrt{2\pi}}
\exp\left[-\frac{\left(\ln\gamma-\mu\right)^{2}}{2\sigma_{\gamma}^{2}}\right].
\end{equation}
Where $\sigma_{\gamma}^{2}\approx n_{s}(\sigma_{\varepsilon}/\overline{\varepsilon})^{2}$ and $\mu=\ln\gamma_{0}+n_{s}\left[\ln\overline{\varepsilon}-
(\sigma_{\varepsilon}/\overline{\varepsilon})^{2}/2\right]$. Therefore, the electron peak energy and curvature in diagram $\log\gamma$-$\log\gamma^{3}N(\gamma)$ can be derived: $\log\gamma_{p}=\log\gamma_{0}+n_{s}\log\overline{\varepsilon}+3/(4r)$ and $r=\ln10/\left[2n_{s}(\sigma_{\varepsilon}/\overline{\varepsilon})^{2}\right]$. Substituting the relation $b_{sy}\approx r/5$ and $\gamma_{b}$, we have $\log\nu_{p}\approx C+2\log\gamma_{0}+2n_{s}\log\overline{\varepsilon}+3/(10b)$. Assuming that $\gamma_{0}$, $n_{s}$, $\overline{\varepsilon}$ and $\sigma_{\varepsilon}$ are independent variables, we obtain $\log\nu_{p}\approx C+3/(10b_{sy})$.

The second scenario is in the framework of stochastic acceleration, provided the Fokker-Planck equation with presence of a momentum-diffusion term. \citet{1962SvA.....6..317K} suggested that a log-parabolic distribution of electron energy can be derived from a `quasi-'monoenergetic particle injection \citep[i.e., $N_{\rm inj}(\gamma)\approx N_{0}\delta(\gamma-\gamma_{0})$, see also][]{2007A&A...466..521T, 2011ApJ...739...66T},
\begin{equation}
N(\gamma)=\frac{N_{0}}{\gamma\sqrt{4\pi a_{1}}}
\exp\left[-\frac{(\ln\gamma-a_{1}-a_{2}-\ln\gamma_{0})^{2}}{4a_{1}}\right].
\end{equation}
Where $a_{1}$ and $a_{2}$ correspond stochastic and systematic acceleration terms, respectively. When represented in diagram $\log\gamma$-$\log\gamma^{3}N(\gamma)$, we obtain the curvature $r=\ln10/4a_{1}$, and the electron peak energy $\log\gamma_{p}=\log\gamma_{0}+5a_{1}/\ln10+a_{2}/\ln10 =\log\gamma_{0}+a_{2}/\ln10+5/(4r)$. Assuming that $\gamma_{0}$, $a_{1}$ and $a_{2}$ are independent variables and substituting the relation $b_{sy}\approx r/5$ and $\gamma_{p}$, we have $\log\nu_{p}\approx C+1/(2b_{sy})$.

\section{The Sample}
\label{TheSample}

\citet{2010ApJ...716...30A} collected broadband quasi-simultaneous spectral data, from radio through $\gamma$-ray, of 48 LBAS \citep[\emph{Fermi} LAT Bright AGN Sample, see][]{2009ApJ...700..597A} blazars based on the first three months operation of \emph{Fermi}/LAT. All these data are properly scaled in $\log\nu$-$\log\nu f_{\nu}$ diagram. The 48 LBAS blazars are taken as our sample in this paper, and the detailed information of these blazars are presented in Table \ref{tabledata}. Column (1) provides the \emph{Fermi}/LAT name of the source. Column (2) is the source redshift. The broadband spectral data of these blazars are also taken from \citet[][]{2010ApJ...716...30A}. Figure \ref{sed} in the Appendix are the SED of these 48 blazars. The red points are the broadband quasi-simultaneous spectral data, while the grey ones represent other observations \citep[see,][for detail description]{2010ApJ...716...30A}.

\section{The Fitting Procedure}
\label{TheFittingProcedure}

In order to get the curvatures of synchrotron and IC components, we adopt a log-parabolic law, i.e., $\log\nu f_{\nu}=-b\left(\log\nu-\log\nu_{p}\right)^{2}+\log\nu_{p}f_{\nu_{p}}$, to fit the quasi-simultaneous SED of the two components, respectively \citep[same as in][]{1986ApJ...308...78L, 1996ApJ...463..444S, 2004A&A...413..489M, 2004A&A...422..103M, 2006A&A...445..441N, 2009RAA.....9..168W, 2008A&A...478..395M, 2011MNRAS.417.1881R, 2011ApJ...739...73M}. The coefficient of the second degree term, $b_{sy}$/$b_{IC}$, measures the curvature around the peak. The least $\chi^{2}$ technique is used to determine the parameters. For some blazars, there are no quasi-simultaneous data (the red points in the Figure \ref{sed}) at radio or microwave bands to constrain the lower energy part of synchrotron fitting. We add other observation data (i.e., the grey points) in the fitting. For some data points, the errors are unavailable. We estimate the errors with average errors of those data points whose errors are available. Since the errors of the data for the synchrotron component are usually much smaller than that for the IC component, we derive the average errors of the data in the two components separately. The best fitting parameter values are listed in Table \ref{tabledata}. Columns (3), (4), (5), (6), and (7) are the (for synchrotron fitting) peak frequency, flux, curvature, the degree of freedom, and the reduced $\chi^{2}$, respectively. Columns (8), (9), (10), (11), and (12) denote the same quantities but for IC fitting.

\begin{figure}
\epsscale{0.8} \plotone{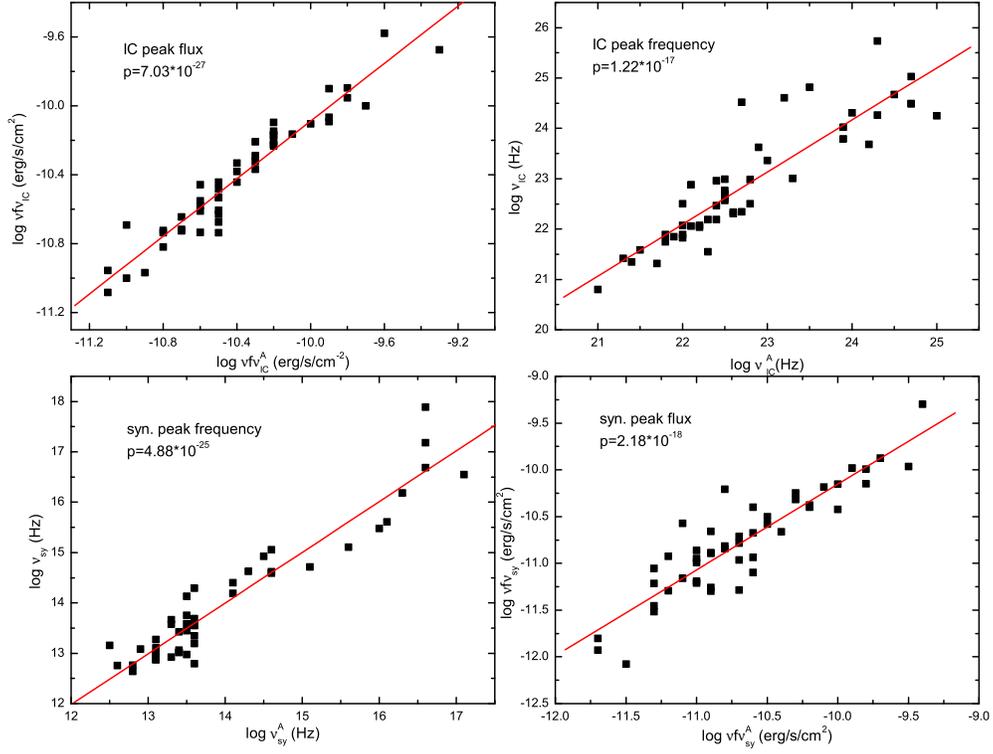} \caption{The comparisons between our fitting results and \citet{2010ApJ...716...30A} results. The top left panel is for synchrotron peak frequency comparison, top right panel for synchrotron peak flux comparison, bottom left panel for IC peak frequency comparison, and the bottom right panel for IC peak flux comparison. The red lines are for best linear fittings. The $p$-value in each panel is the Pearson's probability for a null correlation.}
\label{comparetotal}
\end{figure}

In order to test the validity of our fitting parameters, we compare the peak fluxes and peak frequencies between ours and those obtained from \citet{2010ApJ...716...30A} by fitting a third degree polynomial function for both SEDs of synchrotron and IC components, respectively. The comparison results are shown in Figure \ref{comparetotal}. The top left panel is for synchrotron peak frequency comparison, top right for synchrotron peak flux comparison, bottom left for IC peak frequency comparison, and bottom right for IC peak flux comparison. The red lines are the best linear fits. The Pearson test shows that they are all tightly correlated, with Pearson's probability for a null correlation negligible ($p=1.14\times10^{-24}$, $3.81\times10^{-19}$, $2.24\times10^{-18}$, and $4.23\times10^{-18}$ for synchrotron peak frequency comparison, synchrotron peak flux comparison, IC peak frequency comparison, and IC peak flux comparison, respectively). Therefore, our log-parabolic fittings present similar parameter values as that of \citet[][]{2010ApJ...716...30A}.

The synchrotron fitting curves of total 48 blazars are plotted as red solid lines in Figure \ref{sed} in Appendix. We check all the fitted SEDs (one by one) by naked eyes. From these Figures, it can be seen that the X-ray emissions of 14 blazars belong to the synchrotron components (see Figure \ref{sed} and Table \ref{tabledata} for the name list). The IC components of these blazars cover only $\gamma$-ray band, and therefore the IC fitted parameters may be unreliable because of the narrow SED coverage. These 14 IC fitted curves are plotted as red dashed lines in the Figure \ref{sed}. For blazars J0238.6+1636, J0538.8-4403, J2254.0+1609, and J2345.5-1559, it can be seen that the fitted spectra at \emph{Fermi} $\gamma$-ray band are much harder than the observed spectra. These 4 IC fitted curves are plotted as red dotted lines. For blazars J1719.3+1746 and 2202.4+4217, there are no observational data at the right wing of IC component, and therefore the fitted curvatures ($b_{IC}$) may be unreliable. In fact, the value of $b_{IC}$ of these two blazars are significantly smaller than that of others (see Table \ref{tabledata}). These two IC fitted curves are also plotted as red dashed lines. Because of above reasons, we exclude these 20 ($=14+4+2$) blazars in the following analysis concerning the IC component. The IC fitted curves of remaining 28 blazars are presented as red solid lines. Among these 48 blazars, 43 sources have measured redshifts (see Table \ref{tabledata}). The peak luminosities and frequencies (in AGN frame) can be calculated through $\left(\nu_{p}L_{\nu_{p}}\right)_{s,C}=4\pi d_{L}^{2}\left(\nu_{p}f_{\nu_{p}}\right)_{s,C}$ and $\left(\nu_{p}\right)_{s,C}=\left(1+z\right)\left(\nu_{p}\right)_{s,C}^{obs}$, where $d_{L}$ is the luminosity distance and $z$ is the source redshift. In the following analysis, all values (e.g., peak frequency and curvature) are indicated in AGN frame, noticing that the curvature $b$ is unchanged when transform from observational to AGN frames.

\section{The Results}
\label{TheResults}

\begin{figure}
\epsscale{0.8} \plotone{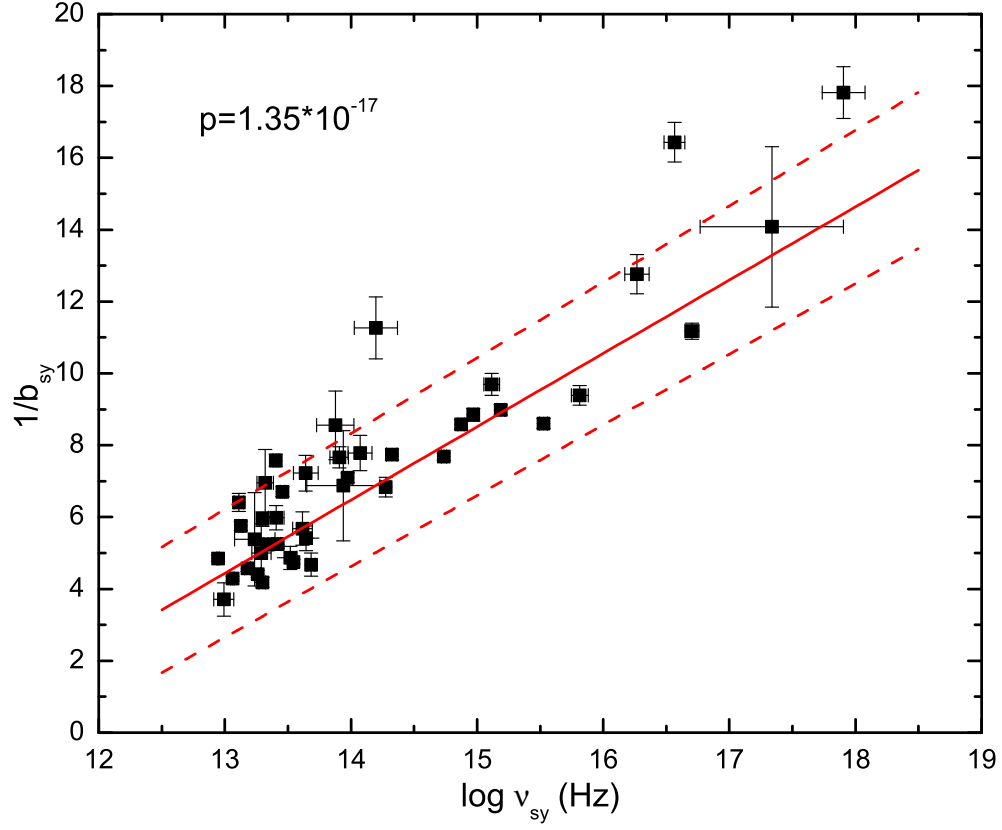} \caption{The synchrotron peak frequency versus curvature (in $1/b_{sy}$). The squares denote the total 43 sources with measured redshift. The Pearson test shows a significant correlation with a $p$-value, $p=1.35\times10^{-17}$. The red solid line is the best linear fitting, which gives $1/b_{sy}=-(22.08\pm0.43)+(2.04\pm0.03)\log\nu_{p}^{sy}$ and the dashed red lines indicate 1$\sigma$ confidence bands.}
\label{synfrecurve}
\end{figure}

From Table \ref{tabledata}, we can see that the value of synchrotron curvature $b_{sy}$ varies between 0.056 and 0.27. The range of the value is larger than that of any previous studies \citep[see,][]{1986ApJ...308...78L, 2004A&A...413..489M, 2004A&A...422..103M, 2008A&A...478..395M, 2011MNRAS.417.1881R, 2011ApJ...739...73M}. We plot the synchrotron curvature versus peak frequency in Figure \ref{synfrecurve}. Here we use $1/b_{sy}$ instead of $b_{sy}$ to represent the synchrotron curvature, since it will be convenient to compare with the theoretical results (see Section \ref{TheoryoftheCorrelation}). The black squares denote these 43 balzars having measured redshift. The Pearson test presents a small $p$-value, $p=1.35\times10^{-17}$. The red solid line is the best linear fitting, which gives $1/b_{sy}=-(22.08\pm0.43)+(2.04\pm0.03)\log\nu_{p}^{sy}$ and the dashed red lines indicate the 1$\sigma$ confidence bands. Hence, the synchrotron peak frequency correlates with its curvature at a high level of confidence.

\begin{figure}
\epsscale{0.8} \plotone{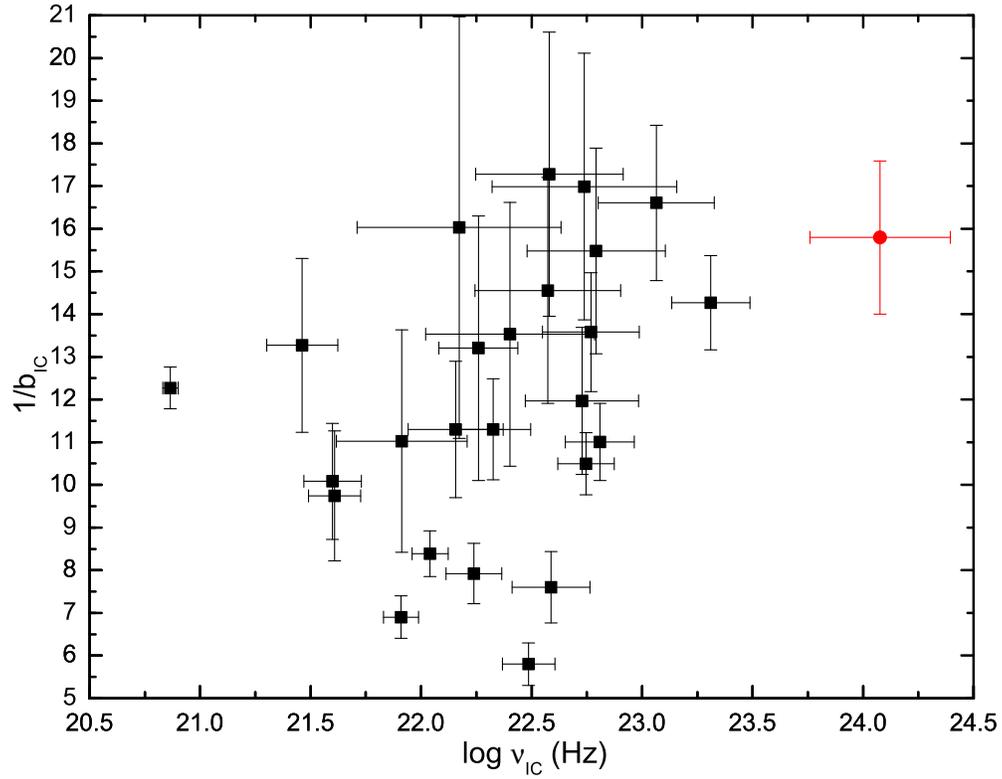} \caption{The IC peak frequency versus curvature (in $1/b_{IC}$). The black squares denote the total 26 blazars having measured redshift and reliable curvature estimated. The Pearson test shows a very weak correlation, with a $p$-value $p=5.16\times10^{-2}$, which is mainly contributed by the object J1504.4+1030 (the red point).}
\label{compfrecurve}
\end{figure}

Among these 28 blazars whose IC curvatures are well estimated, 26 blazars have measured redshift. Similar to the synchrotron one, we plot the IC peak frequency versus IC curvature in Figure \ref{compfrecurve}. The black squares denote these 26 blazars. The Pearson test shows a very weak correlation, with $p$-value $p=5.16\times10^{-2}$, which is mainly contributed by the object J1504.4+1030 (the red point in the Figure \ref{compfrecurve}, through jackknife statistical test\footnote{We estimate the $p$-value for each subsample by omitting the $i$th object. And it is found that the $p$-value reaches maximum, $p=0.118$, when omitting the object J1504.4+1030.}).

\section{Discussion and Conclusions}
\label{DiscussionandConclusions}

Blazars are observationally subdivided into flat spectrum radio quasars (FSRQs) and BL Lacertae objects (BL Lacs), based on presence or absence of emission lines \citep[see,][]{1997A&A...325..109S}. Our sample includes 48 blazars, in which 43 sources have measured redshift. The sample is not large enough to separate them into various subclasses, e.g., FSRQs versus BL Lacs. FSRQs and BL Lacs show continuous properties, although they are divided by some criterions \citep[e.g., the equivalent width of emission line $\gtrless$5 {\AA}, the Eddington ratio $\dot{m}\sim0.01$, see][and references therein]{1997A&A...325..109S, 2009MNRAS.396L.105G,2009ApJ...694L.107X}. Nowadays when discuss \emph{Fermi}/LAT detected blazars, the terms Low Synchrotron Peaked blazars (LSP), Intermediate Synchrotron Peaked blazars (ISP) and High Synchrotron Peaked blazars (HSP) are often used instead of FSRQs and BL Lacs  \citep[see, e.g.,][]{2010ApJ...716...30A, 2010ApJ...715..429A}.

Because of possible importance of the correlation between the synchrotron peak frequency and curvature (see Section \ref{TheoryoftheCorrelation}), many works have studied the correlation \citep{2004A&A...413..489M, 2007A&A...467..501T, 2009A&A...501..879T, 2008A&A...478..395M, 1986ApJ...308...78L, 2011MNRAS.417.1881R}. These works are based on either the X-ray data of single sources or small samples using data from radio to optical. In this paper, we study the correlation using a sample of 48 blazars, whose broadband quasi-simultaneous SEDs are from radio to $\gamma$-ray. The correlation between the synchrotron peak frequency and curvature depicted in this work confirm the results of these previous works.

\begin{figure}
\epsscale{0.8} \plotone{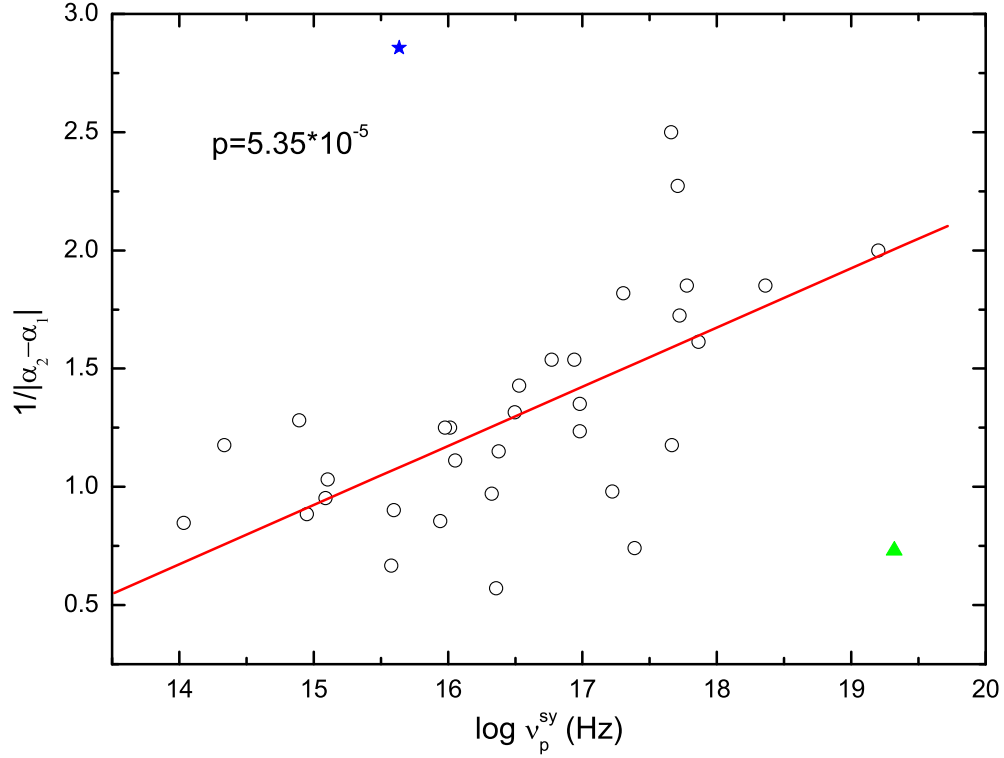} \caption{Synchrotron peak frequency versus the difference between spectral indexes above and below the peak (in $1/|\alpha_{2}-\alpha_{1}|$). Data are taken from \citet{2012ApJ...752..157Z}. The blue star and green triangle represent high states of PKS 2005-489 and Mrk 501. We exclude these two data points because of possibly unreliable values of parameter (see context for detail). The Pearson test presents a significant correlation with $p=5.35\times10^{-5}$.}
\label{a1a2}
\end{figure}

As an alternative to log-parabolic fitting, a broken power-law of electron energy distribution is commonly used to fit the SED in both synchrotron self-Compton (SSC) and external Compton (EC) models \citep[see e.g.,][and references therein]{1998ApJ...509..608T, 2002ApJ...564...86B, 2003ApJ...594L..27G, 2005A&A...432..401G, 2005ChJAS...5..207B, 2009ApJ...699...31A, 2010ApJ...721.1425A, 2012ApJ...748..119C}. Therefore, if the above correlation between the peak frequency and curvature is genuine, there should be a similar correlation even a broken power-law is employed to fit SED. From electron energy distribution to blazar SED, the spectral indexes transform as $\alpha_{1,2}=(p_{1,2}-1)/2$ approximately, where $p_{1}$ and $p_{2}$ are electron energy indexes below and above the break. In this case, the difference between the spectral indexes (i.e., $|\alpha_{2}-\alpha_{1}|=|p_{2}-p_{1}|/2$) can be considered as a ``surrogate" of the curvature ($b_{sy}$), and one expects a correlation between synchrotron peak frequency and this ``curvature". We collect the data of 24 blazars (at both high and low states) from \citet{2012ApJ...752..157Z} to test the correlation, which are plotted in Figure \ref{a1a2} (similar to $1/b_{sy}$, here we use $1/|\alpha_{2}-\alpha_{1}|$). The visual inspection shows a positive correlation between these two quantities, if excluding two obviously departure points (marked as green triangle and blue star). The green triangle represents high state of Mrk 501. From the Fig. 1 of \citet{2012ApJ...752..157Z}, it can be seen that there are no spectral data above the synchrotron peak to constrain the spectral index ($\alpha_{2}$/$p_{2}$), and
the index adopted by the authors is significantly larger than that at lower state \citep[$p_{2}^{\rm high}=4.6$ versus $p_{2}^{\rm low}=3.72$, see Table 1 of][]{2012ApJ...752..157Z}. The blue star represents the high state of PKS 2005-489, whose SED fitting is not very good: the value of synchrotron peak may be underestimated and the value of spectral index below the synchrotron peak may be overestimated \citep[see the Fig. 1 of][]{2012ApJ...752..157Z}. For these reasons, we exclude these two points. The Pearson's test yields a significant correlation with $p=5.35\times10^{-5}$. Therefore, even a broken power-law is employed to fit SED, there is also a correlation between the ``curvature" and the synchrotron peak frequency. Although there are different mathematic functions to describe blazar SED (log-parabolic law versus broken power-law), these two correlations represent an identical property of blazar SED.

\begin{figure}
\epsscale{0.8} \plotone{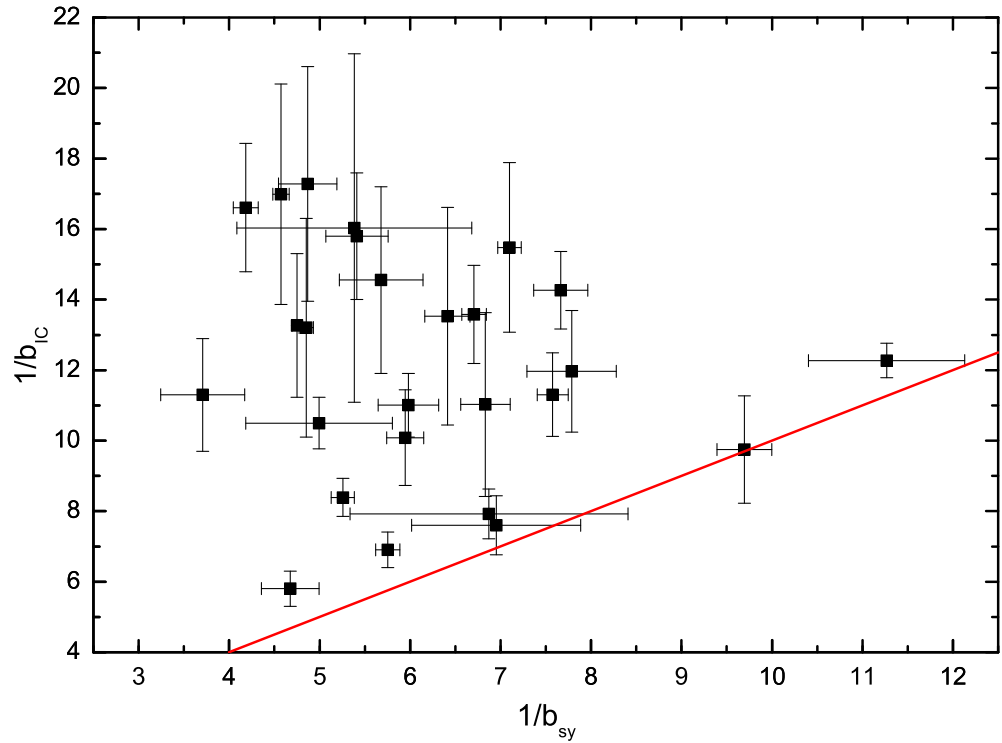} \caption{The synchrotron curvature versus the IC curvature. The red solid line shows a perfect one-to-one relation. It can be seen that the synchrotron curvature ($b_{sy}$) is on average larger than that of IC component ($b_{IC}$), and there is no correlation between these two parameters. See context for detail.}
\label{sycurviccurv}
\end{figure}

Based on SSC model, \citet{2009A&A...504..821P} predicted a relation between curvatures of IC and synchrotron components, $b_{\rm SSC}\approx b_{\rm sy}/2$ at Thomson regime, while $b_{\rm SSC}\approx5b_{\rm sy}$ at Klein-Nishina (KN) regime \citep[see also,][]{2009A&A...508L..31P}. The SSC peak frequency is $\nu_{p}^{SSC}\propto\nu_{p}\gamma_{p}^{2}\propto\gamma_{p}^{4}$ for Thomson scattering, and $\nu_{p}^{SSC}\propto\gamma_{p}$ for KN scattering, respectively. As discussed in Section \ref{TheoryoftheCorrelation} \citep[see also,][]{2007A&A...466..521T, 2009A&A...508L..31P, 2009A&A...504..821P}, the peak energy of electron ($\gamma_{p}$) correlates with the synchrotron curvature ($b_{sy}$). We therefore expect a correlation between SSC peak frequency and curvature at Thomson or KN regimes. For the first time, we study the relation between the IC peak frequency and curvature from observational view, while our results present no significant correlation with a chance probability only $p=0.0516$. This is due to the following reasons:
\begin{itemize}
  \item The IC component is sometimes a composite of SSC and EC emissions.
  \item The IC emission may lie at Thomson regime for some blazars, while others are at KN regime.
  \item The intrinsic spectrum of a single IC scattering is broader than the intrinsic synchrotron spectrum of a single electron \citep[see, Figures 6.6 versus 7.3 in][]{1979rpa..book.....R}, and the SED of seed photons are usually very broad.
\end{itemize}
These reasons will significantly broaden the SED of IC component, and the value of IC curvature is more uncertain relative to the synchrotron one. We note that these reasons lead to another two predictions. Firstly, the synchrotron curvature ($b_{sy}$) would be on average larger than that of IC component ($b_{IC}$) at Thomson regime. Secondly, there would be very weak or no correlation between the two curvatures. These two predictions are confirmed by Figure \ref{sycurviccurv}, which shows IC curvature versus synchrotron curvature (represented by $1/b_{IC}$ versus $1/b_{sy}$), and the red solid line shows a perfect one-to-one relation.

As suggested in Section \ref{TheoryoftheCorrelation}, two scenarios can explain the observed correlation between synchrotron peak frequency and curvature. Theoretical predictions of the slope $B$ (in $1/b_{sy}=A+B\log\nu_{p}$) are $B=$5/2, 10/3, and 2 for models of energy-dependent acceleration probability, fluctuation of fractional acceleration gain, and stochastic acceleration, respectively. Our observational result $B=2.04\pm0.03$ (see Section \ref{TheResults}) is consistent with the stochastic acceleration mechanisms and emission processes.

Some of other evidences seem to support that the electron energy distribution may be log-parabolic. Taking the synchrotron radiation and a three-dimensional turbulent electromagnetic field configuration into account, \citet{2004A&A...423...13N} presented numerical simulations of particle acceleration. They found, in a few cases, energy spectral index characterized by a steepening spectra at high energies. \citet{2006A&A...448..861M} verified that these spectra can be represented well by a log-parabolic law or by a combination of a power law and a parabola. \citet{2013ApJ...771L...4C} presented a detailed SED modeling of 3C 454.3 and finds that its GeV $\gamma$-ray break could be well reproduced if electron energy distribution is log-parabolic. In order to fit the SED of MeV blazars, \citet{2002ApJ...577...78S} assumed that electrons are accelerated via a two-step process with a broken power-law energy distribution as injection. Taking the cooling effect into account, \citet{2006A&A...448..861M} verified that these resulting electron spectra can be well described by a log-parabola over a range wider than three decades. The log-parabolic law of synchrotron component of SED are also found to be consistent with some empirical relations/correlations (see Appendix for detail).

\section{Summary}
\label{Summary}
The curvature, in addition to peak frequency and peak flux, is the third important parameter to characterize a broadband SED, which may shine out the hidden electron energy distribution, particle acceleration mechanism, energy dissipation mechanism, and many other blazar properties. A larger sample of blazars with high quality SED is needed to check these results in detail, especially for the IC components. We summarize the main results of this paper as follows.
\begin{itemize}
  \item We found a significant correlation between the curvature (in $1/b_{sy}$) and peak frequency for synchrotron component and no significant correlation between same quantities for IC component. It is also found that the synchrotron curvatures are on average larger than that of IC curvatures, and there is no correlation between the two parameters. This may be caused by complicated seed photon field.
  \item The difference between spectral indexes above and below the SED peak of 24 blazars (at both high and low states) are calculated \citep[i.e., $|\alpha_{2}-\alpha_{1}|$, data from][]{2012ApJ...752..157Z}. We found a significant correlation between $1/|\alpha_{2}-\alpha_{1}|$ and the synchrotron peak frequency. Parameter $|\alpha_{2}-\alpha_{1}|$ can be considered as a ``surrogate" of the curvature. Therefore, this result confirms the correlation between the synchrotron peak frequency and curvature, even provided a broken power-law fit of SED.
  \item We found that the slop of correlation between synchrotron curvature of peak frequency ($B=2.04\pm0.03$) is consistent with the prediction of stochastic scenario.
  \item Some of other evidences also seem to support that the electron energy distribution (and/or synchrotron SED) may be log-parabolic, which include SED modeling, particle acceleration simulation, and comparisons between some predictions and empirical relations/correlations.
\end{itemize}

\acknowledgments

We thank the anonymous referee for insightful comments and constructive suggestions. We are grateful to the help from Xinwu Cao, Jinming Bai, Jin Zhang and Wentao Luo. This work is supported by the NSFC (grants 11233006, 11133006, 11173043, 11103054 and 11103060), and XTP project XDA04060604.

\begin{deluxetable}{llllllllllll}
\tabletypesize{\tiny}
\setlength{\tabcolsep}{0.04in}
\tablecolumns{12}
\tablewidth{0pc}
\tablecaption{The SED Fitting Parameters
\label{tabledata}
}
\tablehead{
\colhead{Name(0FGL)}&
\colhead{redshift}&
\colhead{$\nu_{p}^{sy}$}&
\colhead{$(\nu f_{\nu})_{p}^{sy}$}&
\colhead{$b_{sy}$}&
\colhead{dof$_{sy}$}&
\colhead{$\chi^{2}_{sy.r}$}&
\colhead{$\nu_{p}^{IC}$}&
\colhead{$(\nu f_{\nu})_{p}^{IC}$}&
\colhead{$b_{IC}$}&
\colhead{dof$_{IC}$}&
\colhead{$\chi^{2}_{IC.r}$}\\
\colhead{(1)}&
\colhead{(2)}&
\colhead{(3)}&
\colhead{(4)}&
\colhead{(5)}&
\colhead{(6)}&
\colhead{(7)}&
\colhead{(8)}&
\colhead{(9)}&
\colhead{(10)}&
\colhead{(11)}&
\colhead{(12)}
}
\startdata
J0033.6-1921*  &  0.610  &  15.607$\pm$ 0.066  &  -11.160$\pm$  0.029  &  0.106$\pm$0.003  &  11  &   0.4145  &  24.265$\pm$ 0.335*  &  -10.955$\pm$  0.120*  &  0.480$\pm$0.297*  &   1  &   0.7803\\
J0050.5-0928*  &  ---    &  14.626$\pm$ 0.059  &  -10.842$\pm$  0.063  &  0.101$\pm$0.005  &  11  &   2.6429  &  22.962$\pm$ 0.448*  &  -10.578$\pm$  0.043*  &  0.178$\pm$0.169*  &   1  &   0.0001\\
J0137.1+4751   &  0.859  &  13.347$\pm$ 0.078  &  -10.502$\pm$  0.083  &  0.176$\pm$0.014  &   8  &   0.6019  &  22.304$\pm$ 0.330   &  -10.550$\pm$  0.075   &  0.069$\pm$0.012   &   7  &   1.7003\\
J0210.8-5100   &  1.003  &  13.156$\pm$ 0.019  &  -10.965$\pm$  0.017  &  0.149$\pm$0.003  &  35  &  17.7985  &  22.467$\pm$ 0.219   &  -10.232$\pm$  0.035   &  0.074$\pm$0.008   &  11  &   4.8359\\
J0222.6+4302*  &  0.444  &  14.714$\pm$ 0.025  &  -10.400$\pm$  0.018  &  0.117$\pm$0.002  &  40  &   2.6078  &  23.682$\pm$ 0.038*  &  -10.147$\pm$  0.028*  &  0.237$\pm$0.013*  &  13  &   4.3426\\
J0229.5-3640   &  2.115  &  13.446$\pm$ 0.290  &  -11.928$\pm$  0.206  &  0.146$\pm$0.033  &   7  &   0.8482  &  21.745$\pm$ 0.126   &  -10.332$\pm$  0.069   &  0.126$\pm$0.011   &   8  &   2.1163\\
J0238.4+2855   &  1.213  &  12.765$\pm$ 0.037  &  -11.285$\pm$  0.043  &  0.156$\pm$0.006  &  12  &   1.1031  &  22.058$\pm$ 0.382   &  -10.737$\pm$  0.097   &  0.074$\pm$0.017   &   5  &   1.9191\\
J0238.6+1636$\dag$  &  0.940  &  12.973$\pm$ 0.021  &  -10.421$\pm$  0.026  &  0.227$\pm$0.005  &  47  &  10.7721  &  24.604$\pm$ 1.085$\dag$  &   -9.899$\pm$  0.074$\dag$  &  0.031$\pm$0.009$\dag$  &  16  &  34.6859\\
J0349.8-2102   &  2.944  &  13.087$\pm$ 0.032  &  -11.055$\pm$  0.111  &  0.214$\pm$0.015  &   8  &   0.4579  &  21.891$\pm$ 0.119   &  -10.205$\pm$  0.061   &  0.172$\pm$0.015   &   2  &   0.4180\\
J0423.1-0112   &  0.915  &  13.017$\pm$ 0.023  &  -10.893$\pm$  0.028  &  0.168$\pm$0.006  &  35  &   1.0149  &  21.317$\pm$ 0.130   &  -10.368$\pm$  0.096   &  0.099$\pm$0.013   &   7  &   2.6917\\
J0428.7-3755   &  1.112  &  13.582$\pm$ 0.073  &  -11.211$\pm$  0.030  &  0.130$\pm$0.005  &  20  &   2.7216  &  22.986$\pm$ 0.177   &  -10.223$\pm$  0.024   &  0.070$\pm$0.005   &  12  &   6.1359\\
J0449.7-4348*  &  0.205  &  15.105$\pm$ 0.032  &  -10.375$\pm$  0.020  &  0.111$\pm$0.002  &  24  &   9.9664  &  24.025$\pm$ 0.895*  &  -10.531$\pm$  0.045*  &  0.037$\pm$0.011*  &   4  &   1.2272\\
J0457.1-2325   &  1.003  &  13.104$\pm$ 0.062  &  -10.992$\pm$  0.068  &  0.167$\pm$0.009  &  10  &   1.1554  &  22.507$\pm$ 0.156   &  -10.091$\pm$  0.058   &  0.091$\pm$0.007   &  10  &   5.7873\\
J0507.9+6739*  &  0.416  &  17.186$\pm$ 0.568  &  -10.783$\pm$  0.078  &  0.071$\pm$0.011  &  17  &   0.0690  &  25.738$\pm$ 0.735*  &  -10.607$\pm$  0.144*  &  0.088$\pm$0.054*  &   5  &   2.1029\\
J0516.2-6200   &  ---    &  13.550$\pm$ 0.177  &  -11.515$\pm$  0.107  &  0.141$\pm$0.017  &  14  &   0.4809  &  22.612$\pm$ 0.298   &  -10.726$\pm$  0.024   &  0.067$\pm$0.008   &  20  &   4.6043\\
J0531.0+1331   &  2.070  &  12.640$\pm$ 0.015  &  -11.295$\pm$  0.021  &  0.174$\pm$0.004  &  39  &   7.5104  &  21.422$\pm$ 0.079   &   -9.953$\pm$  0.066   &  0.145$\pm$0.011   &  10  &   4.8000\\
J0538.8-4403$\dag$  &  0.892  &  13.062$\pm$ 0.033  &  -10.400$\pm$  0.042  &  0.191$\pm$0.006  &  18  &   1.8883  &  24.520$\pm$ 0.919$\dag$  &  -10.164$\pm$  0.053$\dag$  &  0.032$\pm$0.008$\dag$  &  17  &   1.2905\\
J0712.9+5034   &  ---    &  13.557$\pm$ 0.647  &  -11.213$\pm$  0.477  &  0.155$\pm$0.101  &   3  &   0.0340  &  23.365$\pm$ 1.375   &  -11.000$\pm$  0.069   &  0.052$\pm$0.025   &   4  &   0.7682\\
J0722.0+7120*  &  0.310  &  14.620$\pm$ 0.021  &   -9.978$\pm$  0.016  &  0.130$\pm$0.002  &  62  &   5.8716  &  23.003$\pm$ 0.353*  &  -10.380$\pm$  0.025*  &  0.046$\pm$0.007*  &   3  &   1.2265\\
J0730.4-1142   &  1.589  &  12.876$\pm$ 0.076  &  -10.572$\pm$  0.226  &  0.200$\pm$0.033  &   6  &  17.0701  &  22.333$\pm$ 0.128   &  -10.104$\pm$  0.036   &  0.095$\pm$0.007   &  10  &   6.9697\\
J0855.4+2009   &  0.306  &  13.429$\pm$ 0.014  &  -10.148$\pm$  0.011  &  0.210$\pm$0.003  &  38  &  16.4465  &  21.346$\pm$ 0.161   &  -10.482$\pm$  0.086   &  0.075$\pm$0.012   &  13  &   1.2470\\
J0921.2+4437   &  2.190  &  13.016$\pm$ 0.051  &  -10.926$\pm$  0.082  &  0.205$\pm$0.014  &   5  &   2.2666  &  22.077$\pm$ 0.334   &  -10.734$\pm$  0.064   &  0.058$\pm$0.011   &   9  &   1.8761\\
J1015.2+4927*  &  0.212  &  16.183$\pm$ 0.097  &  -10.580$\pm$  0.024  &  0.078$\pm$0.003  &  17  &   3.3939  &  24.676$\pm$ 0.280*  &  -10.610$\pm$  0.071*  &  0.067$\pm$0.025*  &   4  &   2.0654\\
J1058.9+5629   &  0.143  &  15.057$\pm$ 0.063  &  -10.884$\pm$  0.040  &  0.103$\pm$0.003  &  11  &   1.8680  &  21.551$\pm$ 0.118   &  -10.692$\pm$  0.059   &  0.103$\pm$0.016   &  18  &   2.3297\\
J1057.8+0138   &  0.888  &  12.961$\pm$ 0.160  &  -10.813$\pm$  0.302  &  0.186$\pm$0.045  &   5  &   0.1860  &  21.896$\pm$ 0.462   &  -10.819$\pm$  0.136   &  0.062$\pm$0.019   &   5  &   0.9898\\
J1104.5+3811*  &  0.030  &  16.689$\pm$ 0.056  &   -9.299$\pm$  0.021  &  0.089$\pm$0.002  &  47  &  25.3150  &  24.249$\pm$ 0.057*  &  -10.066$\pm$  0.033*  &  0.147$\pm$0.011*  &  13  &   3.1334\\
J1159.2+2912   &  0.729  &  12.942$\pm$ 0.023  &  -10.712$\pm$  0.027  &  0.219$\pm$0.004  &  16  &   6.4871  &  22.501$\pm$ 0.418   &  -10.737$\pm$  0.059   &  0.059$\pm$0.011   &  11  &   2.7382\\
J1221.7+2814*  &  0.102  &  14.927$\pm$ 0.034  &  -10.675$\pm$  0.023  &  0.113$\pm$0.002  &  33  &  11.1198  &  24.312$\pm$ 0.089*  &  -10.458$\pm$  0.057*  &  0.156$\pm$0.031*  &  15  &   4.6653\\
J1229.1+0202   &  0.158  &  14.134$\pm$ 0.172  &   -9.990$\pm$  0.034  &  0.089$\pm$0.007  &  18  &   1.9384  &  20.802$\pm$ 0.035   &   -9.579$\pm$  0.023   &  0.081$\pm$0.003   &  26  &   2.1859\\
J1248.7+5811*  &  ---    &  14.590$\pm$ 0.043  &  -10.946$\pm$  0.038  &  0.136$\pm$0.004  &  10  &   1.0594  &  22.885$\pm$ 1.426*  &  -10.969$\pm$  0.094*  &  0.117$\pm$0.285*  &   1  &   0.2037\\
J1256.1-0548   &  0.536  &  12.760$\pm$ 0.008  &  -10.247$\pm$  0.019  &  0.206$\pm$0.003  &  30  &  13.6854  &  22.073$\pm$ 0.179   &  -10.208$\pm$  0.099   &  0.076$\pm$0.018   &  12  &   0.7205\\
J1310.6+3220   &  0.997  &  12.996$\pm$ 0.025  &  -10.659$\pm$  0.051  &  0.239$\pm$0.008  &  10  &   2.4370  &  22.764$\pm$ 0.263   &  -10.442$\pm$  0.023   &  0.060$\pm$0.007   &  14  &   3.5332\\
J1457.6-3538   &  1.424  &  13.688$\pm$ 0.095  &  -11.256$\pm$  0.044  &  0.128$\pm$0.008  &   9  &   5.0165  &  22.344$\pm$ 0.256   &  -10.095$\pm$  0.080   &  0.084$\pm$0.012   &   6  &   2.3088\\
J1504.4+1030   &  1.839  &  13.191$\pm$ 0.049  &  -10.863$\pm$  0.069  &  0.185$\pm$0.012  &  11  &   0.9404  &  23.624$\pm$ 0.318   &   -9.895$\pm$  0.032   &  0.063$\pm$0.007   &  12  &   5.2857\\
J1512.7-0905   &  0.360  &  13.271$\pm$ 0.024  &  -11.095$\pm$  0.013  &  0.132$\pm$0.003  &  30  &  17.7053  &  22.193$\pm$ 0.170   &   -9.999$\pm$  0.061   &  0.088$\pm$0.009   &  14  &   2.6933\\
J1522.2+3143   &  1.487  &  12.925$\pm$ 0.065  &  -12.078$\pm$  0.124  &  0.144$\pm$0.019  &  10  &   4.7695  &  22.192$\pm$ 0.176   &  -10.172$\pm$  0.097   &  0.132$\pm$0.014   &   3  &   2.8494\\
J1543.1+6130*  &  ---    &  14.404$\pm$ 0.042  &  -11.292$\pm$  0.048  &  0.137$\pm$0.004  &  13  &   5.0679  &  24.819$\pm$ 3.734*  &  -11.082$\pm$  0.287*  &  0.035$\pm$0.032*  &   2  &   0.0362\\
J1653.9+3946*  &  0.033  &  16.551$\pm$ 0.083  &  -10.317$\pm$  0.026  &  0.061$\pm$0.002  &  44  &   0.7394  &  25.033$\pm$ 0.220*  &  -10.674$\pm$  0.110*  &  0.097$\pm$0.041*  &   6  &   0.1482\\
J1719.3+1746*  &  0.137  &  13.585$\pm$ 0.098  &  -11.453$\pm$  0.057  &  0.138$\pm$0.010  &  10  &   0.5019  &  24.493$\pm$ 0.387*  &  -10.644$\pm$  0.033*  &  0.036$\pm$0.004*  &  10  &   2.7840\\
J1751.5+0935   &  0.322  &  12.871$\pm$ 0.079  &  -10.207$\pm$  0.224  &  0.270$\pm$0.034  &   5  &   0.6216  &  22.036$\pm$ 0.215   &  -10.324$\pm$  0.083   &  0.088$\pm$0.013   &   6  &   0.4860\\
J1849.4+6706   &  0.657  &  13.753$\pm$ 0.039  &  -10.938$\pm$  0.021  &  0.141$\pm$0.003  &  14  &  45.2431  &  22.573$\pm$ 0.313   &  -10.443$\pm$  0.055   &  0.065$\pm$0.010   &   8  &   0.9814\\
J2000.2+6506*  &  0.047  &  17.886$\pm$ 0.171  &  -10.154$\pm$  0.023  &  0.056$\pm$0.002  &  29  &   2.0107  &  24.492$\pm$ 0.103*  &  -10.629$\pm$  0.078*  &  0.177$\pm$0.045*  &   6  &   2.3534\\
J2143.2+1741   &  0.213  &  14.192$\pm$ 0.036  &  -10.661$\pm$  0.074  &  0.146$\pm$0.006  &   8  &   3.1776  &  21.829$\pm$ 0.296   &  -10.531$\pm$  0.127   &  0.091$\pm$0.021   &   3  &   0.1145\\
J2158.8-3014*  &  0.116  &  15.479$\pm$ 0.032  &   -9.874$\pm$  0.018  &  0.116$\pm$0.002  &  40  &   6.1386  &  23.789$\pm$ 0.057*  &  -10.167$\pm$  0.026*  &  0.203$\pm$0.013*  &  11  &   1.8782\\
J2202.4+4217*  &  0.069  &  14.295$\pm$ 0.042  &  -10.185$\pm$  0.009  &  0.129$\pm$0.003  &  40  &   9.5177  &  21.848$\pm$ 0.331*  &  -10.723$\pm$  0.055*  &  0.034$\pm$0.009*  &  23  &   1.4534\\
J2254.0+1609$\dag$  &  0.859  &  12.791$\pm$ 0.013  &   -9.966$\pm$  0.015  &  0.233$\pm$0.003  &  33  &  23.2350  &  22.698$\pm$ 0.155$\dag$  &   -9.674$\pm$  0.028$\dag$  &  0.048$\pm$0.004$\dag$  &  22  &   6.1458\\
J2327.3+0947   &  1.843  &  12.965$\pm$ 0.025  &  -11.191$\pm$  0.031  &  0.190$\pm$0.005  &   7  &  10.4132  &  21.587$\pm$ 0.082   &  -10.289$\pm$  0.045   &  0.119$\pm$0.008   &  15  &   1.6895\\
J2345.5-1559$\dag$  &  0.621  &  13.666$\pm$ 0.148  &  -11.803$\pm$  0.068  &  0.117$\pm$0.013  &   7  &   2.6984  &  22.993$\pm$ 0.536$\dag$  &  -10.717$\pm$  0.029$\dag$  &  0.069$\pm$0.014$\dag$  &   5  &   9.0312\\

\enddata
\tablecomments{Column (1) provides the LAT name of the source. Column (2) is the source redshift. Columns (3), (4), (5), (6), and (7) are the (for the synchrotron fitting) peak frequency, flux, curvature, the degree of freedom, and the reduced $\chi^{2}$, respectively. Columns (8), (9), (10), (11), and (12) denote the same quantities but for IC fitting. The primes `*' represent 14 blazars whose IC components cover only $\gamma$-ray band, and J1719.3+1746 and 2202.4+4217 whose fitting values of curvature $b_{IC}$ are obviously smaller than that of other blazars. The IC fitting curves of these 16 blazars are showed as red dashed lines in Figure \ref{sed} in Appendix. The prime `$\dag$' represent 4 blazars whose IC fitting curve significantly departure from the observational SED, especially inconsistent with the $\gamma$-ray spectral index. The IC fitting curves of these 4 blazars are presented as red dotted lines. See context for detail.
}
\end{deluxetable}

\appendix

\section{Figures}

\begin{figure*}
\centering
\includegraphics[height=4cm]{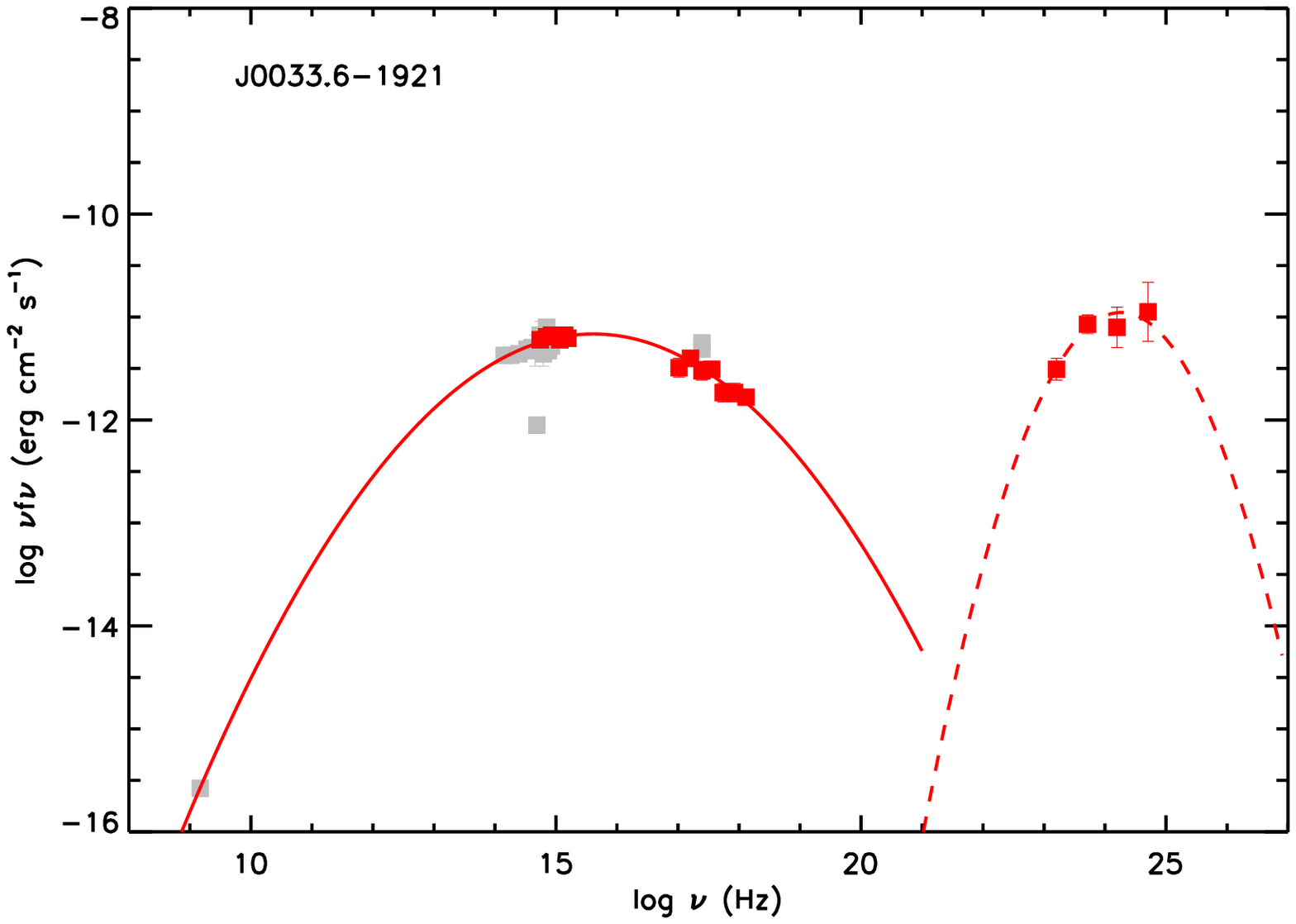}
\includegraphics[height=4cm]{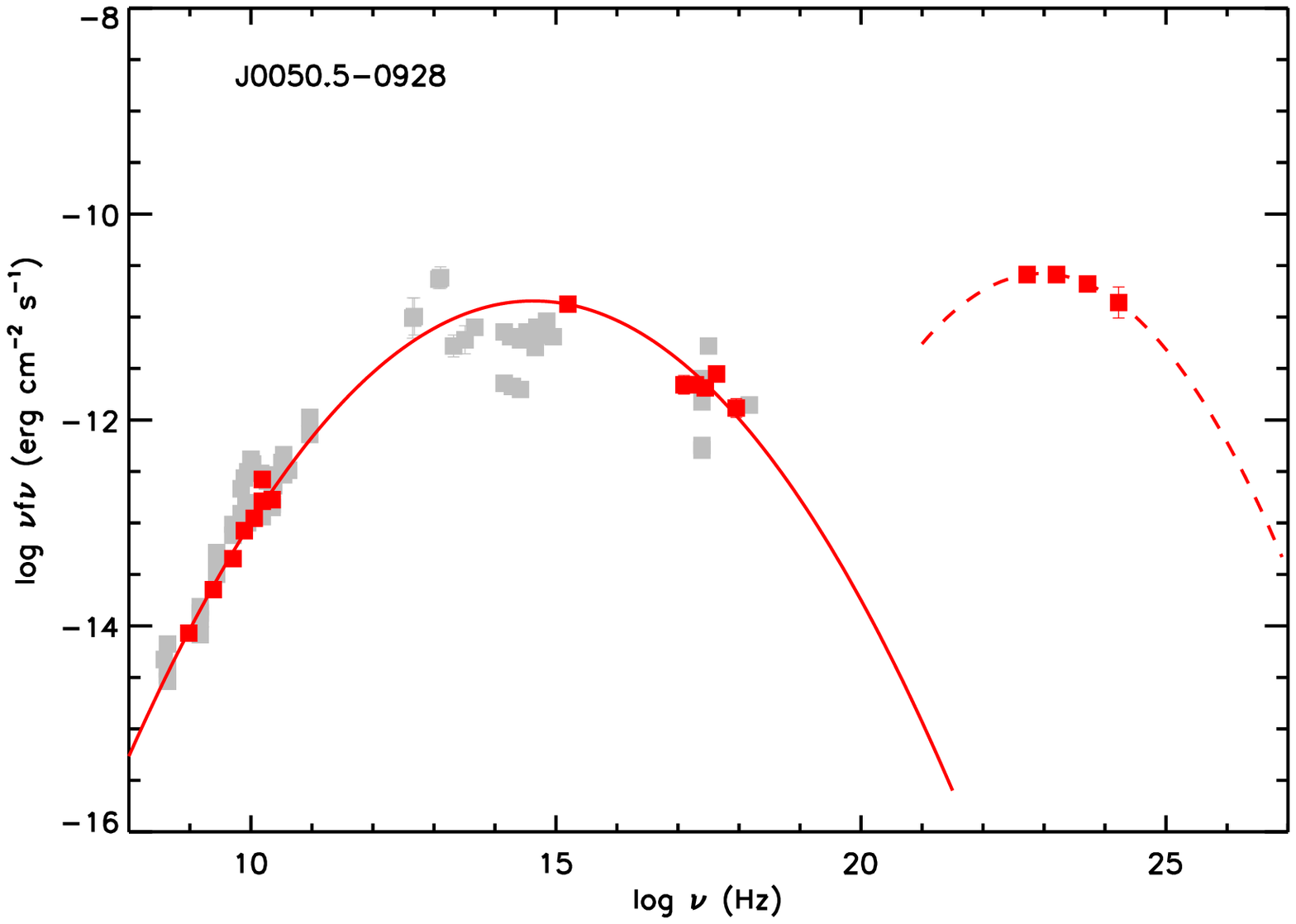}
\includegraphics[height=4cm]{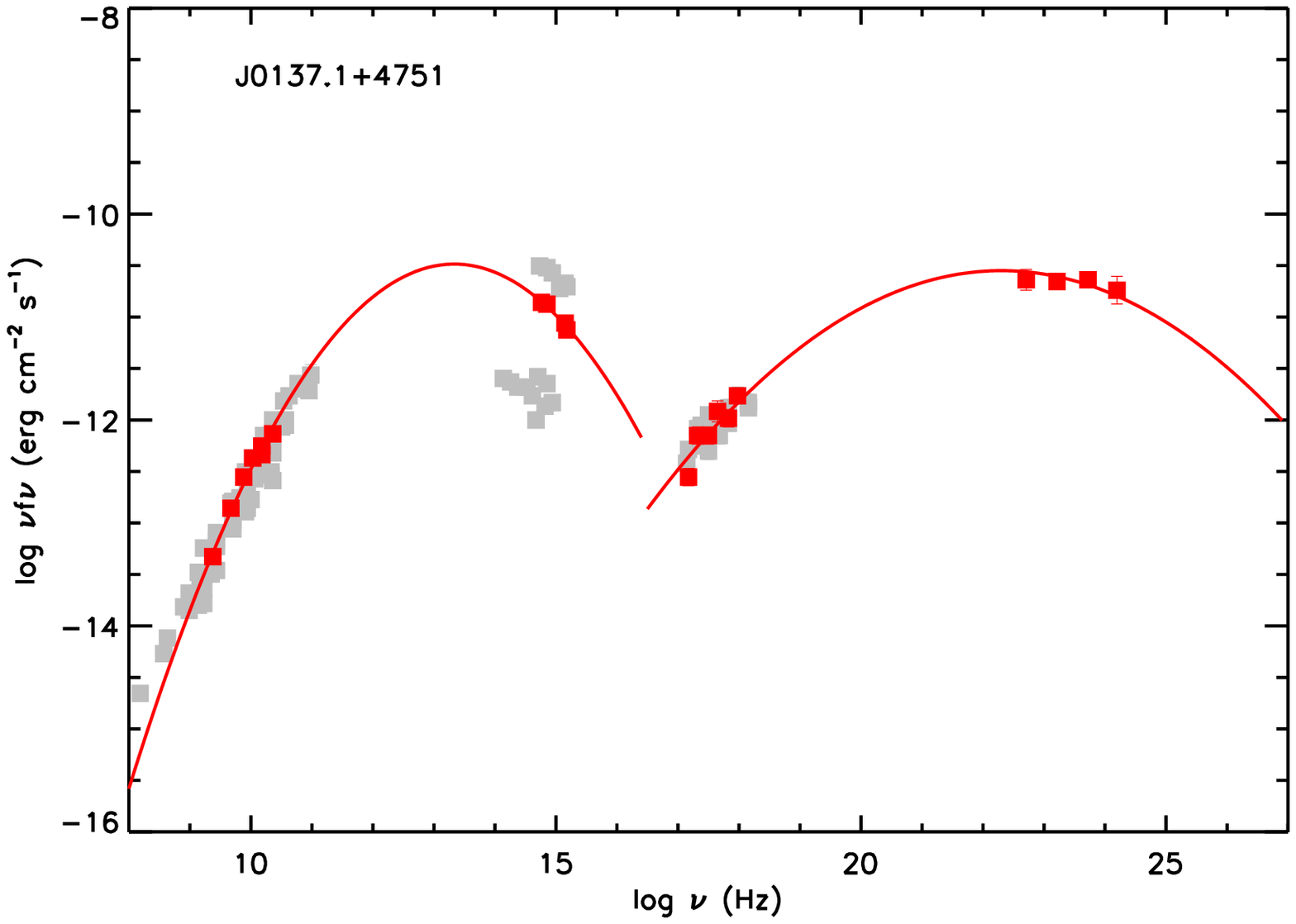}
\includegraphics[height=4cm]{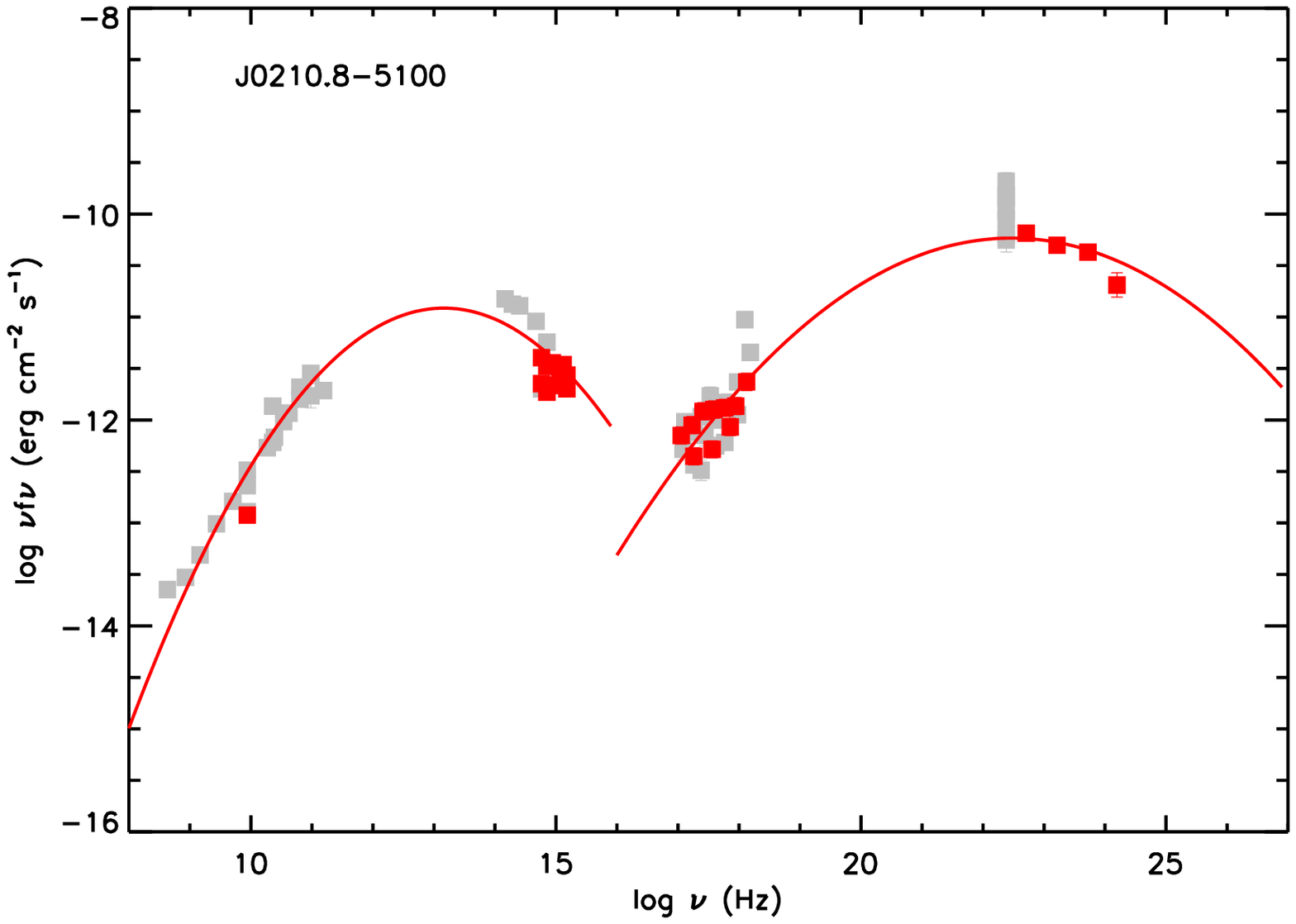}
\includegraphics[height=4cm]{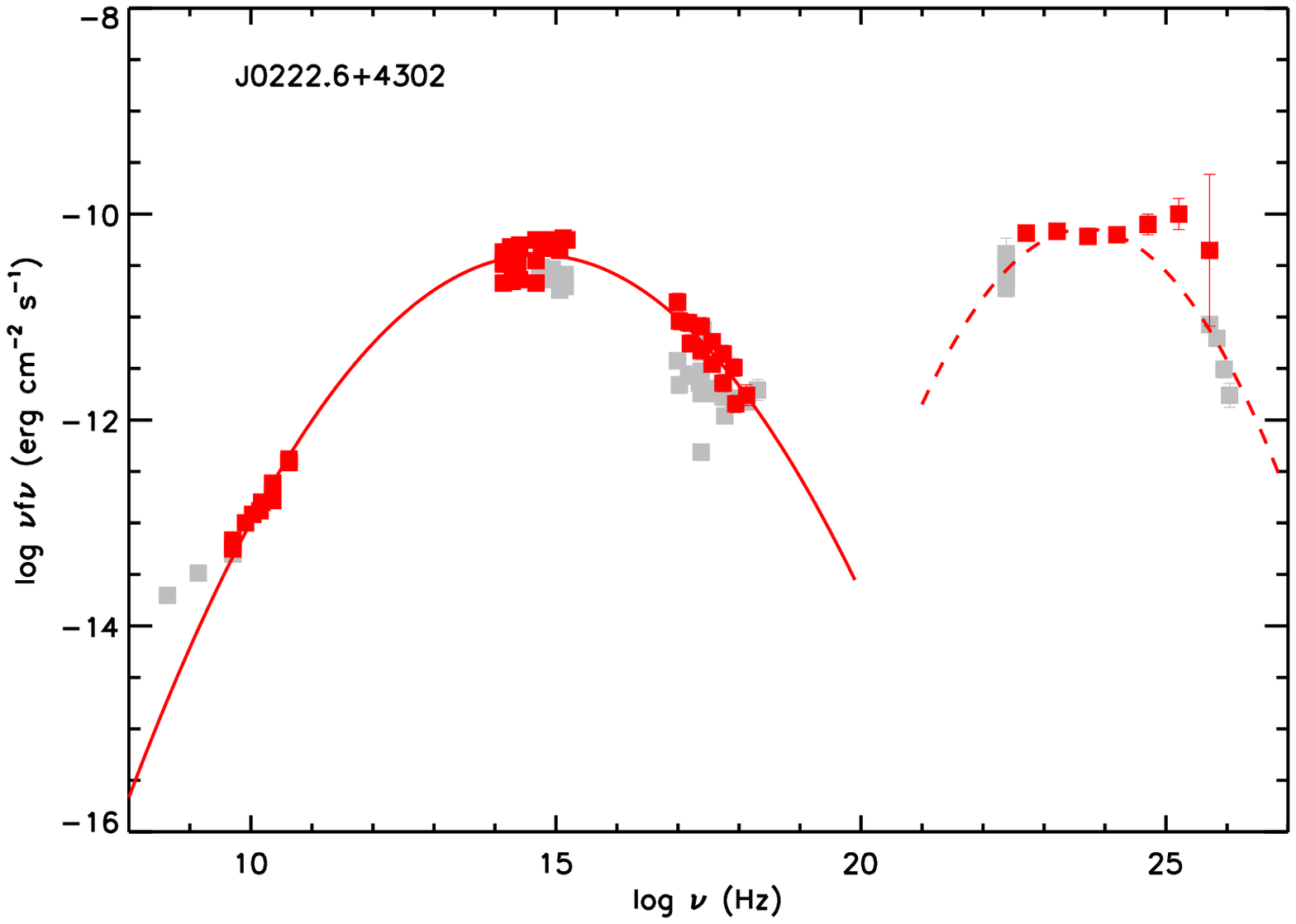}
\includegraphics[height=4cm]{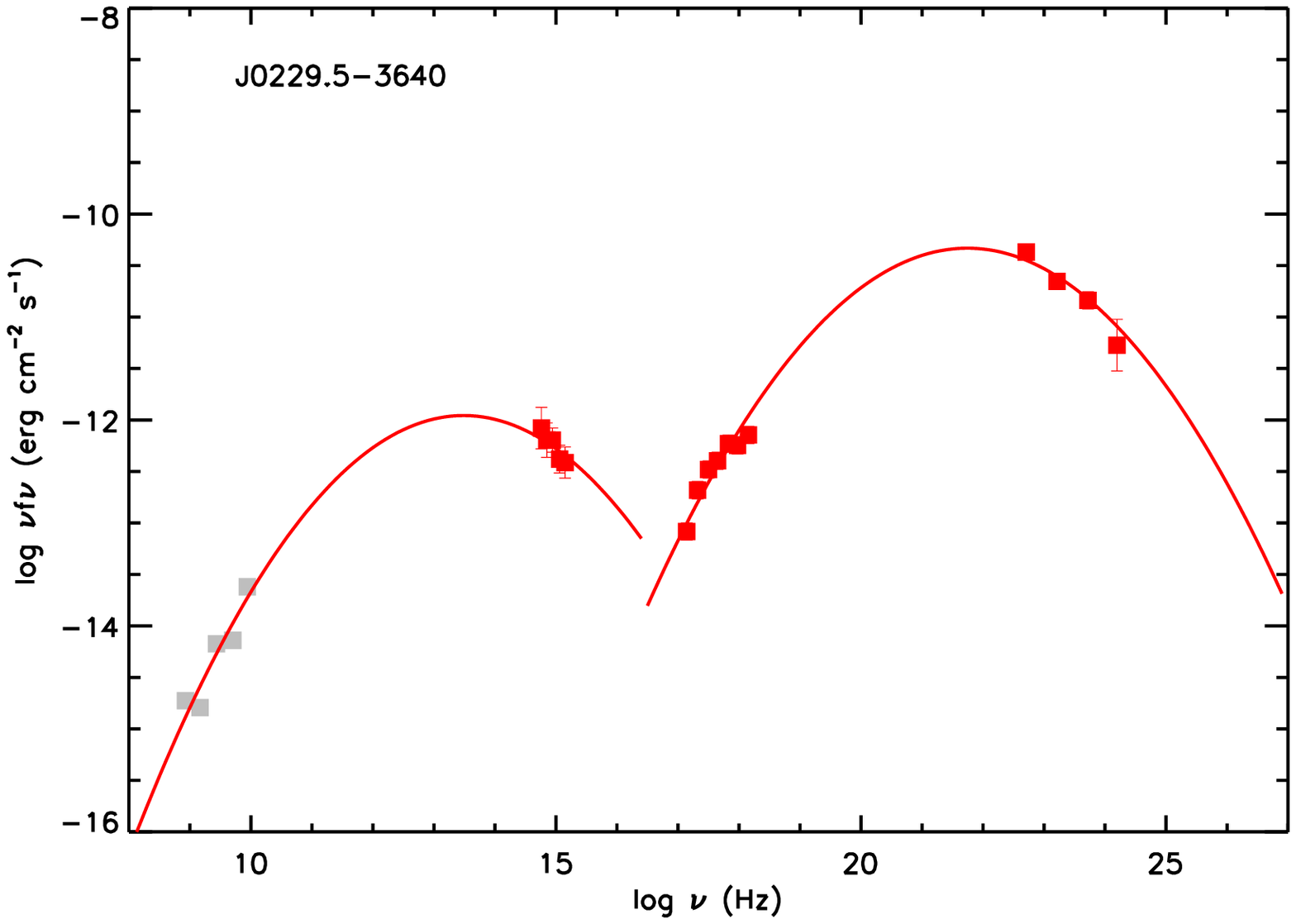}
\includegraphics[height=4cm]{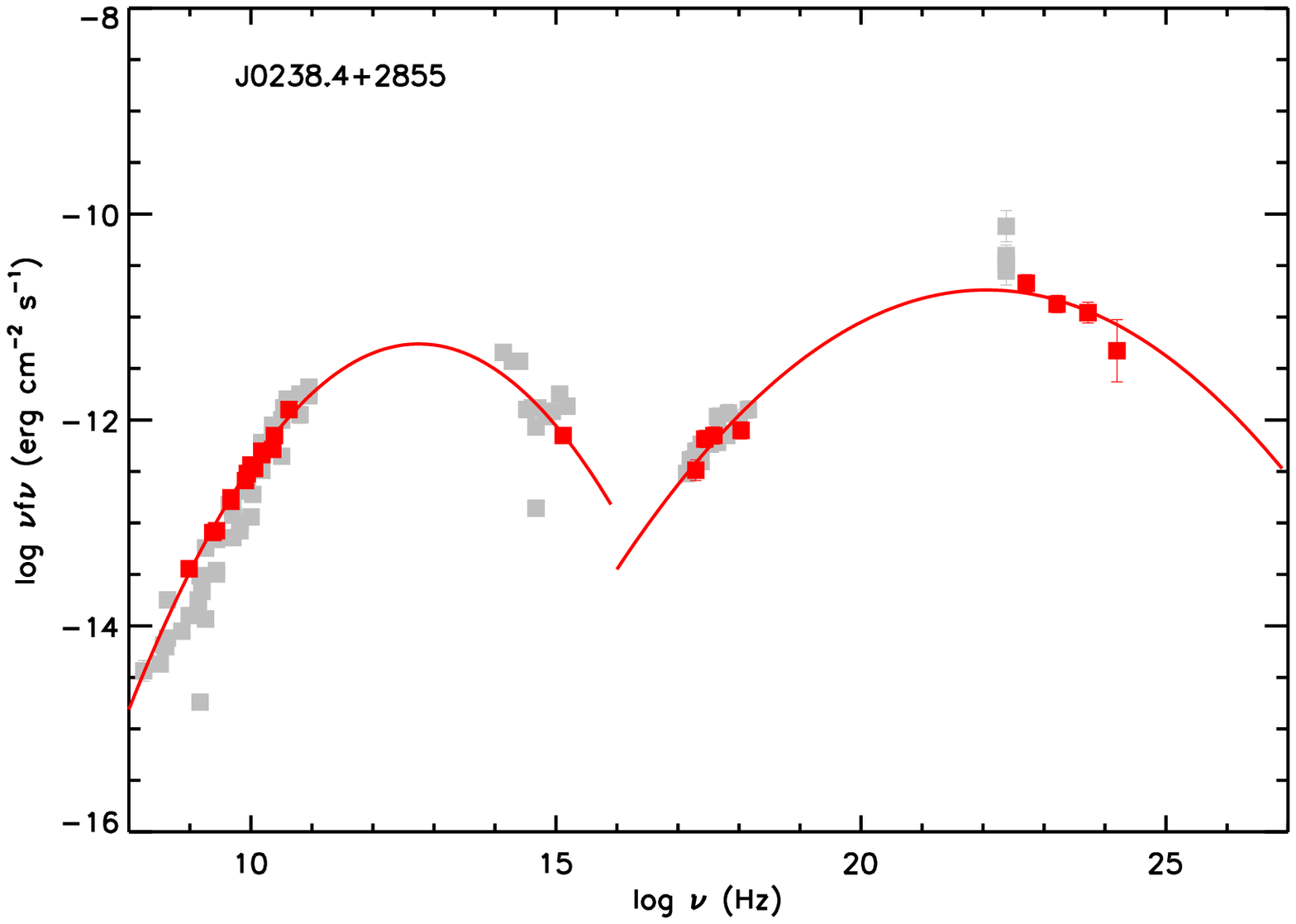}
\includegraphics[height=4cm]{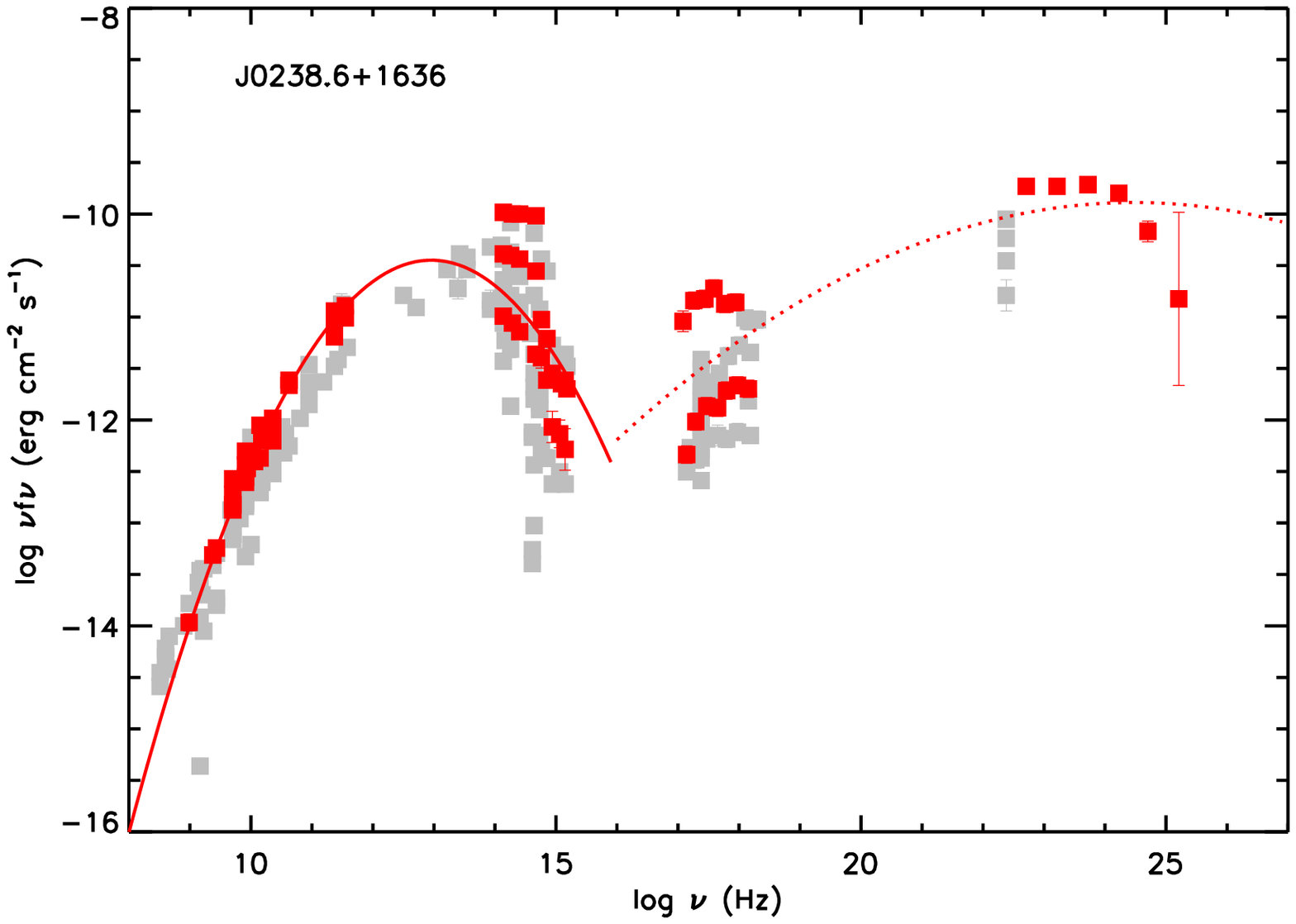}
\caption{The red points for the quasi-simultaneous data, while the grey ones represent other observations \citep[see,][for detail description]{2010ApJ...716...30A}. Both SEDs of synchrotron and IC components are fitted by a log-parabolic law, i.e., $\log\nu f_{\nu}=-b\left(\log\nu-\log\nu_{p}\right)^{2}+\log\nu_{p}f_{\nu}^{p}$. The IC component of 14 out of 48 blazars covers only $\gamma$-ray band (see Table \ref{tabledata} for name list), and the IC curvature of J1719.3+1746 and 2202.4+4217 are obviously smaller than that of other blazars. These 16 IC fitting curves are shown as red dashed lines. For J0238.6+1636, J0538.8-4403, J2254.0+1609, and J2345.5-1559, the IC fittings significantly departure from the observational SED, especially inconsistent with the $\gamma$-ray spectral index. These 4 IC fitting curves are presented as red dotted lines. All fitting parameters are presented in Table \ref{tabledata}. See context for detail.}
\label{sed}
\end{figure*}

\begin{figure*}
\centering
\includegraphics[height=5cm]{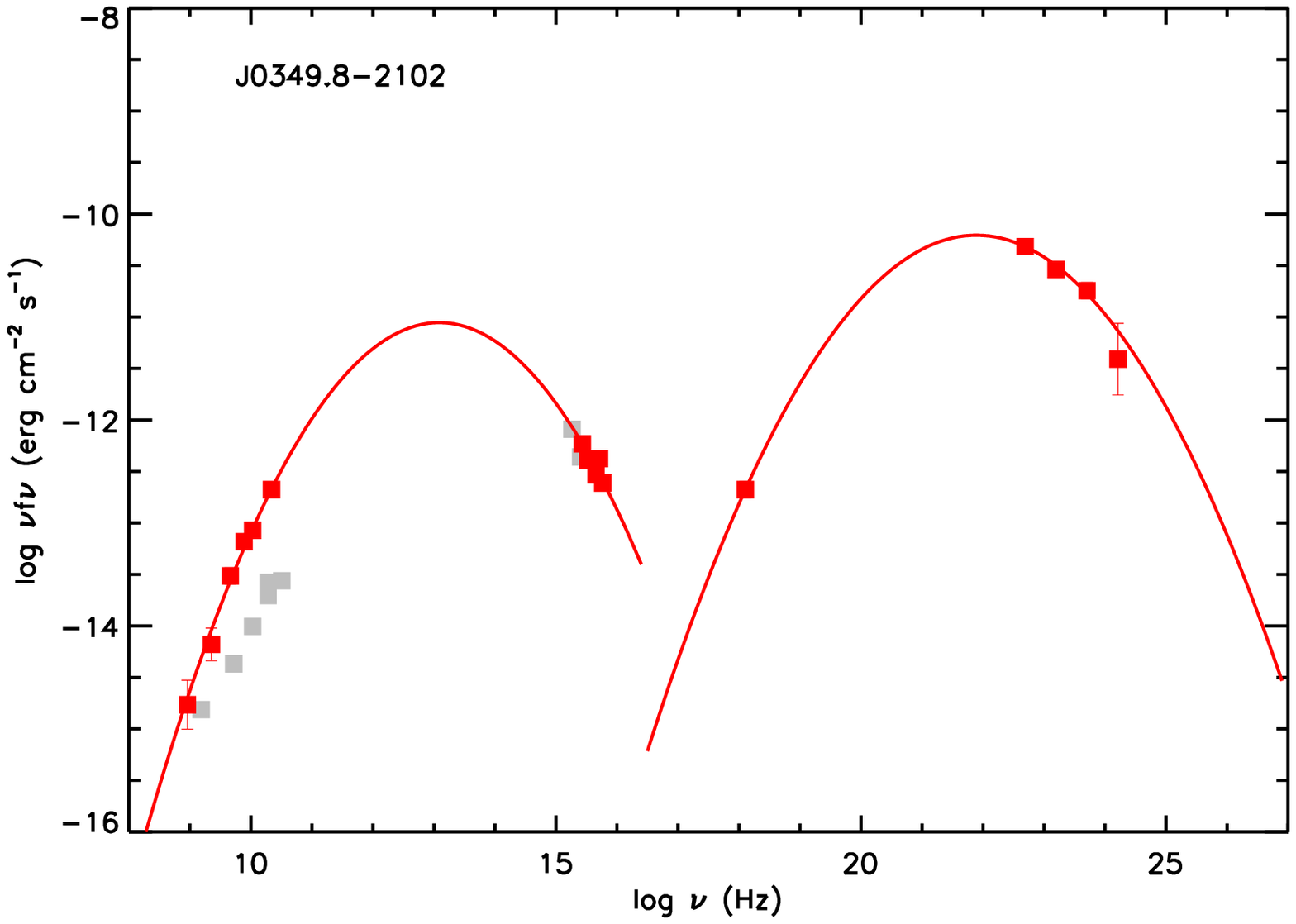}
\includegraphics[height=5cm]{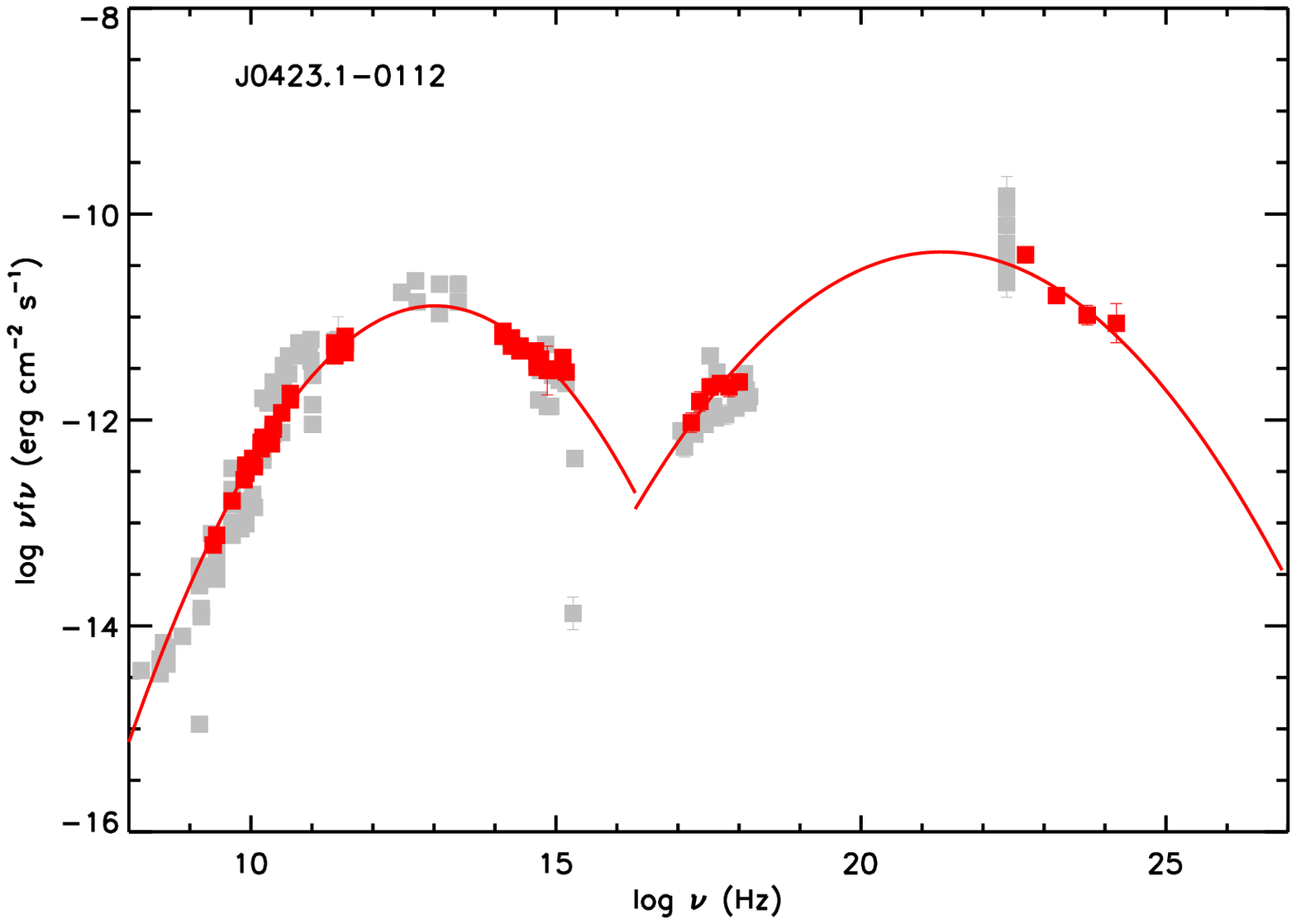}
\includegraphics[height=5cm]{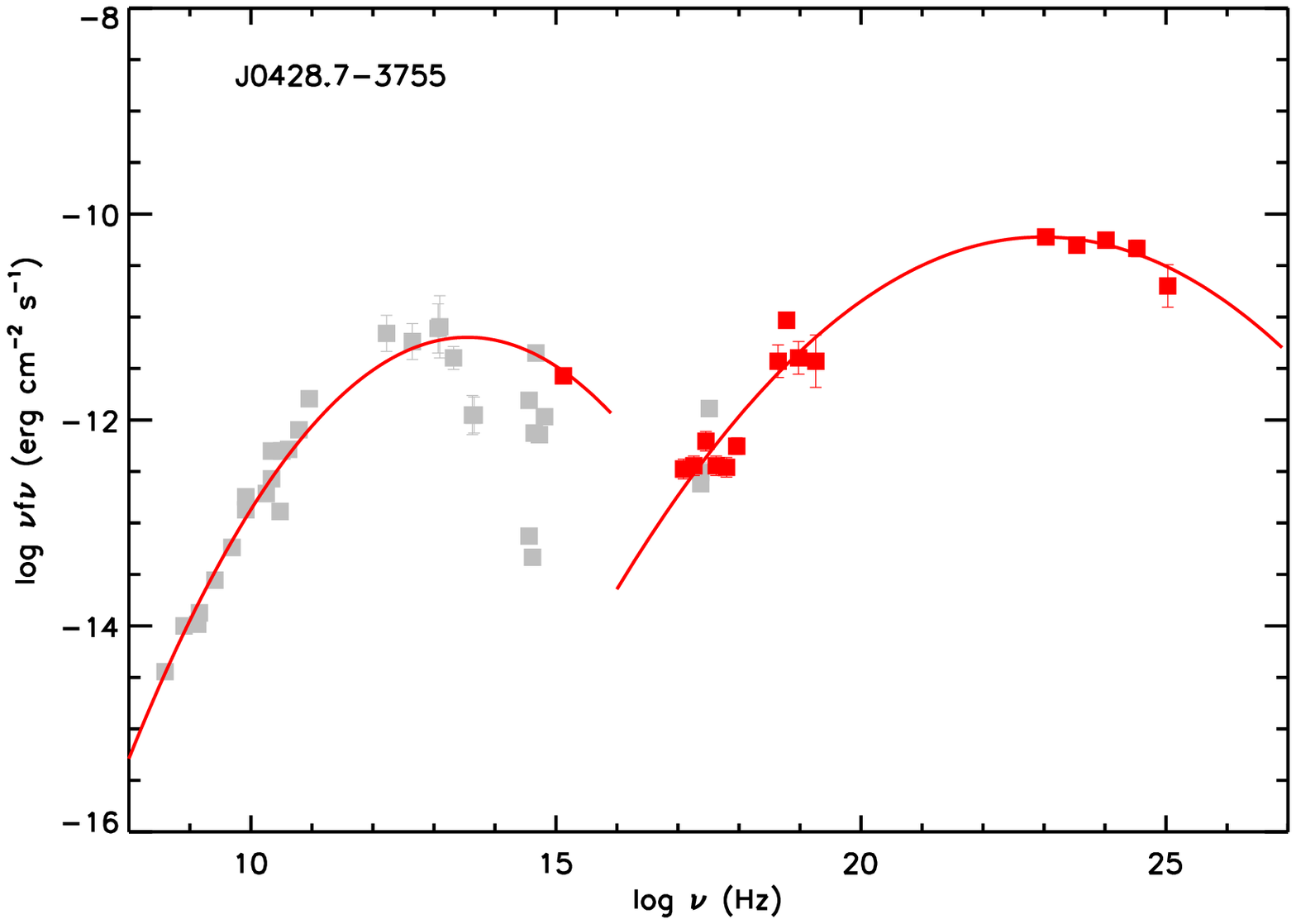}
\includegraphics[height=5cm]{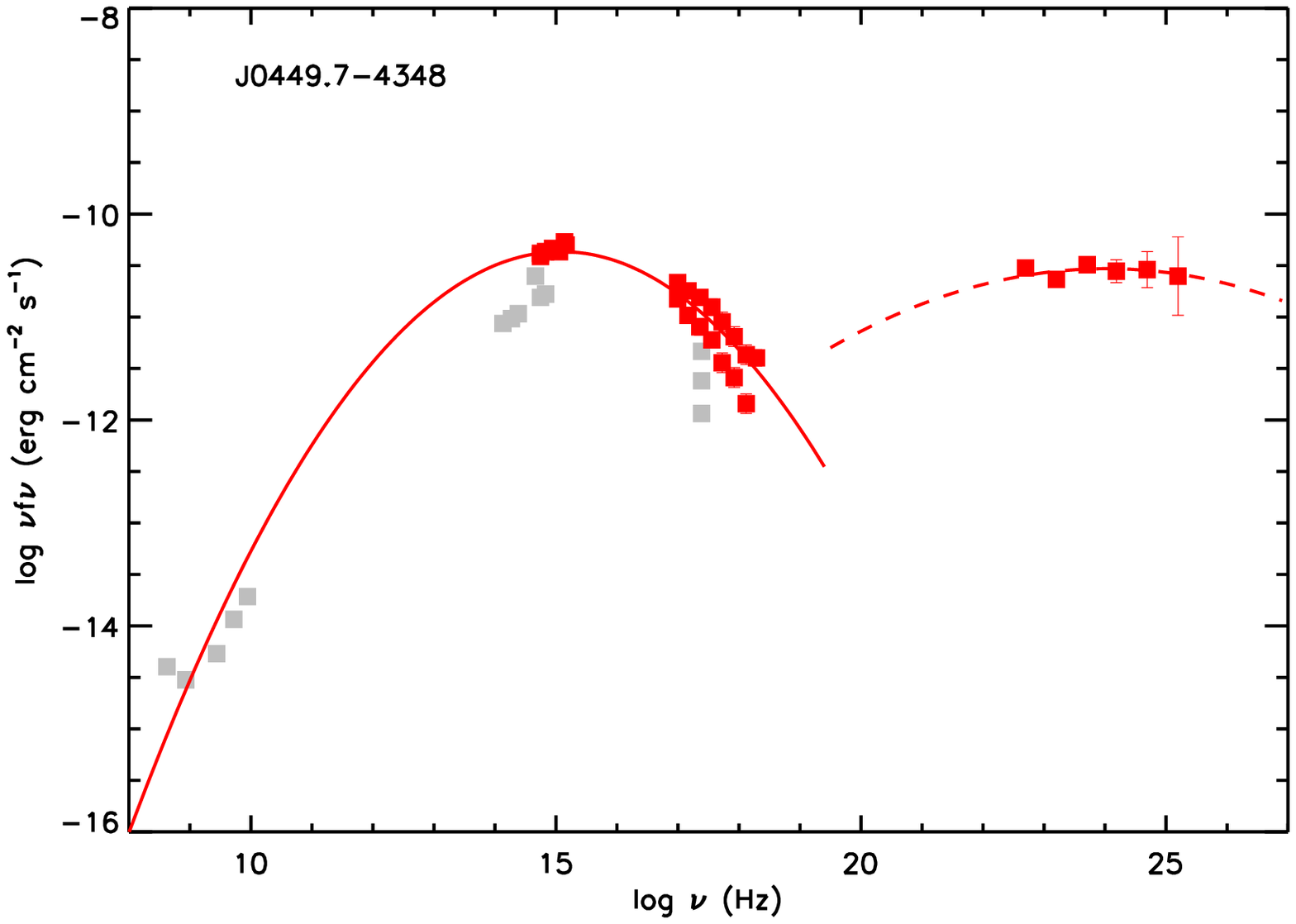}
\includegraphics[height=5cm]{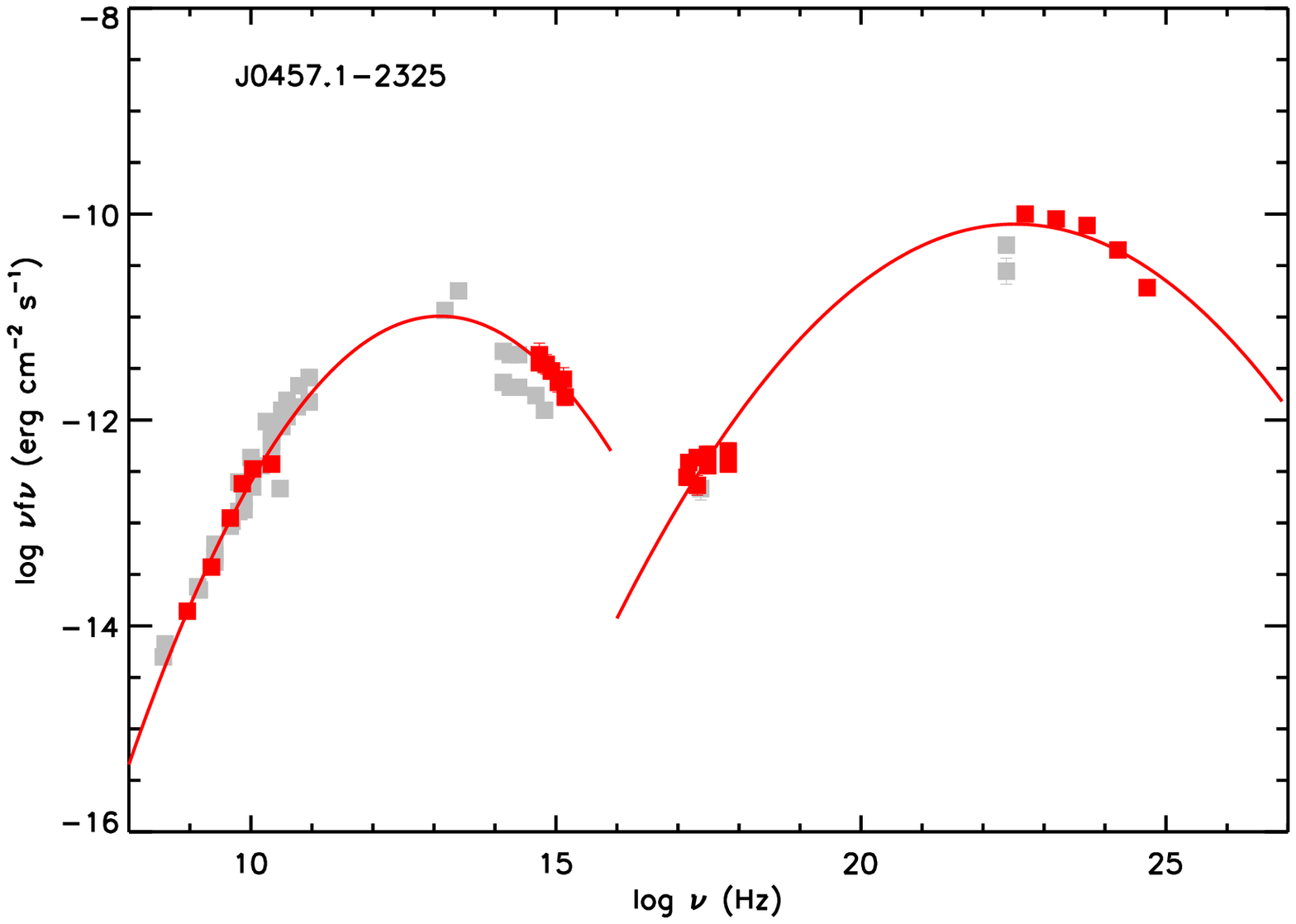}
\includegraphics[height=5cm]{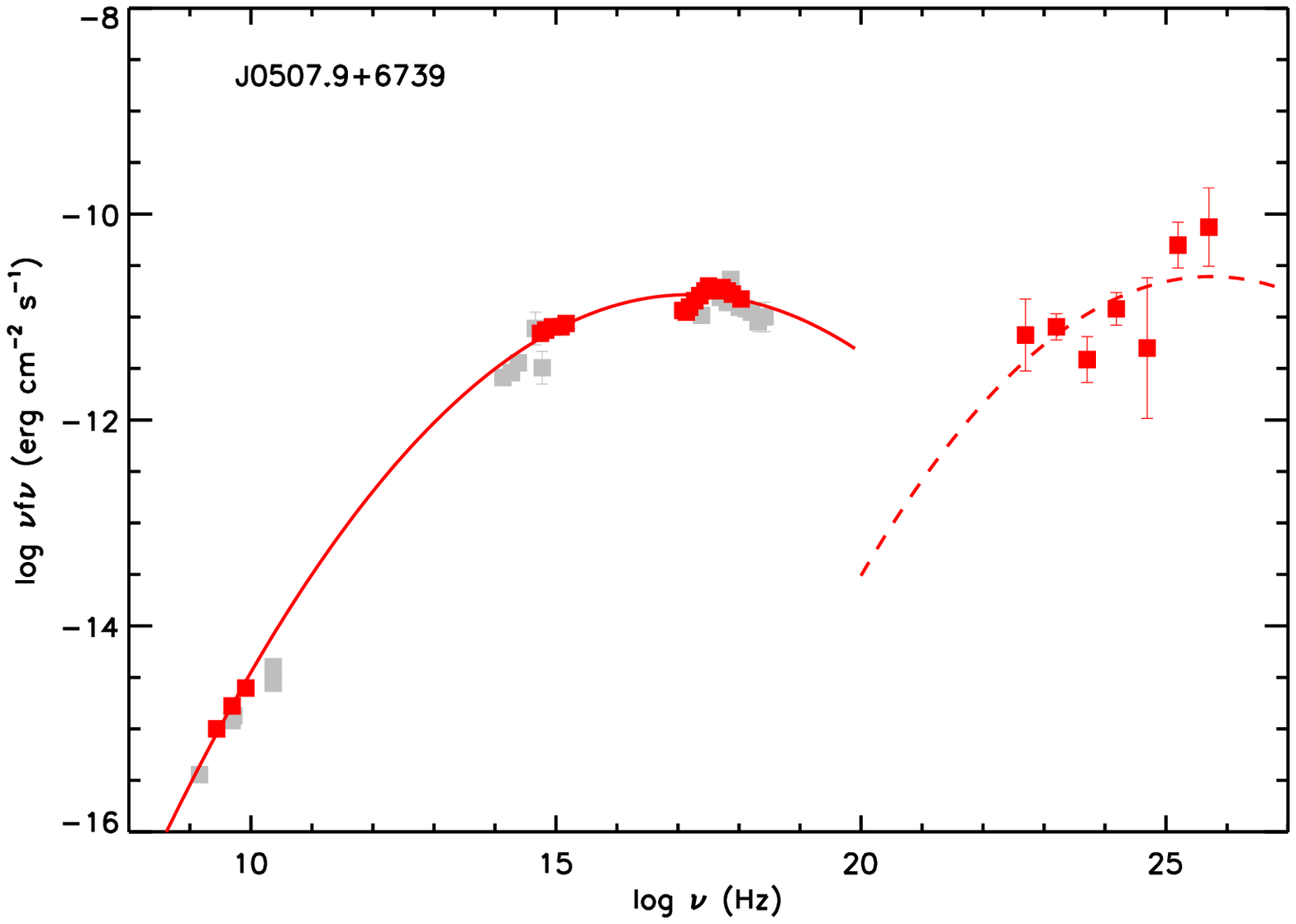}
\includegraphics[height=5cm]{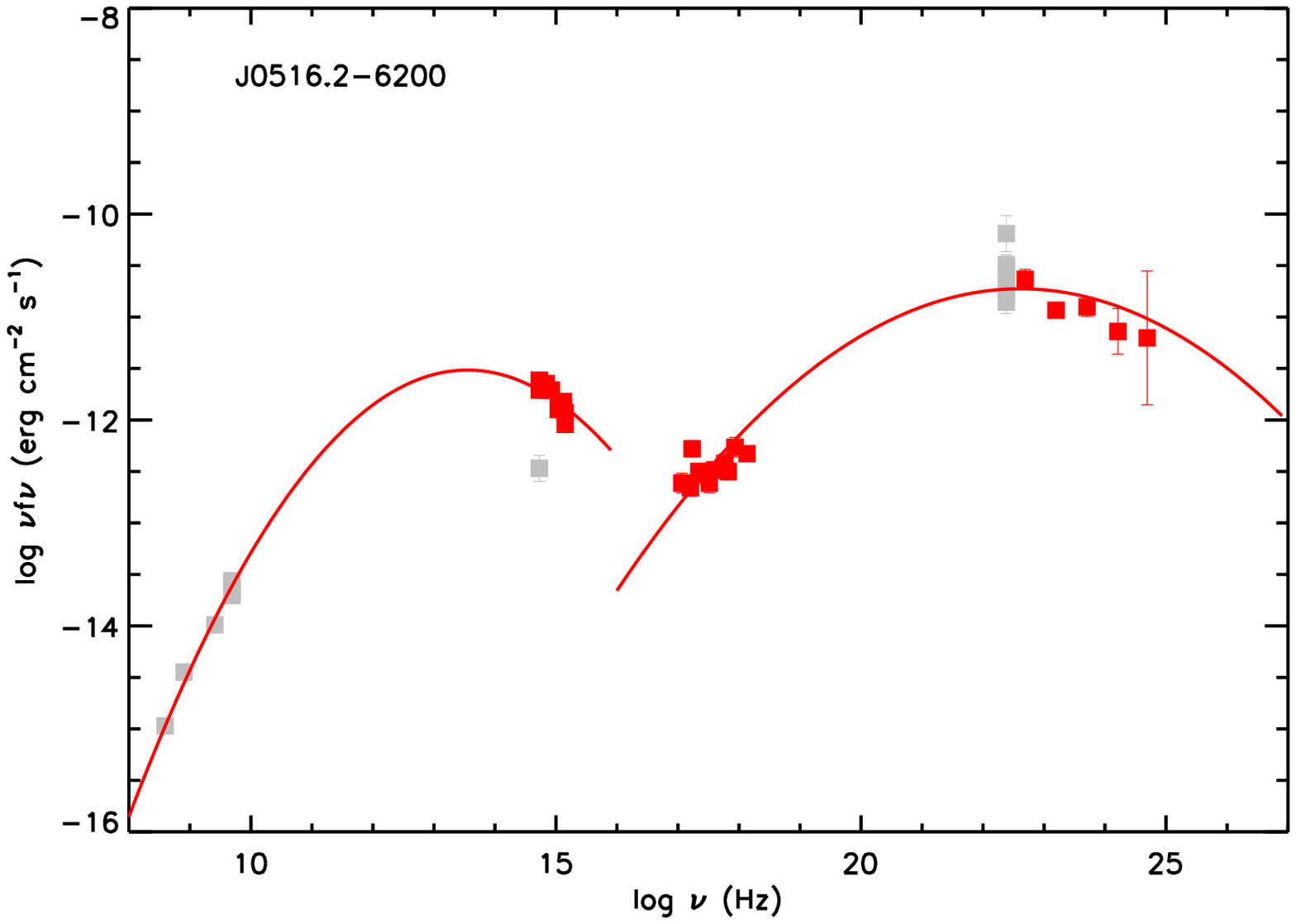}
\includegraphics[height=5cm]{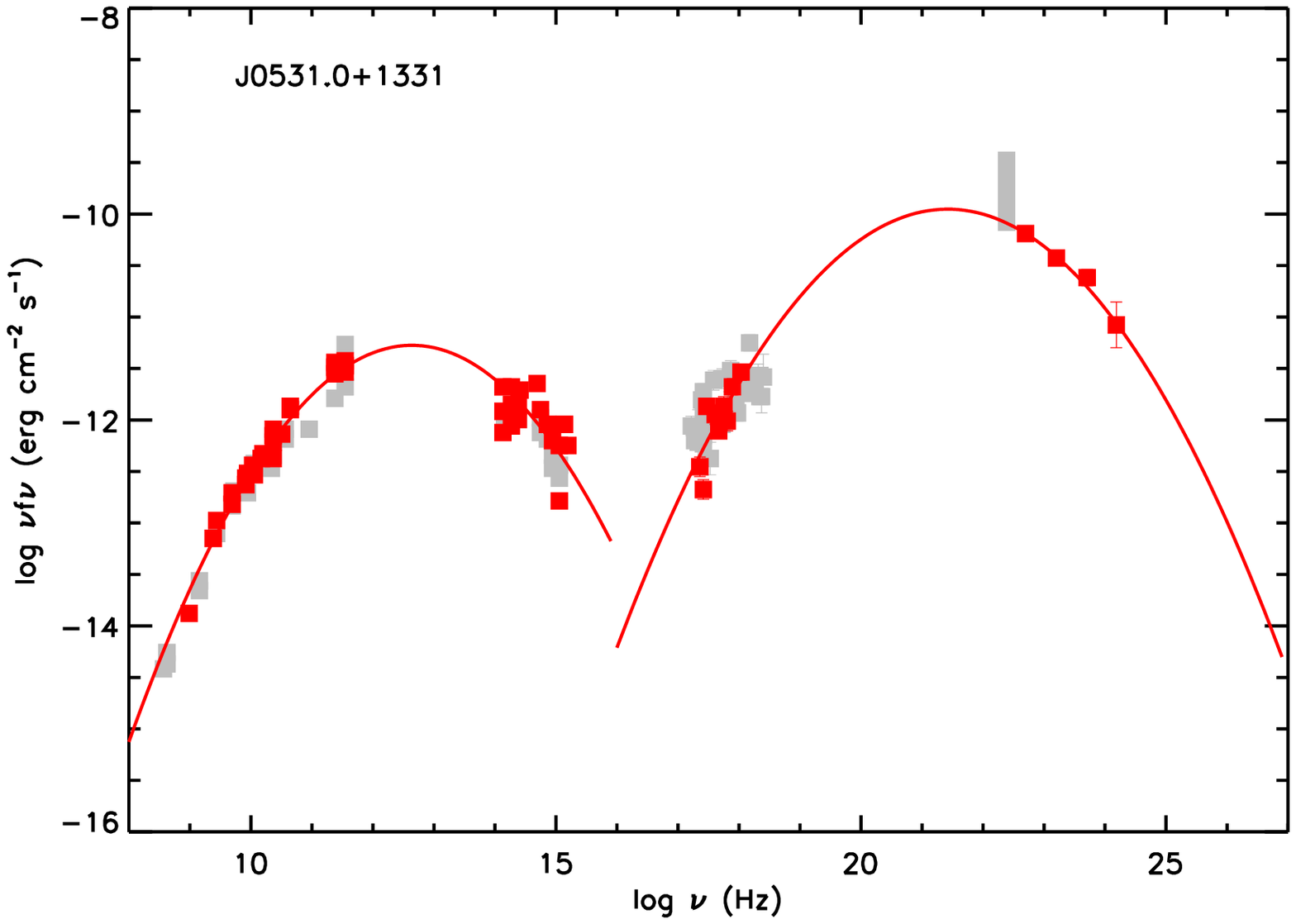}
\center{Fig. \ref{sed} --- continued.}
\end{figure*}

\begin{figure*}
\centering
\includegraphics[height=5cm]{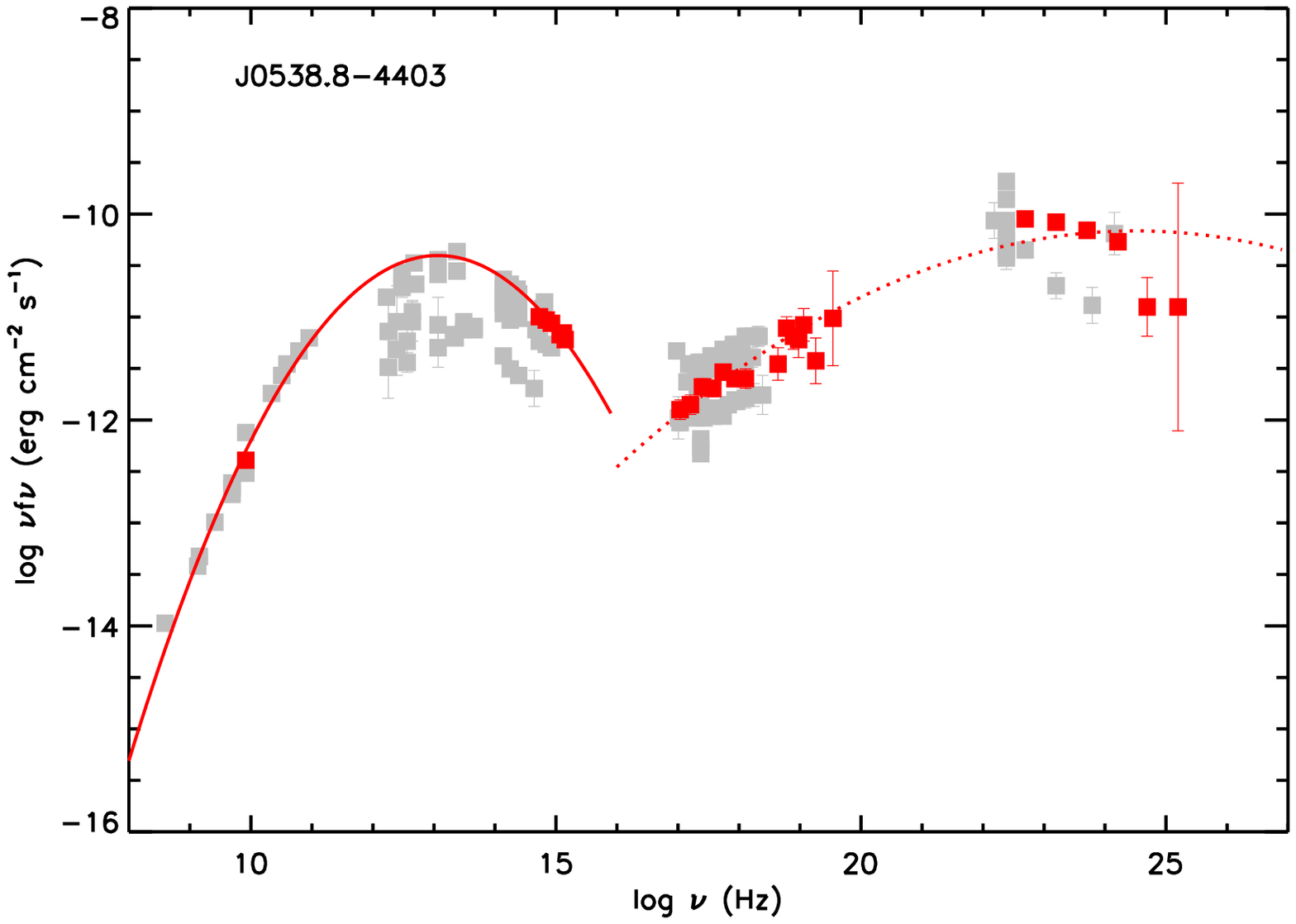}
\includegraphics[height=5cm]{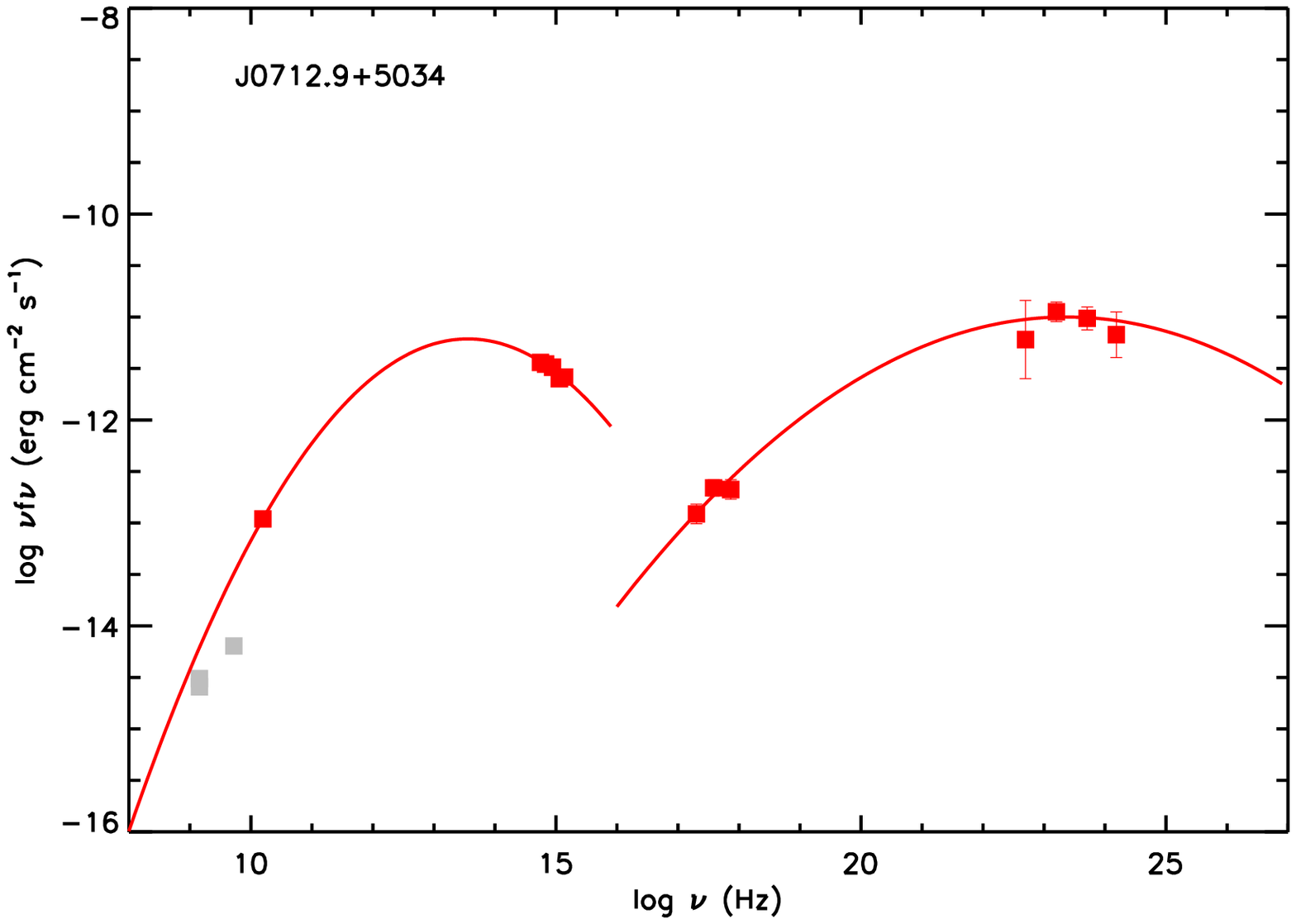}
\includegraphics[height=5cm]{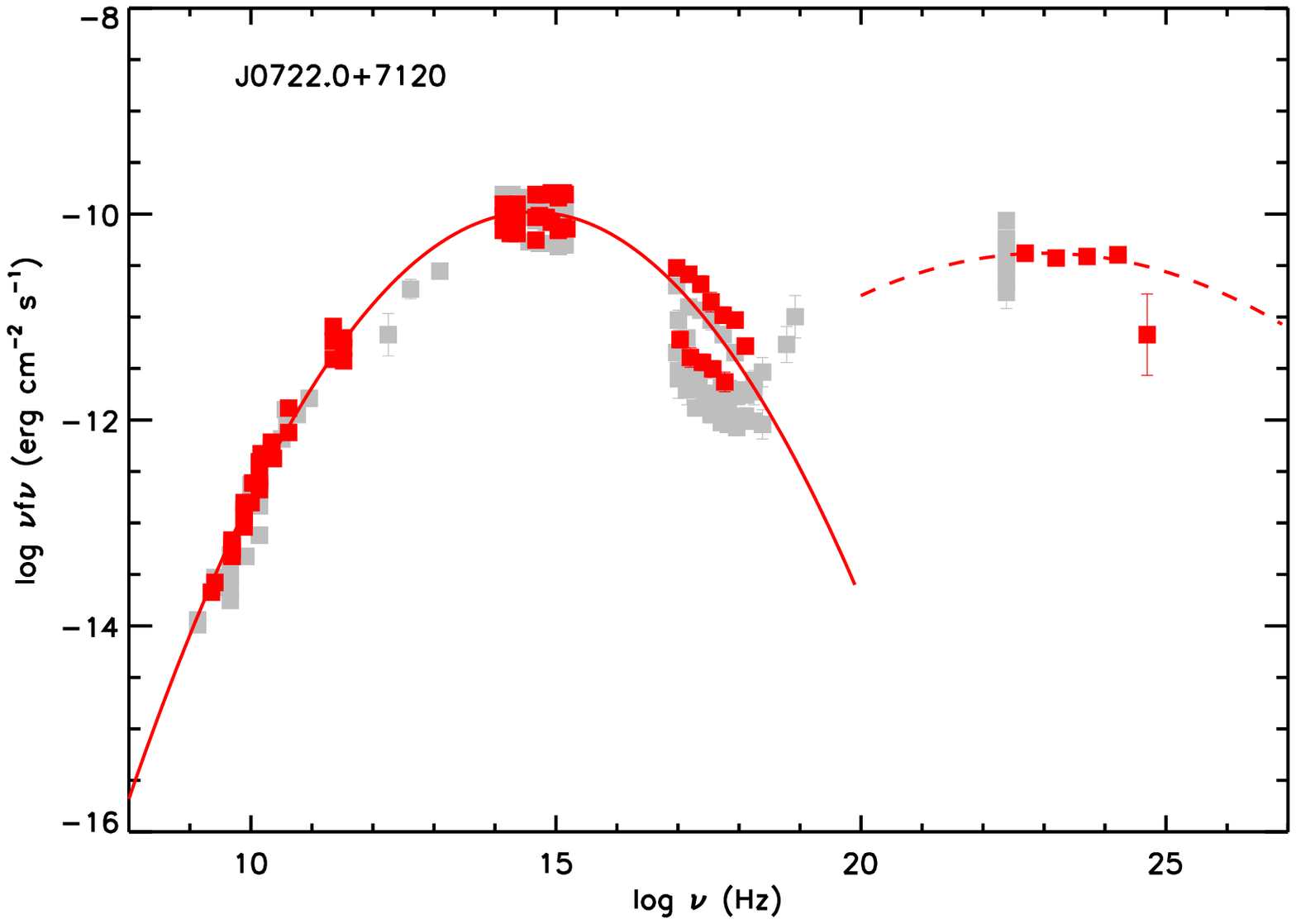}
\includegraphics[height=5cm]{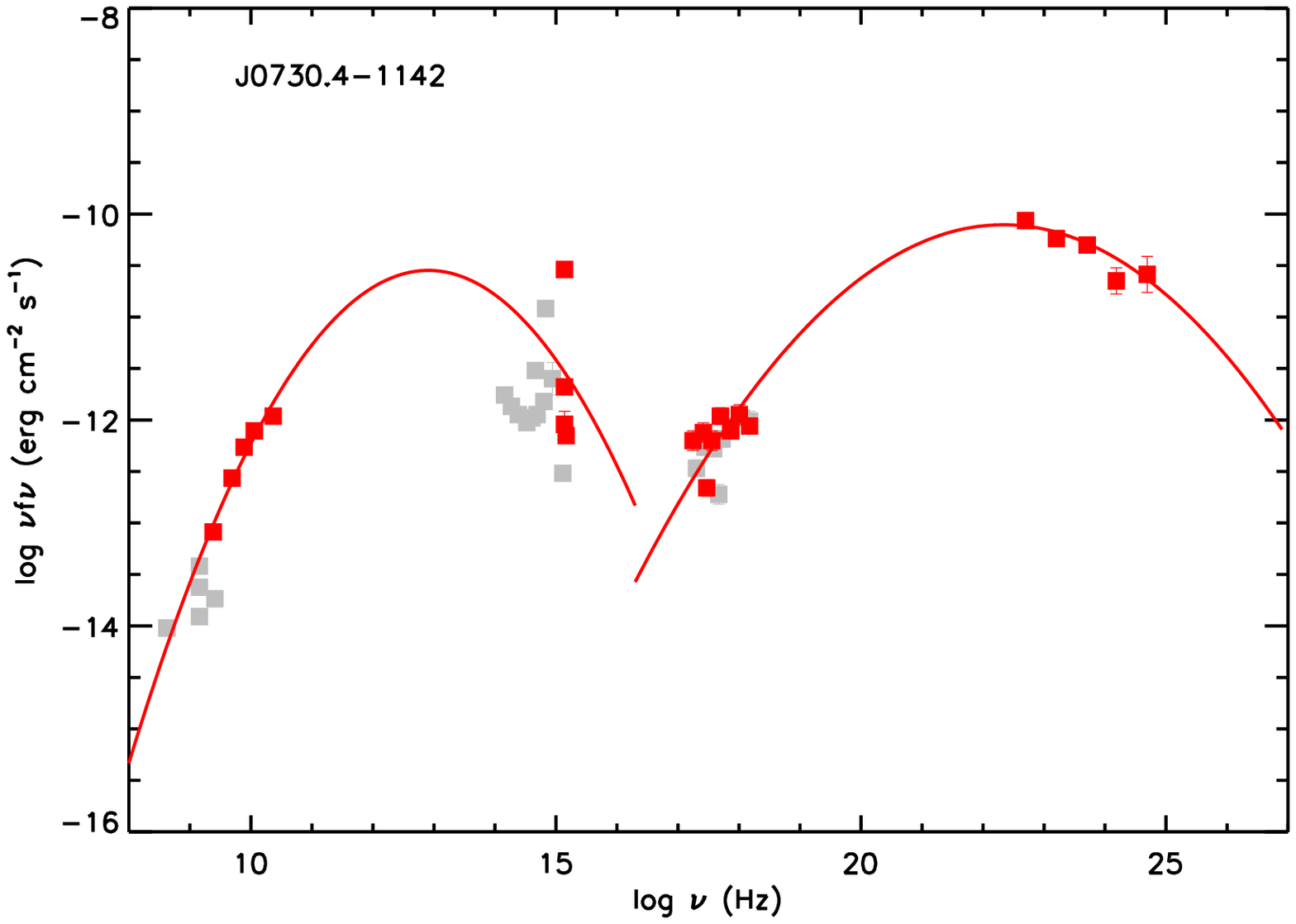}
\includegraphics[height=5cm]{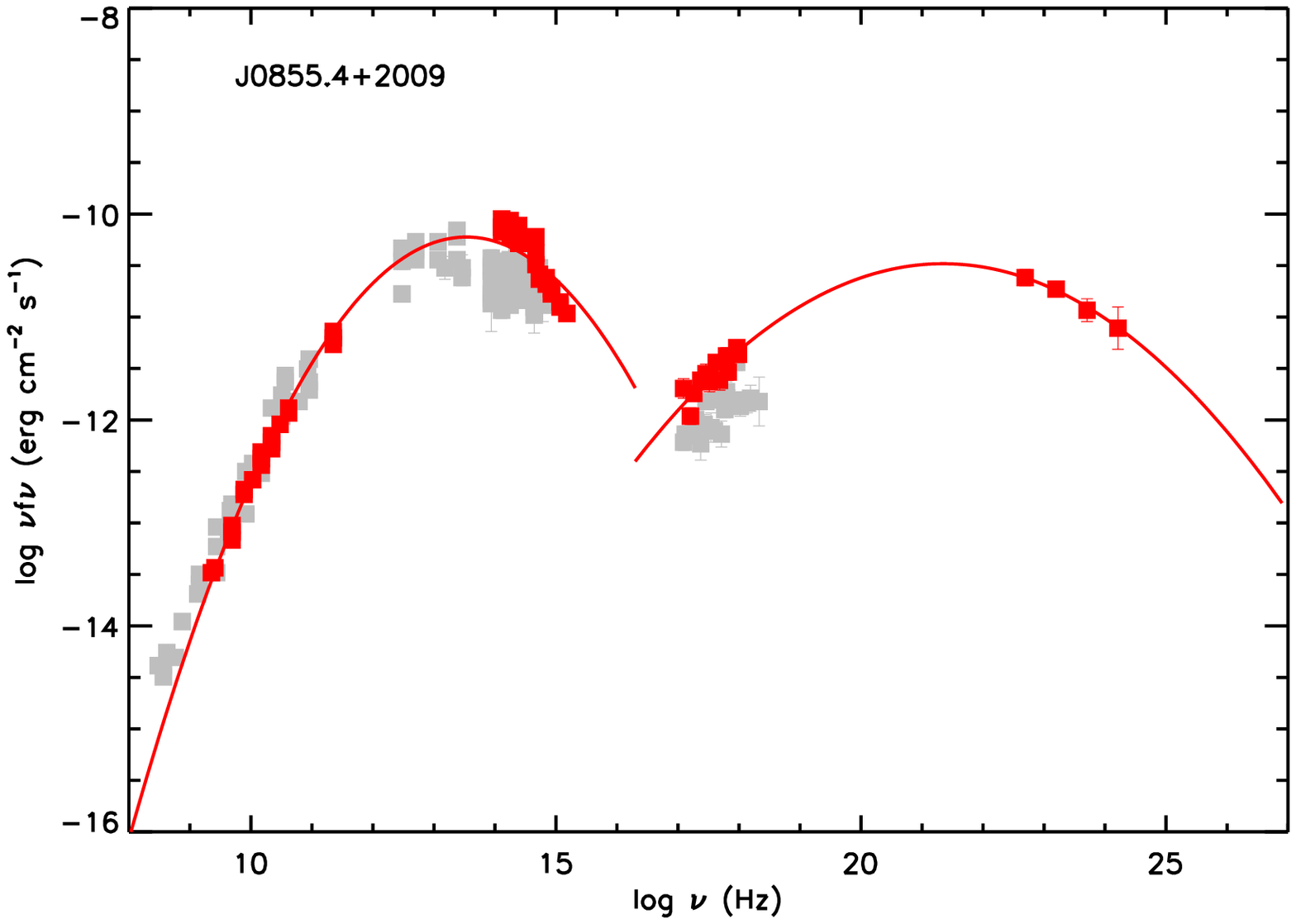}
\includegraphics[height=5cm]{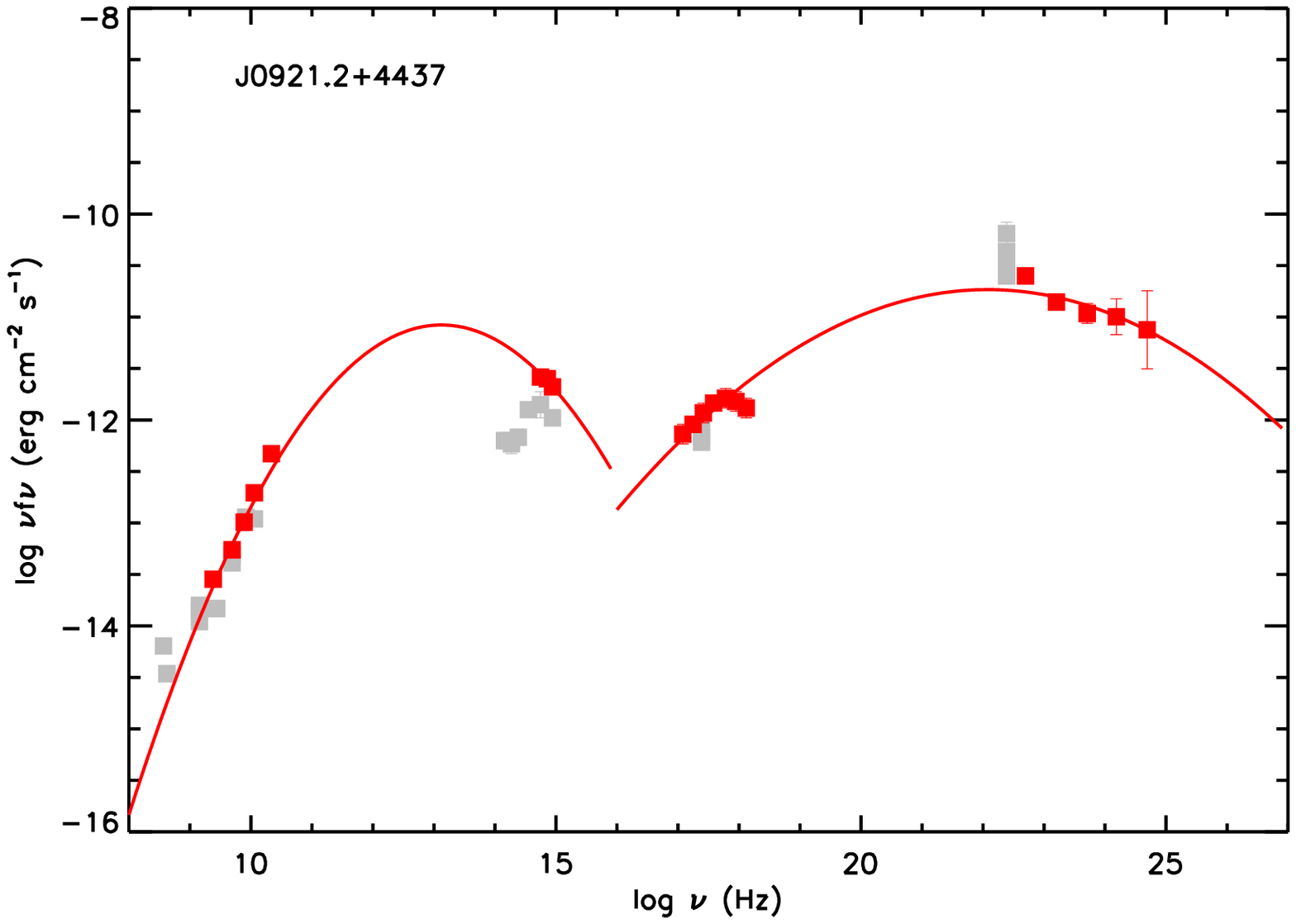}
\includegraphics[height=5cm]{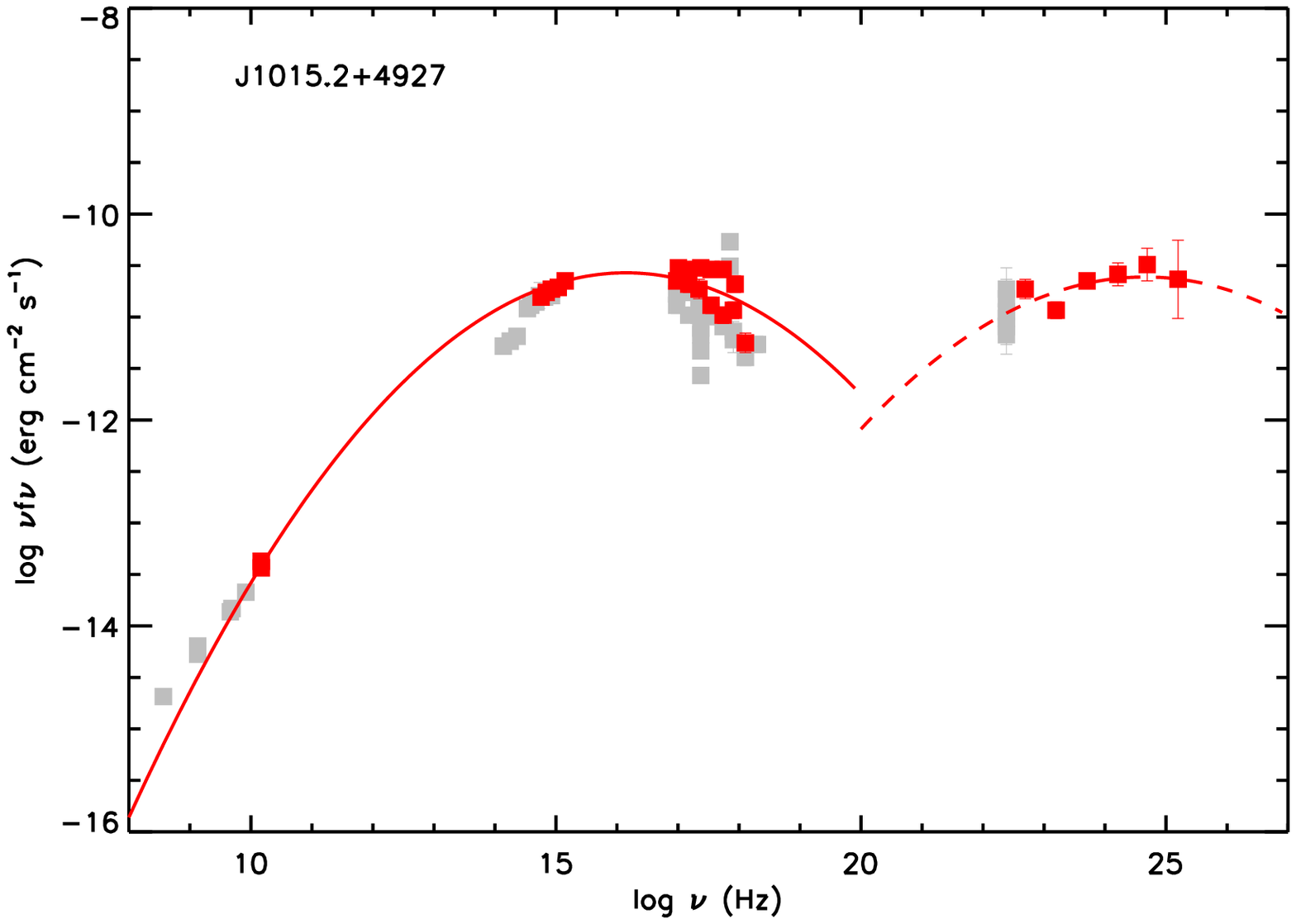}
\includegraphics[height=5cm]{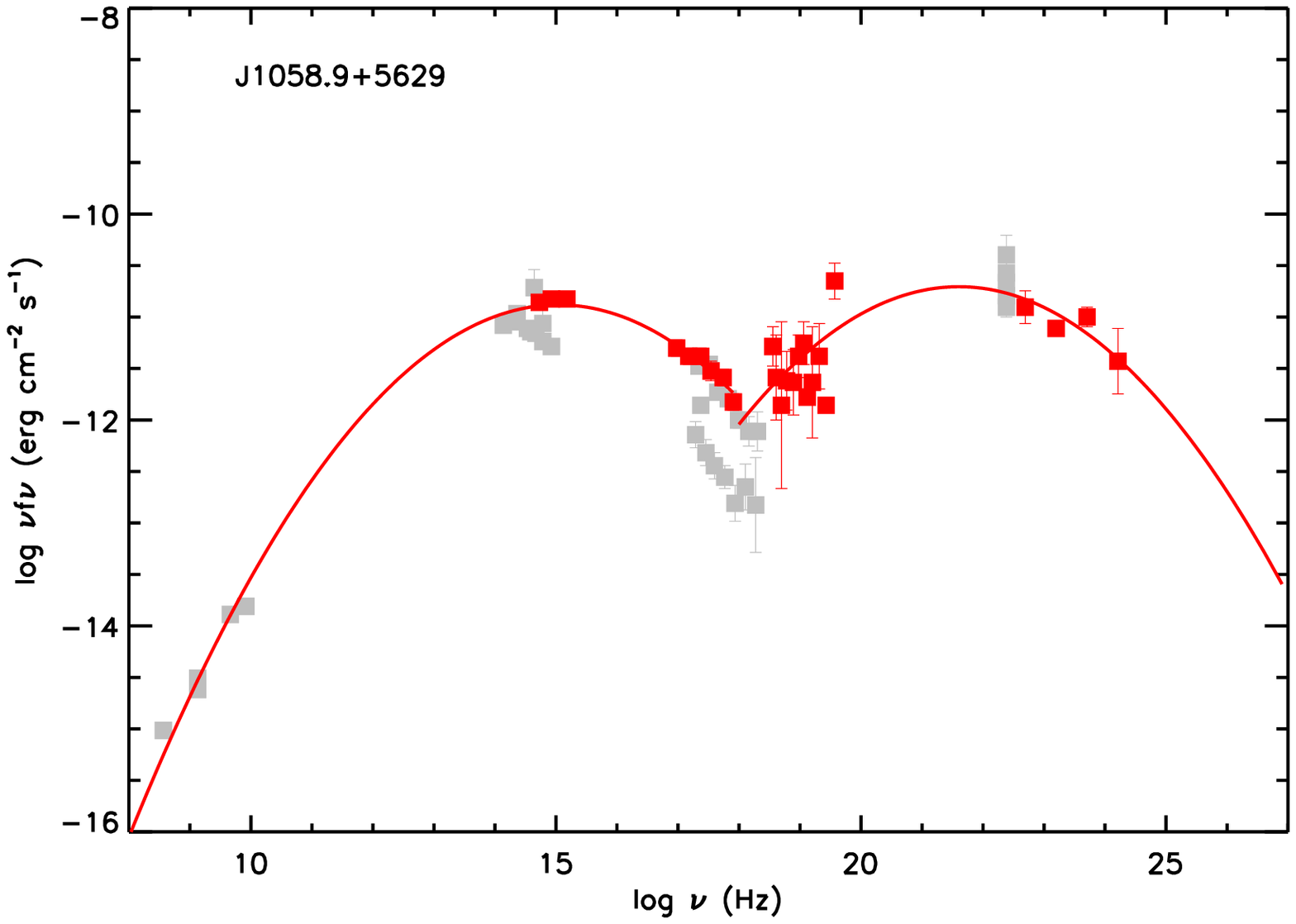}
\center{Fig. \ref{sed} --- continued.}
\end{figure*}

\begin{figure*}
\centering
\includegraphics[height=5cm]{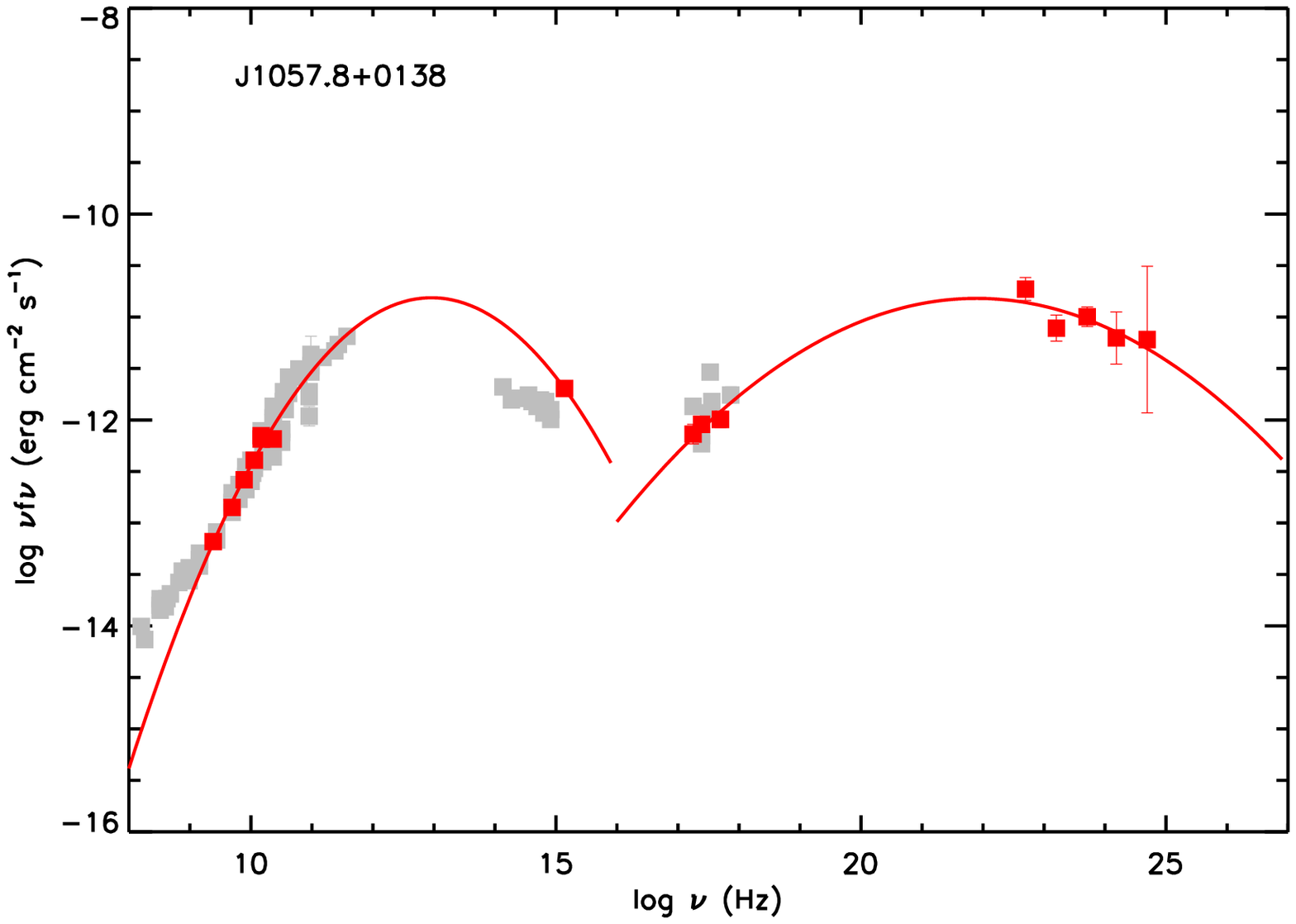}
\includegraphics[height=5cm]{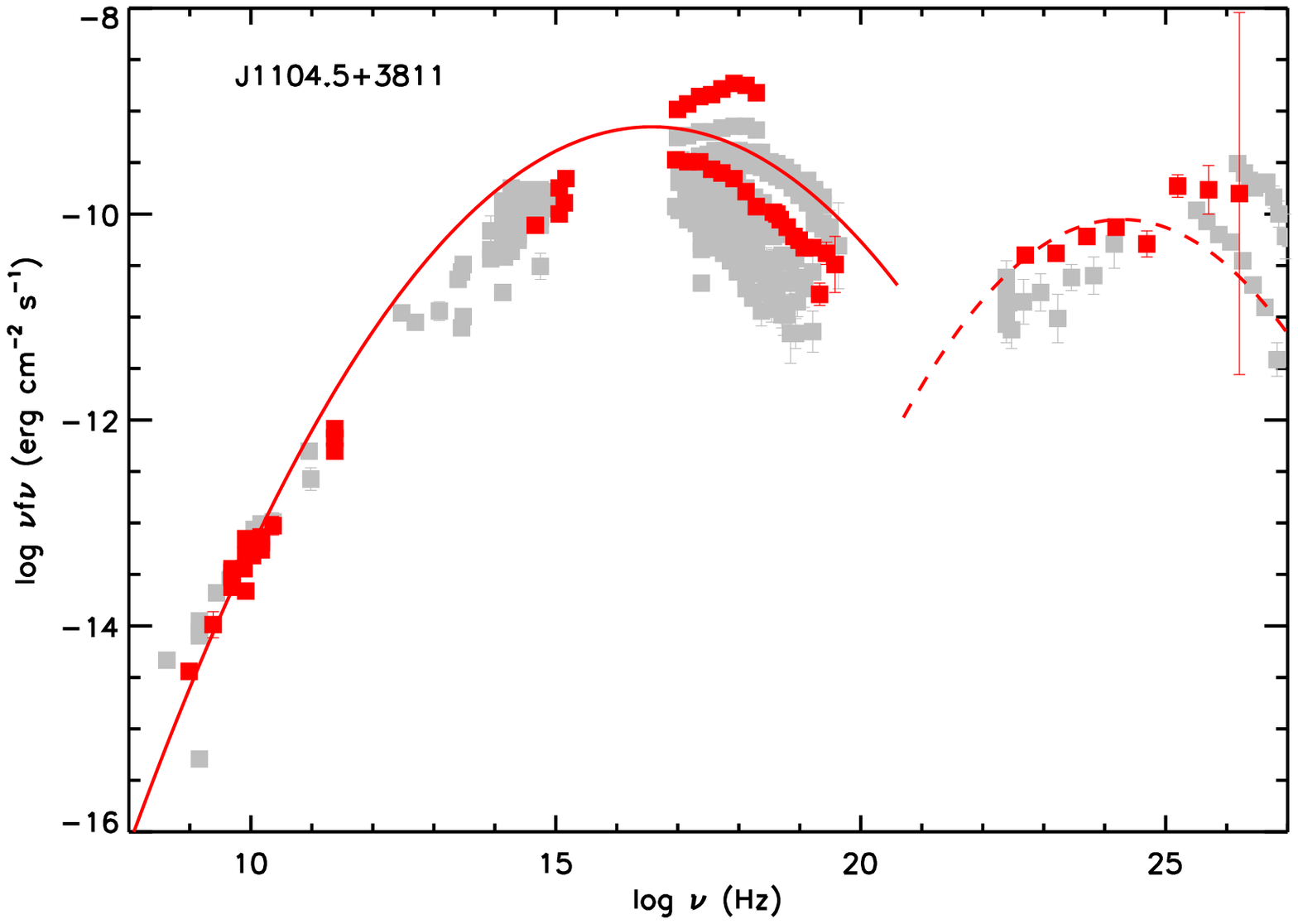}
\includegraphics[height=5cm]{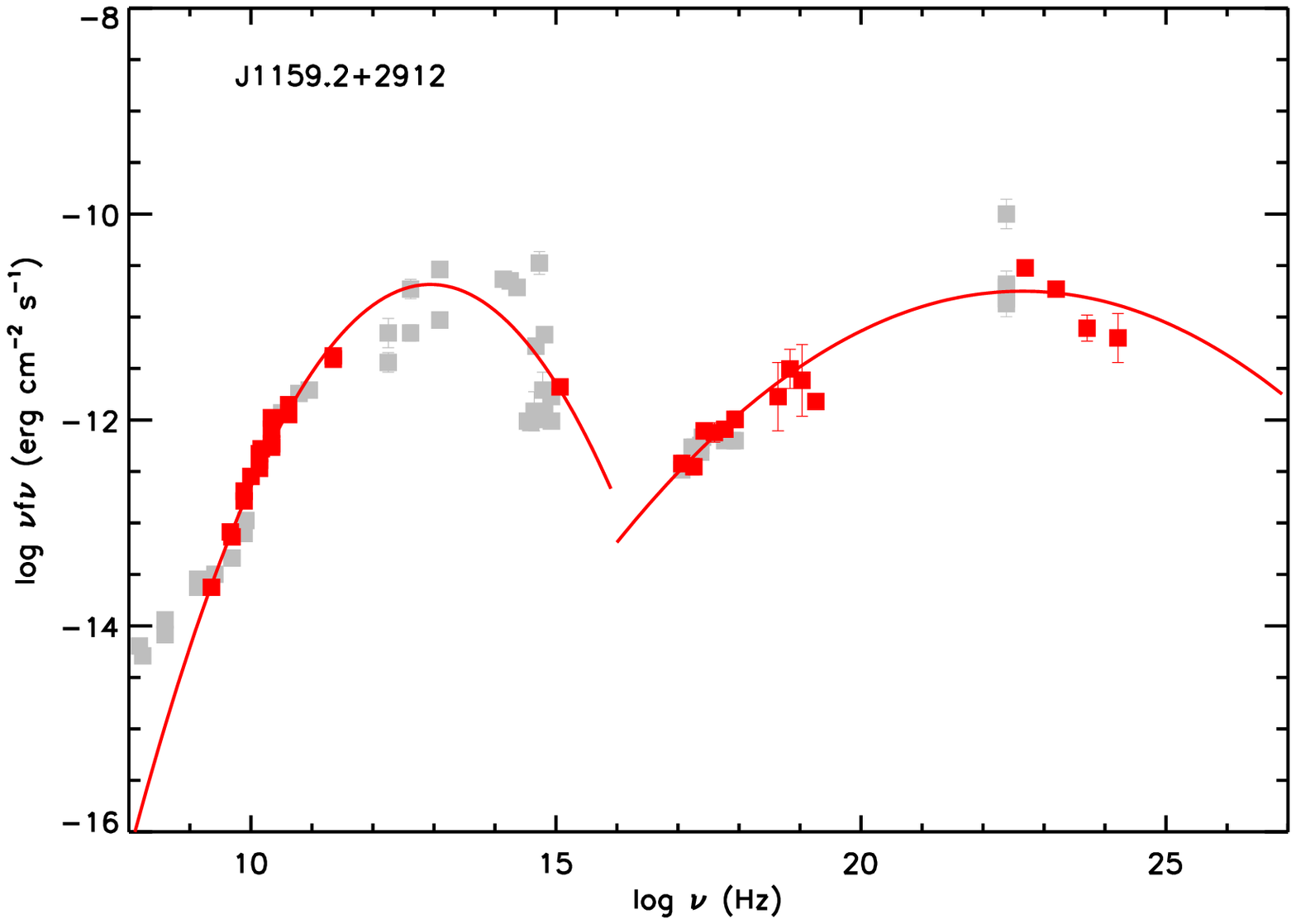}
\includegraphics[height=5cm]{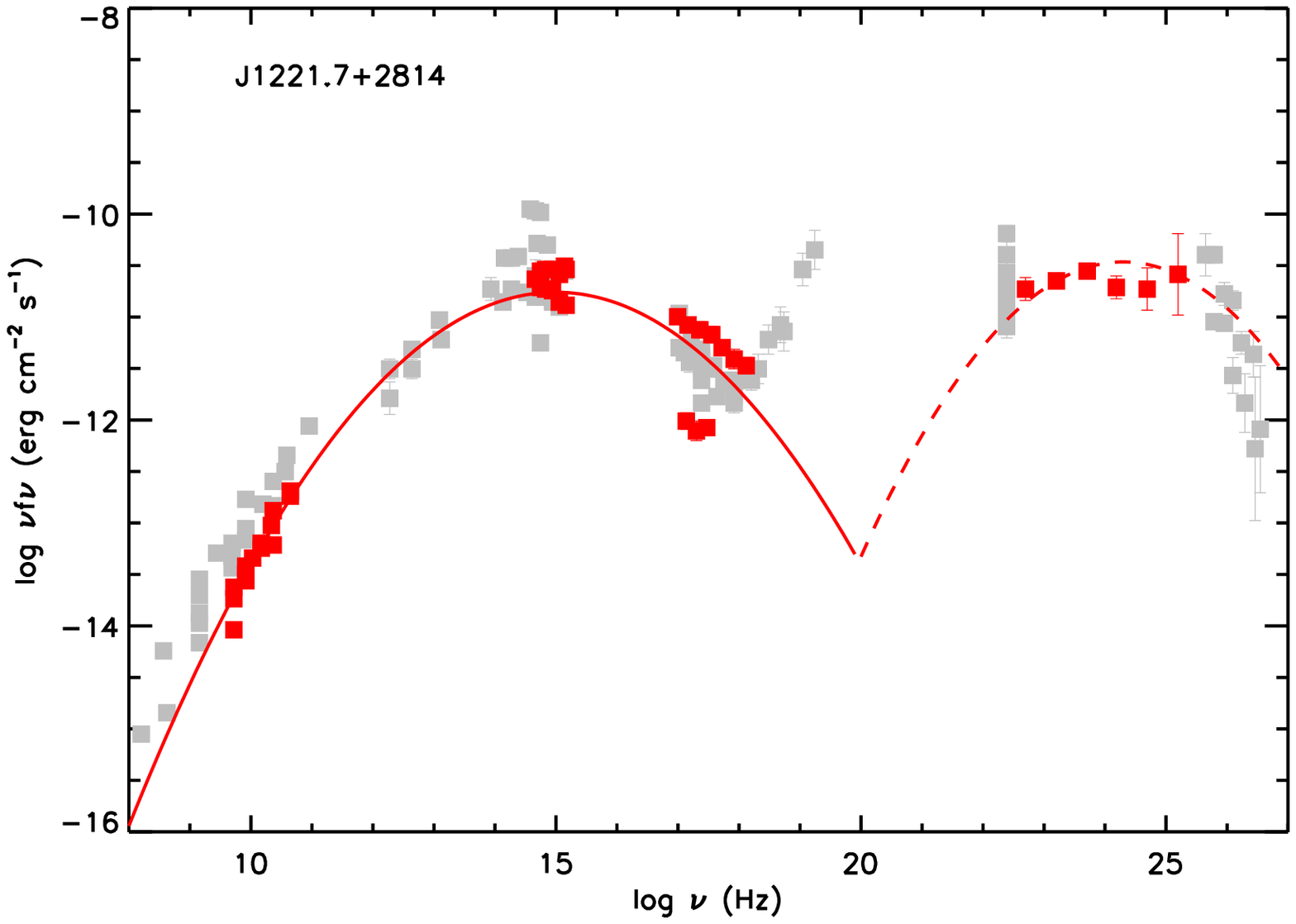}
\includegraphics[height=5cm]{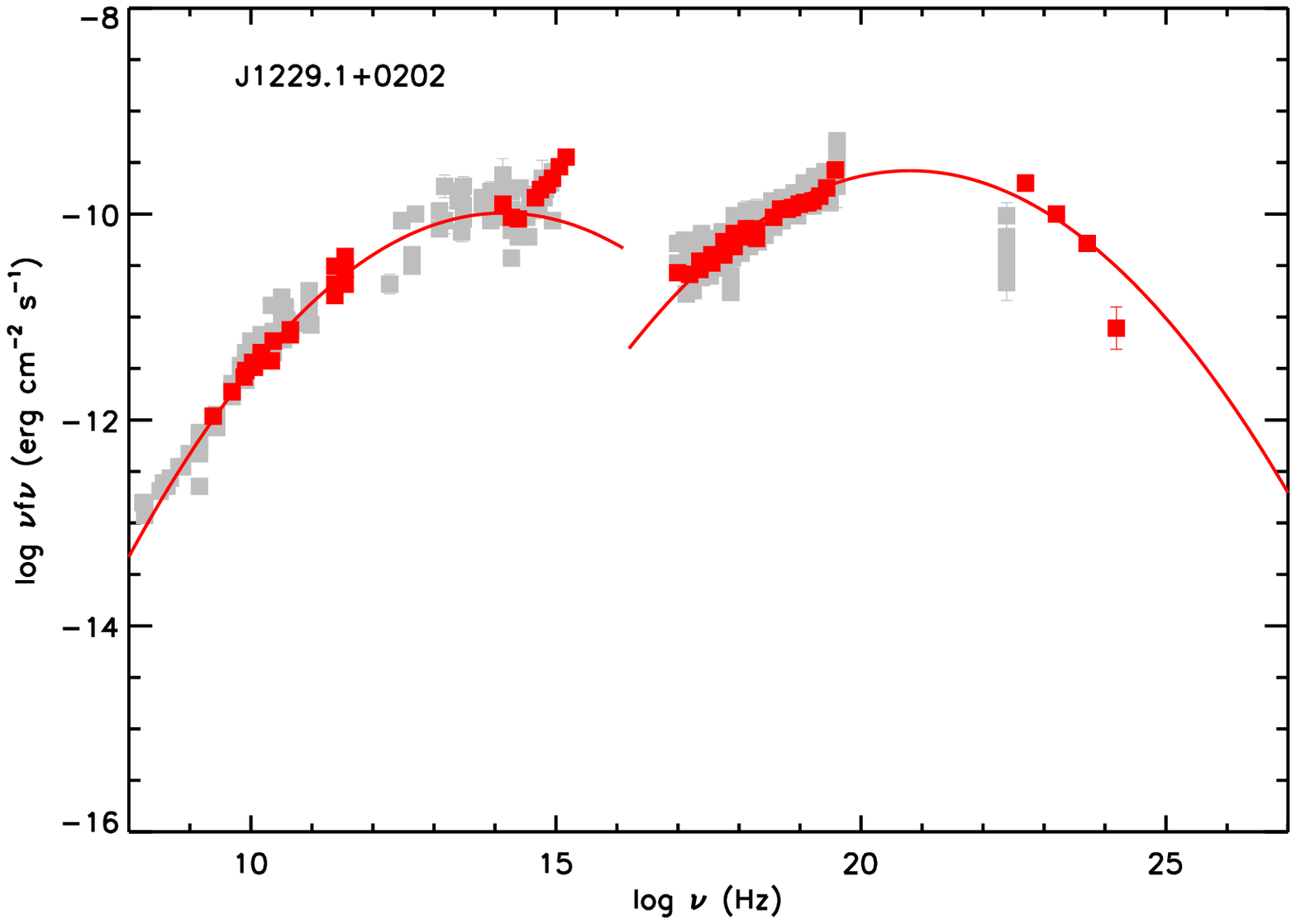}
\includegraphics[height=5cm]{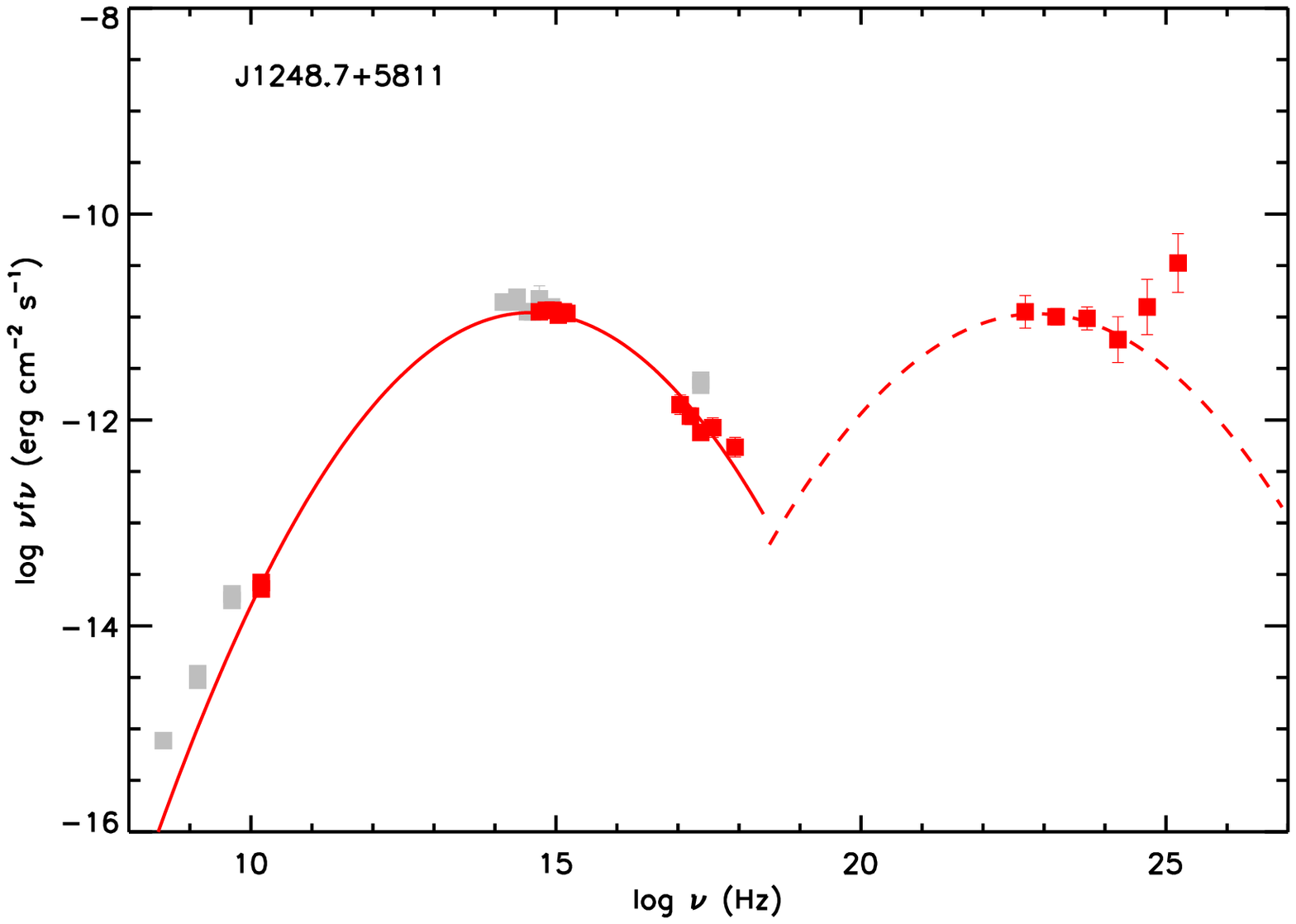}
\includegraphics[height=5cm]{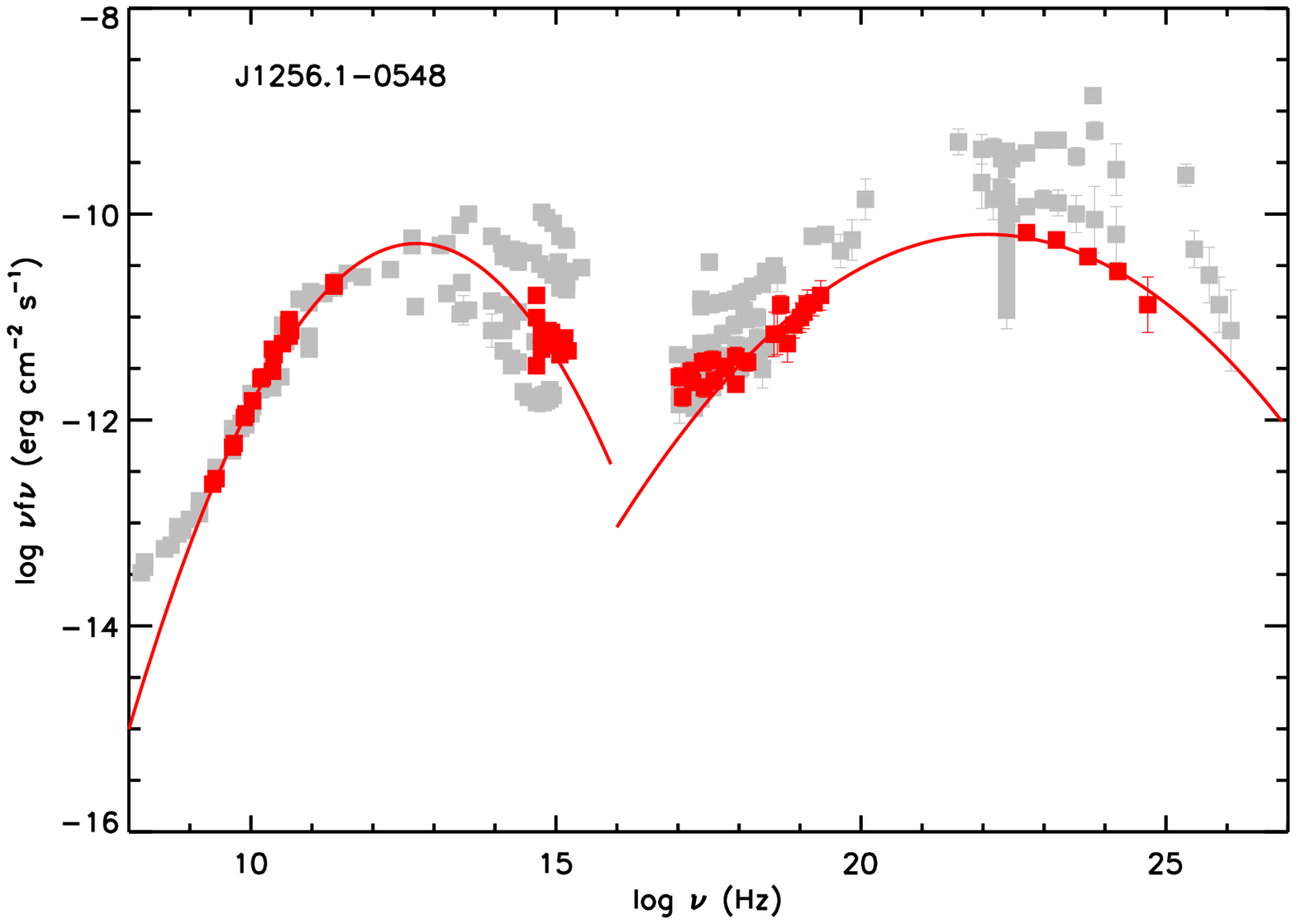}
\includegraphics[height=5cm]{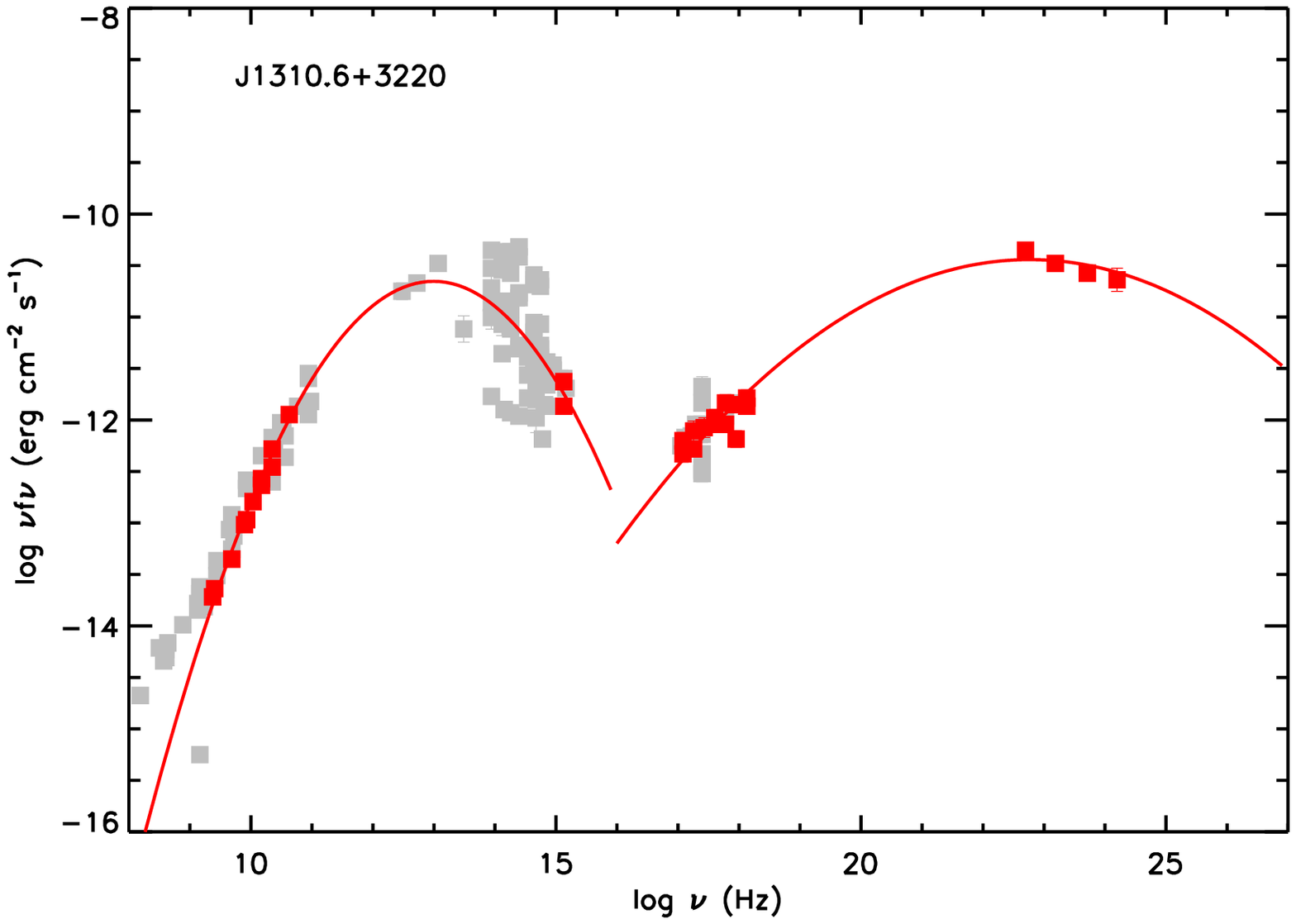}
\center{Fig. \ref{sed} --- continued}
\end{figure*}

\begin{figure*}
\centering
\includegraphics[height=5cm]{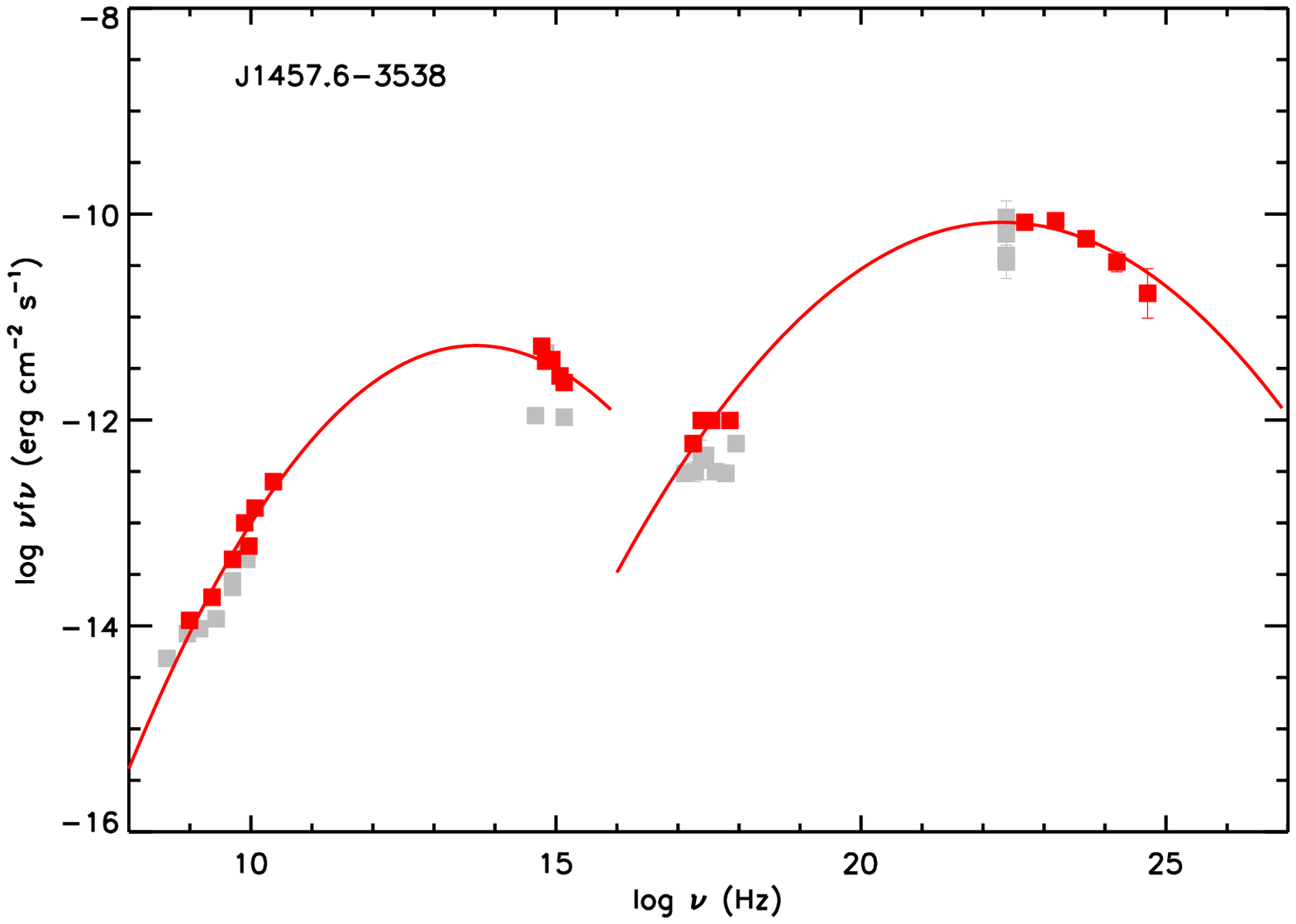}
\includegraphics[height=5cm]{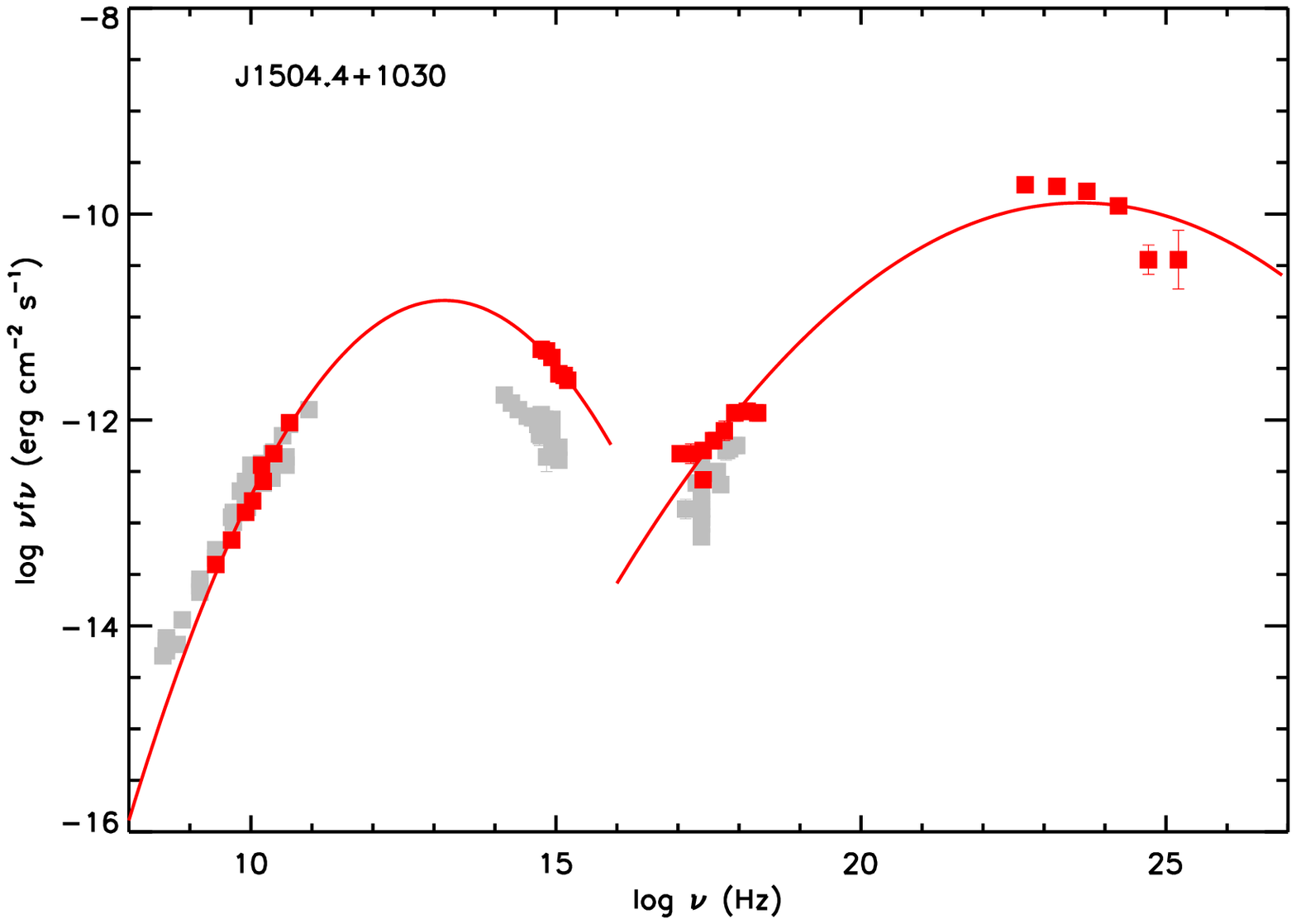}
\includegraphics[height=5cm]{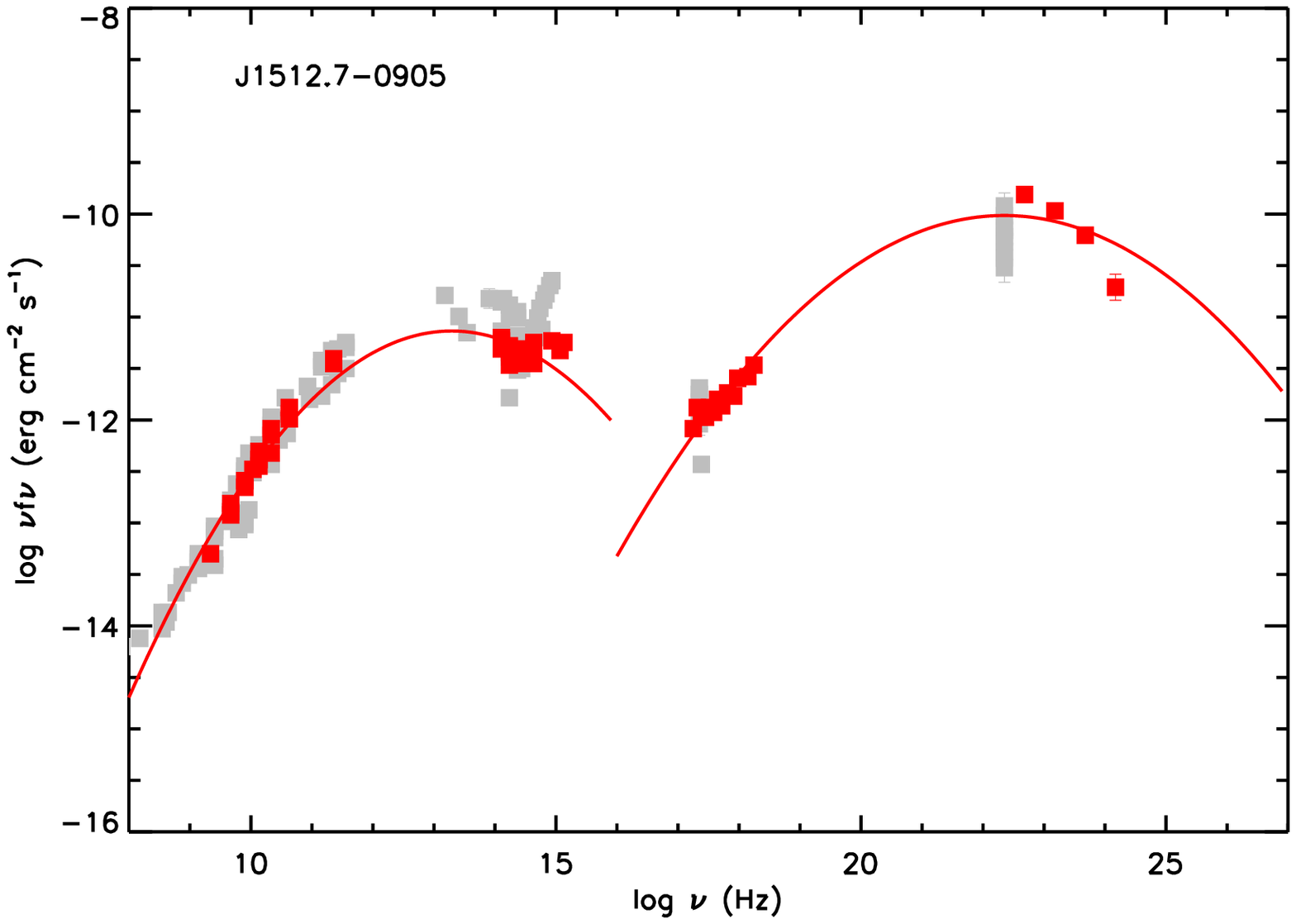}
\includegraphics[height=5cm]{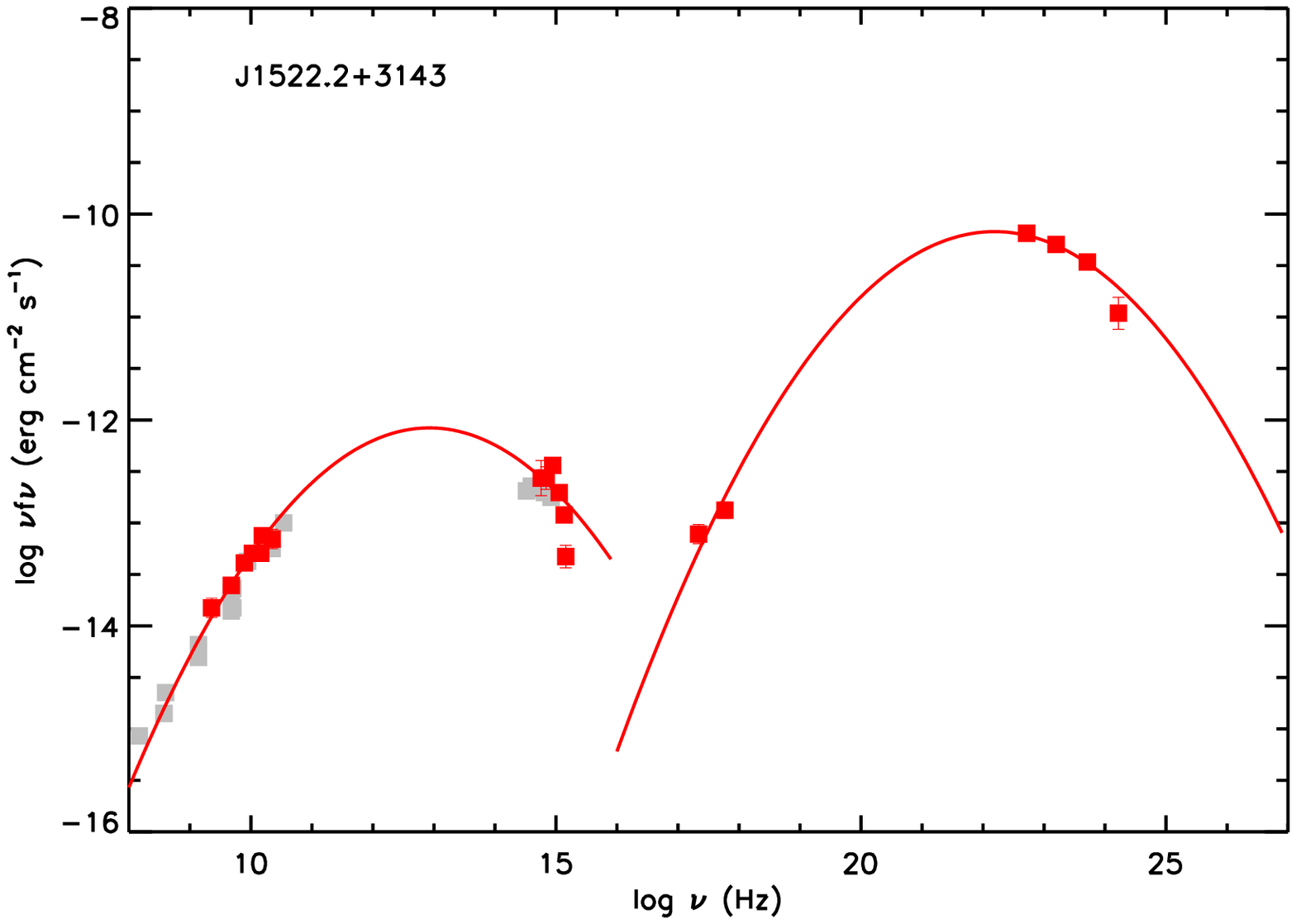}
\includegraphics[height=5cm]{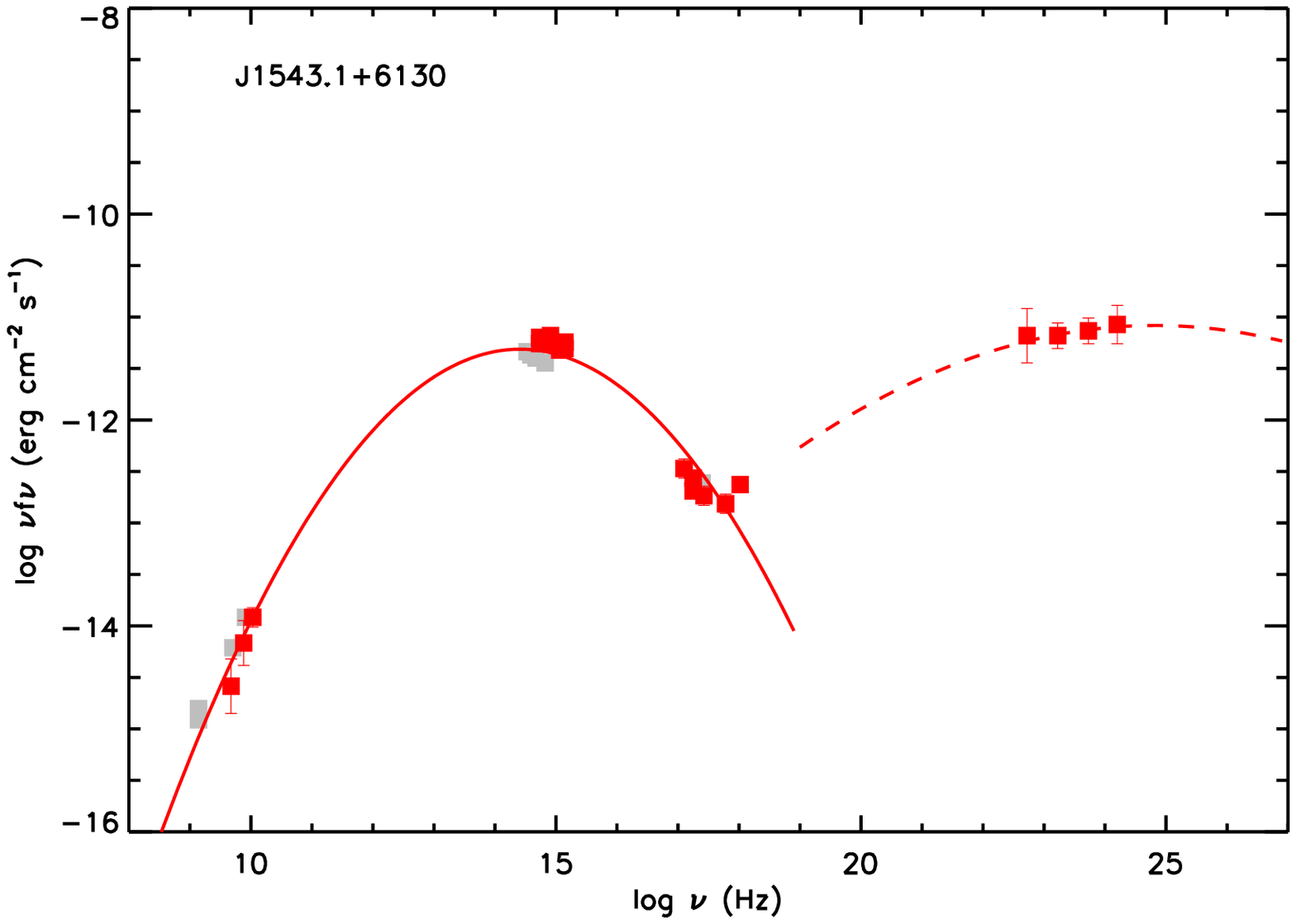}
\includegraphics[height=5cm]{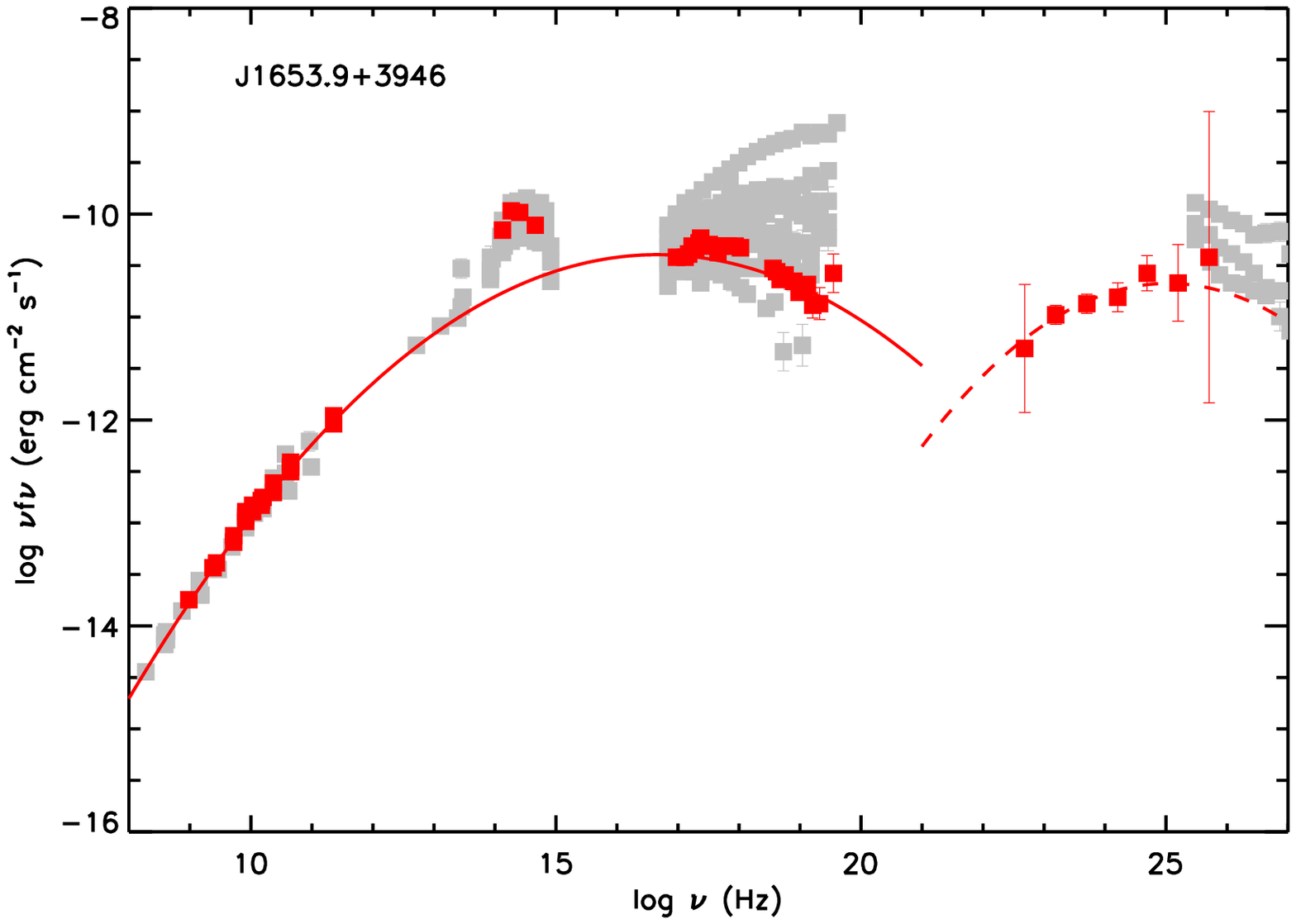}
\includegraphics[height=5cm]{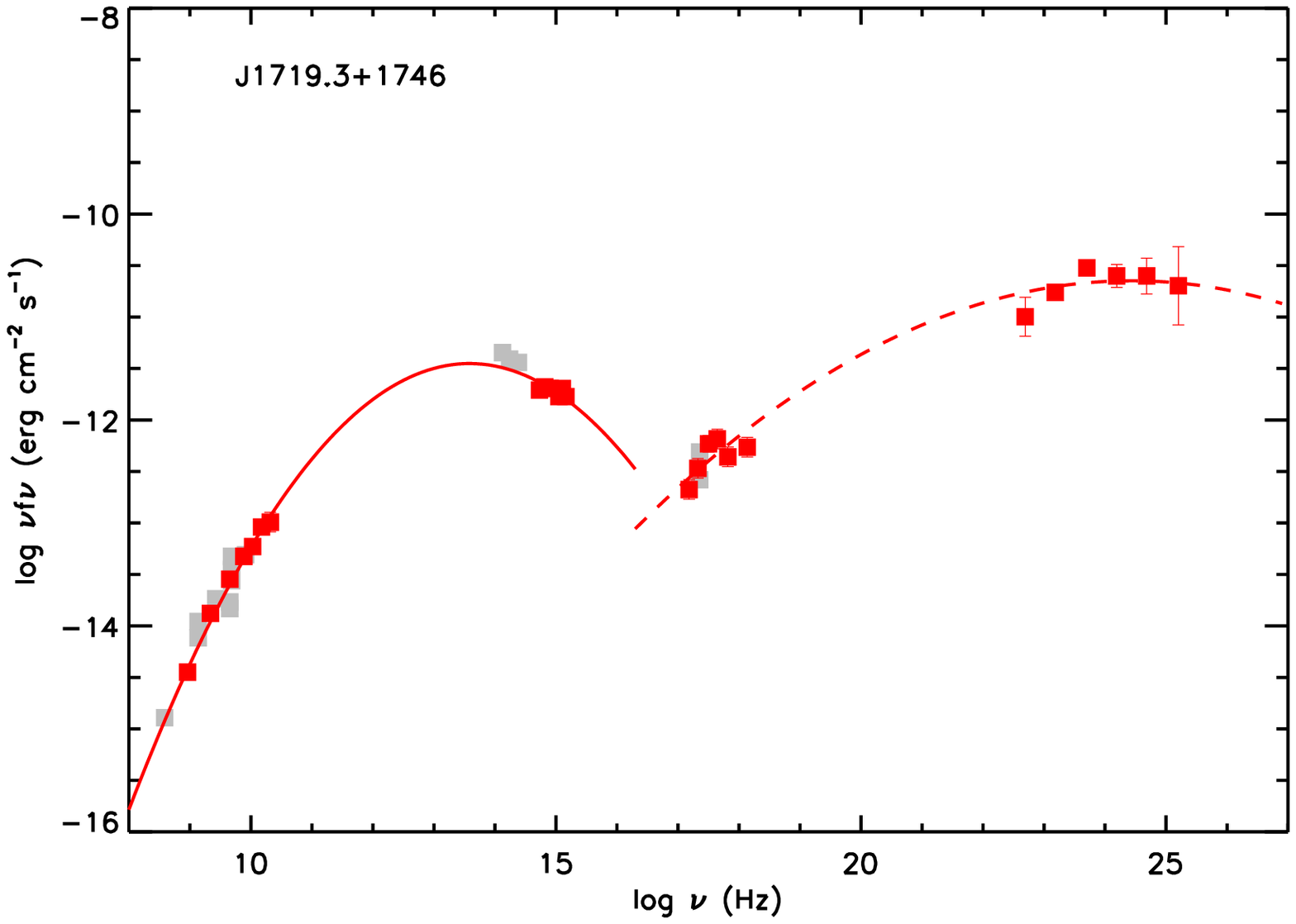}
\includegraphics[height=5cm]{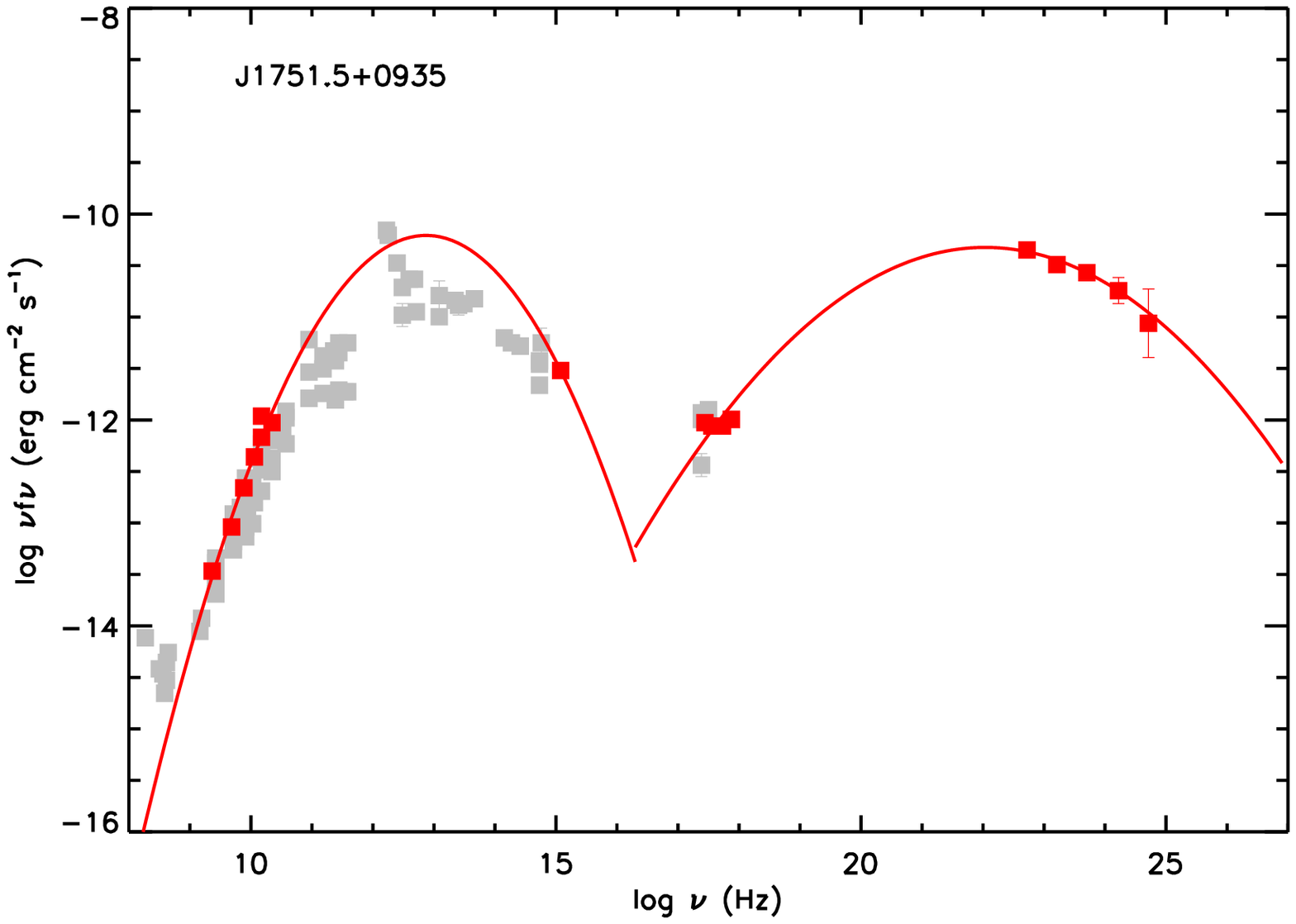}
\center{Fig. \ref{sed} --- continued.}
\end{figure*}

\begin{figure*}
\centering
\includegraphics[height=5cm]{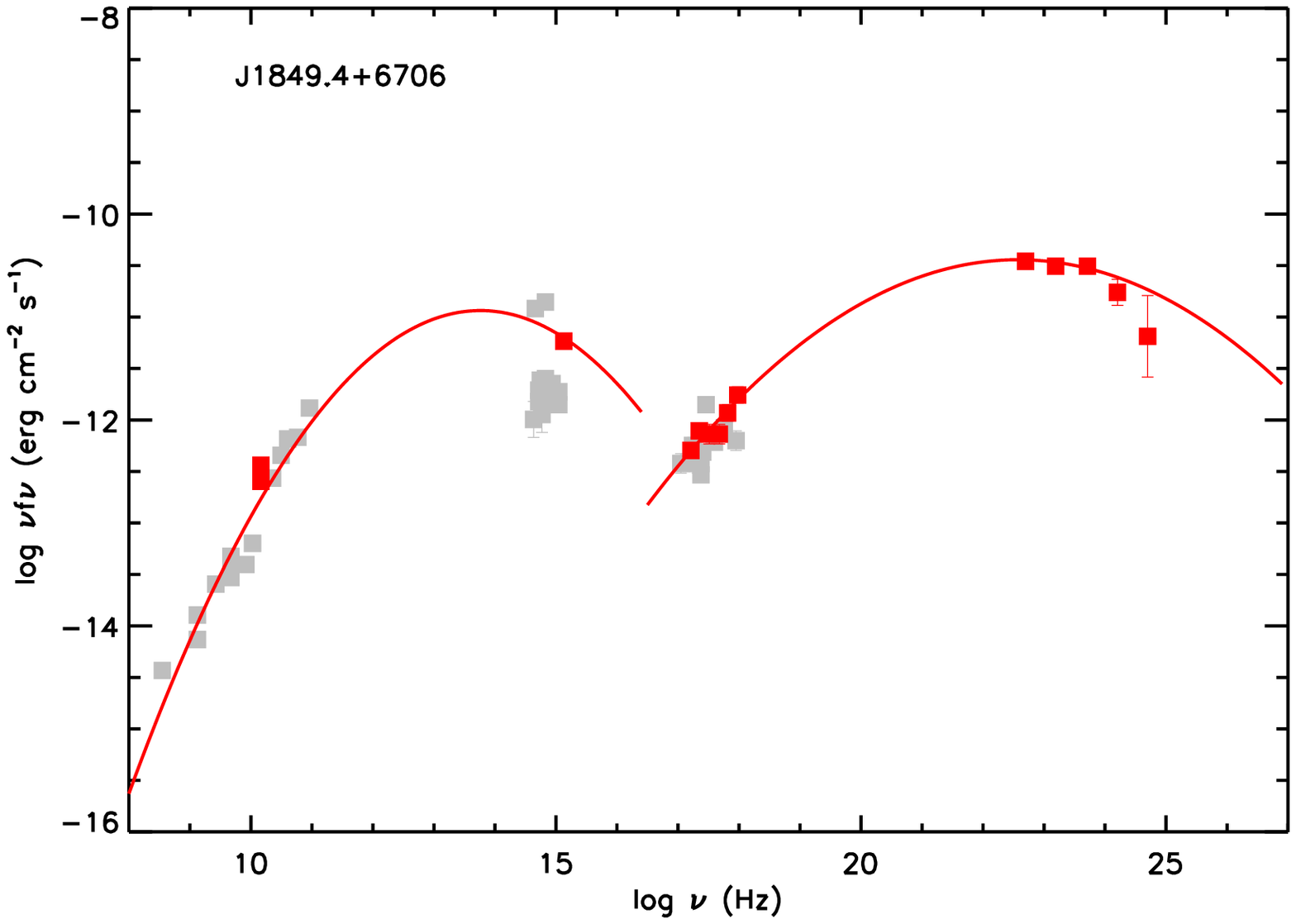}
\includegraphics[height=5cm]{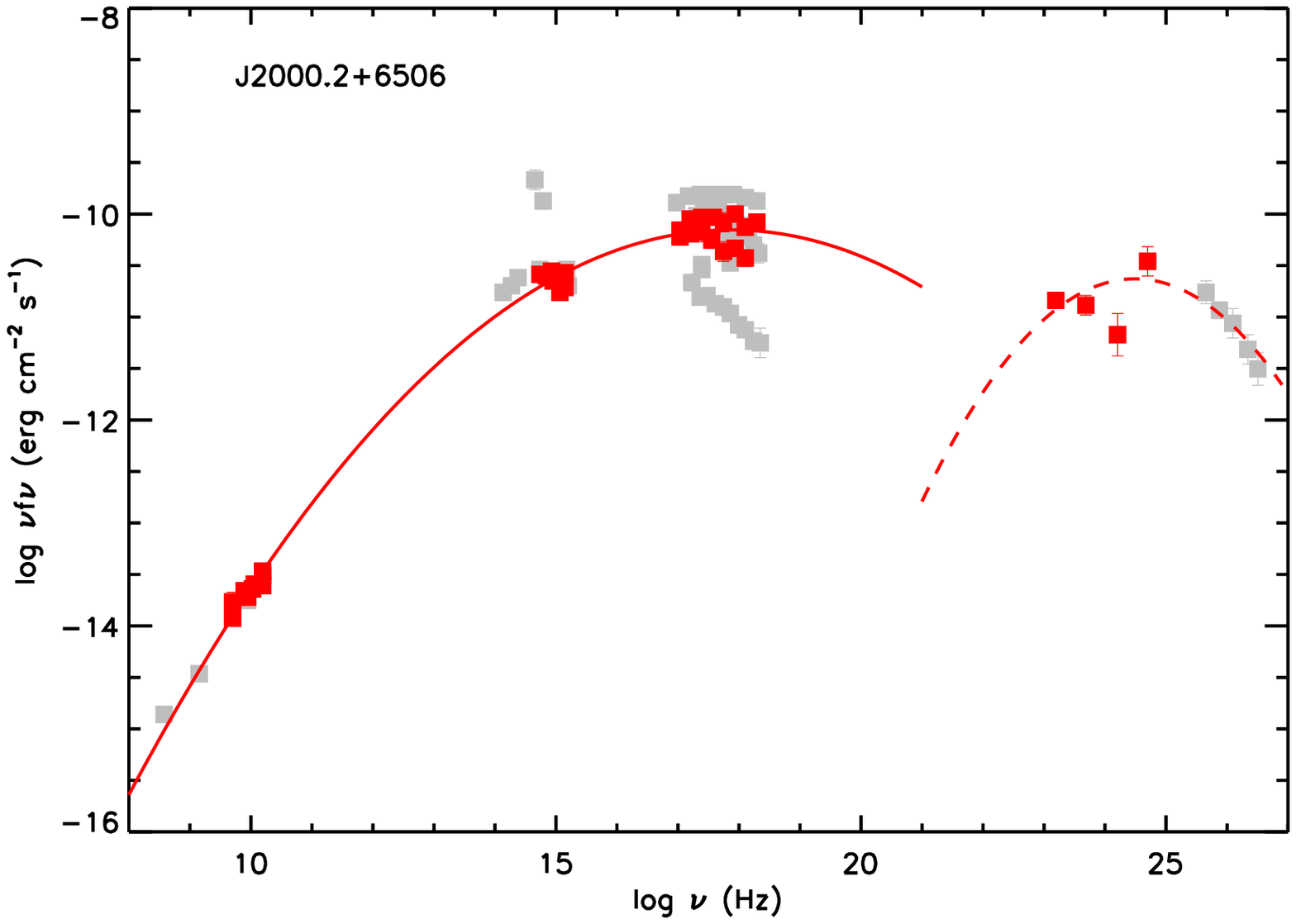}
\includegraphics[height=5cm]{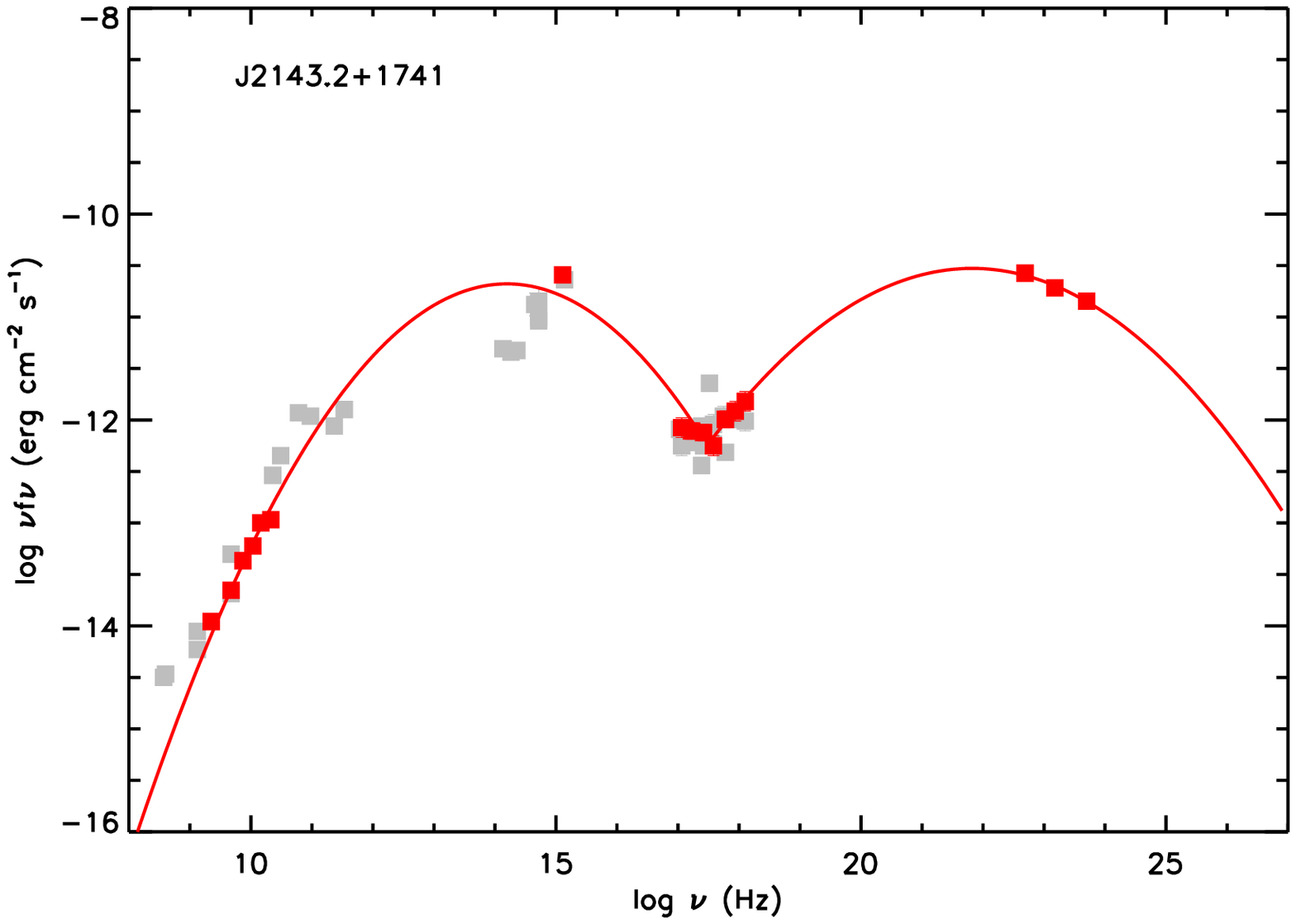}
\includegraphics[height=5cm]{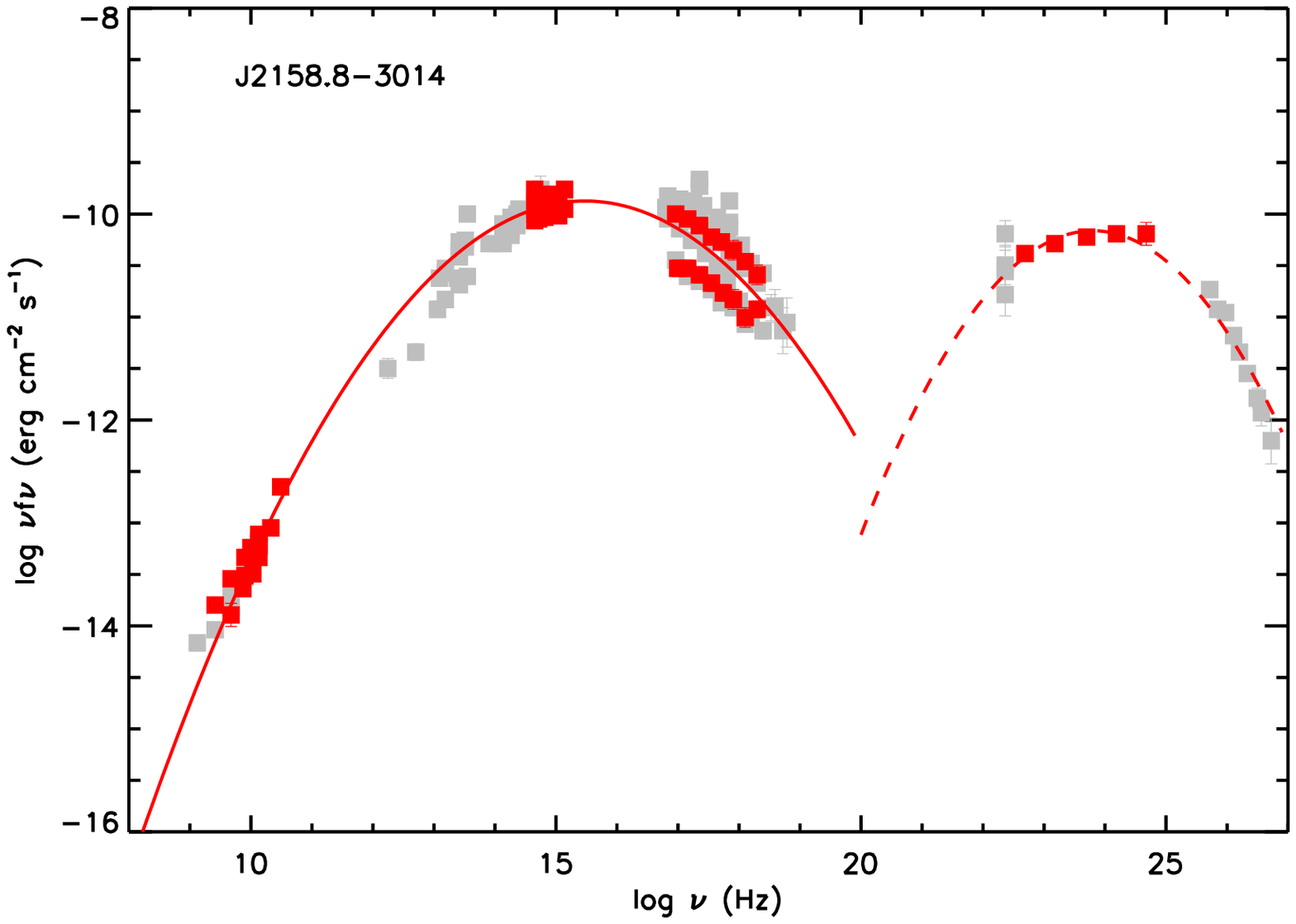}
\includegraphics[height=5cm]{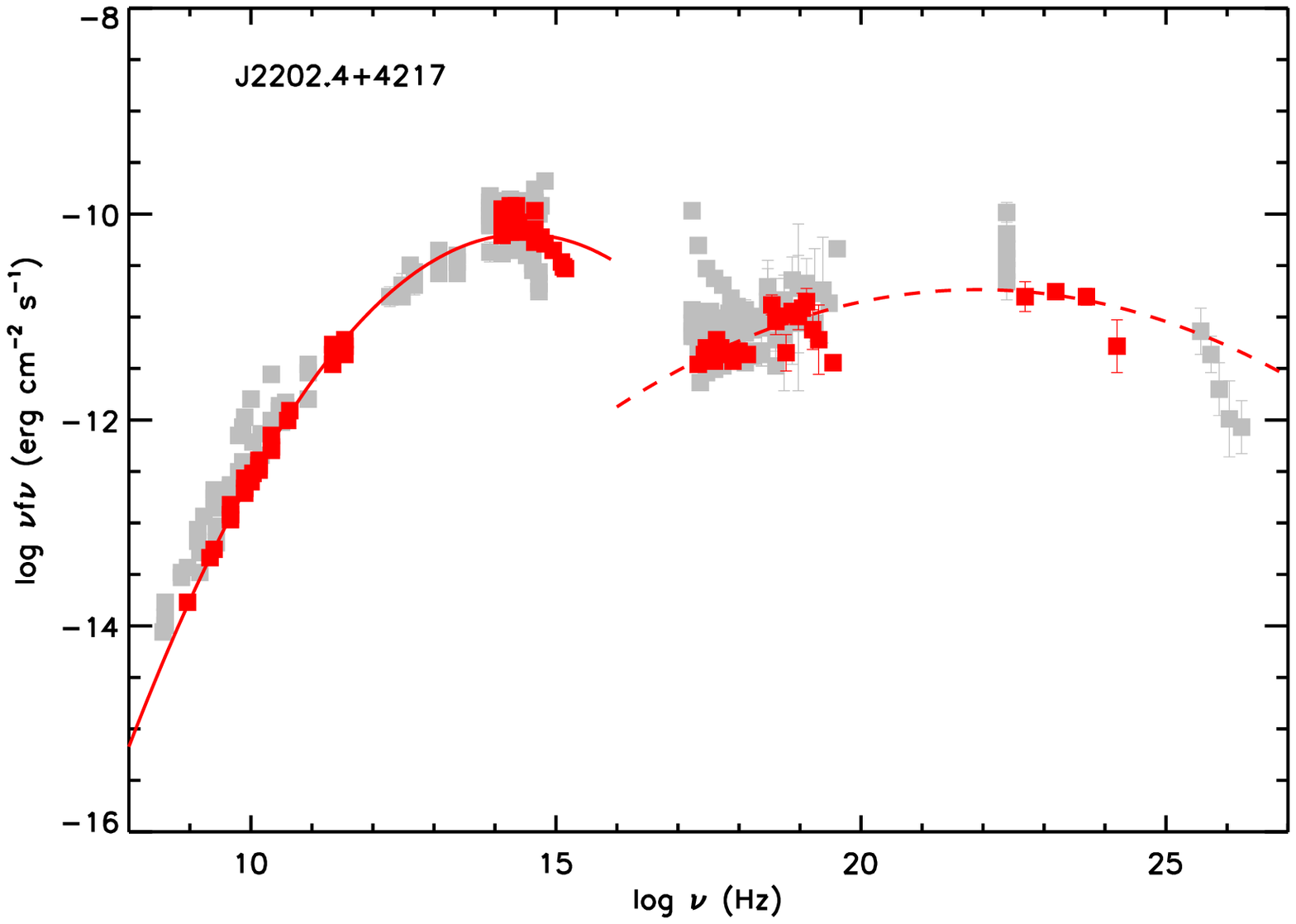}
\includegraphics[height=5cm]{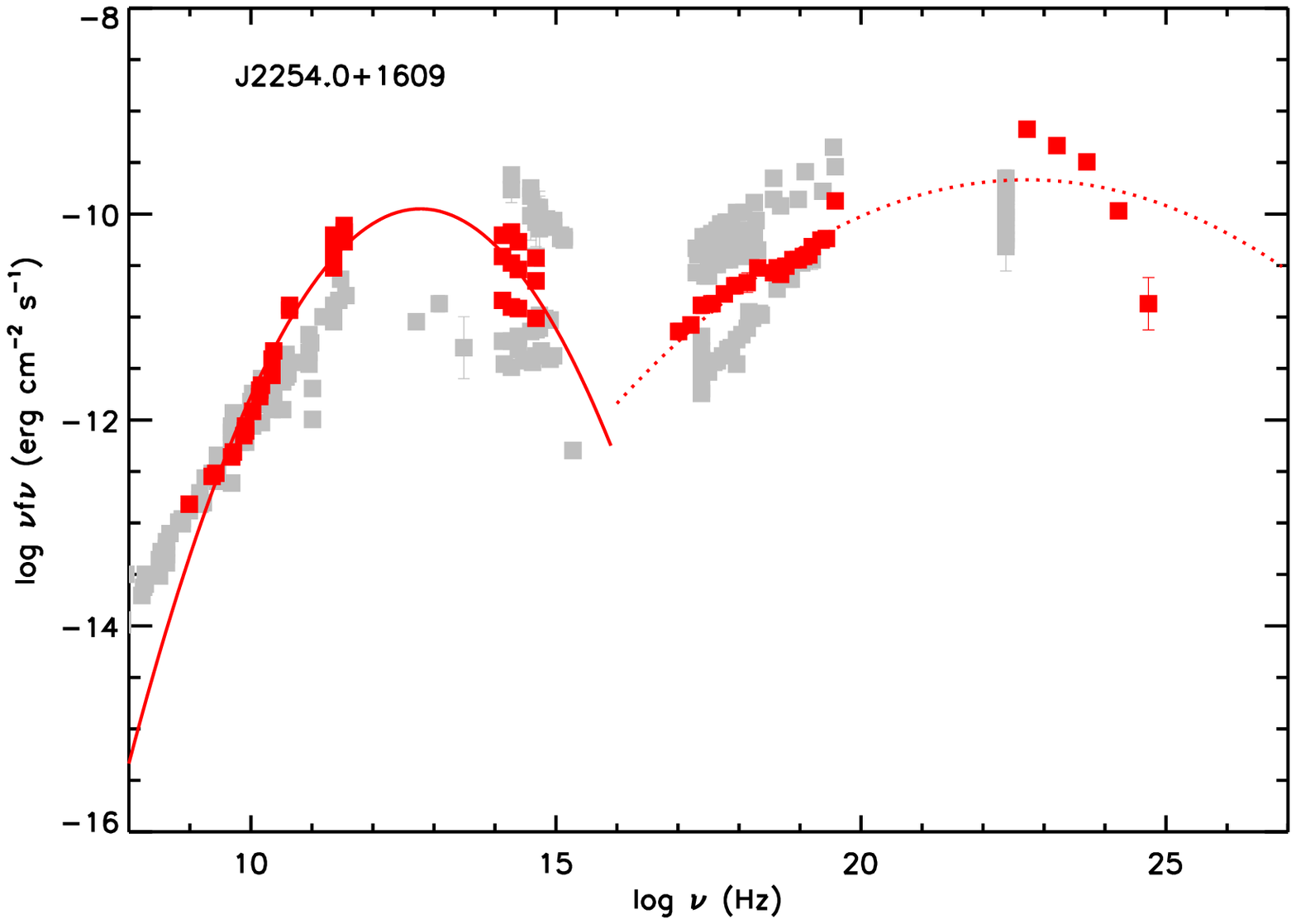}
\includegraphics[height=5cm]{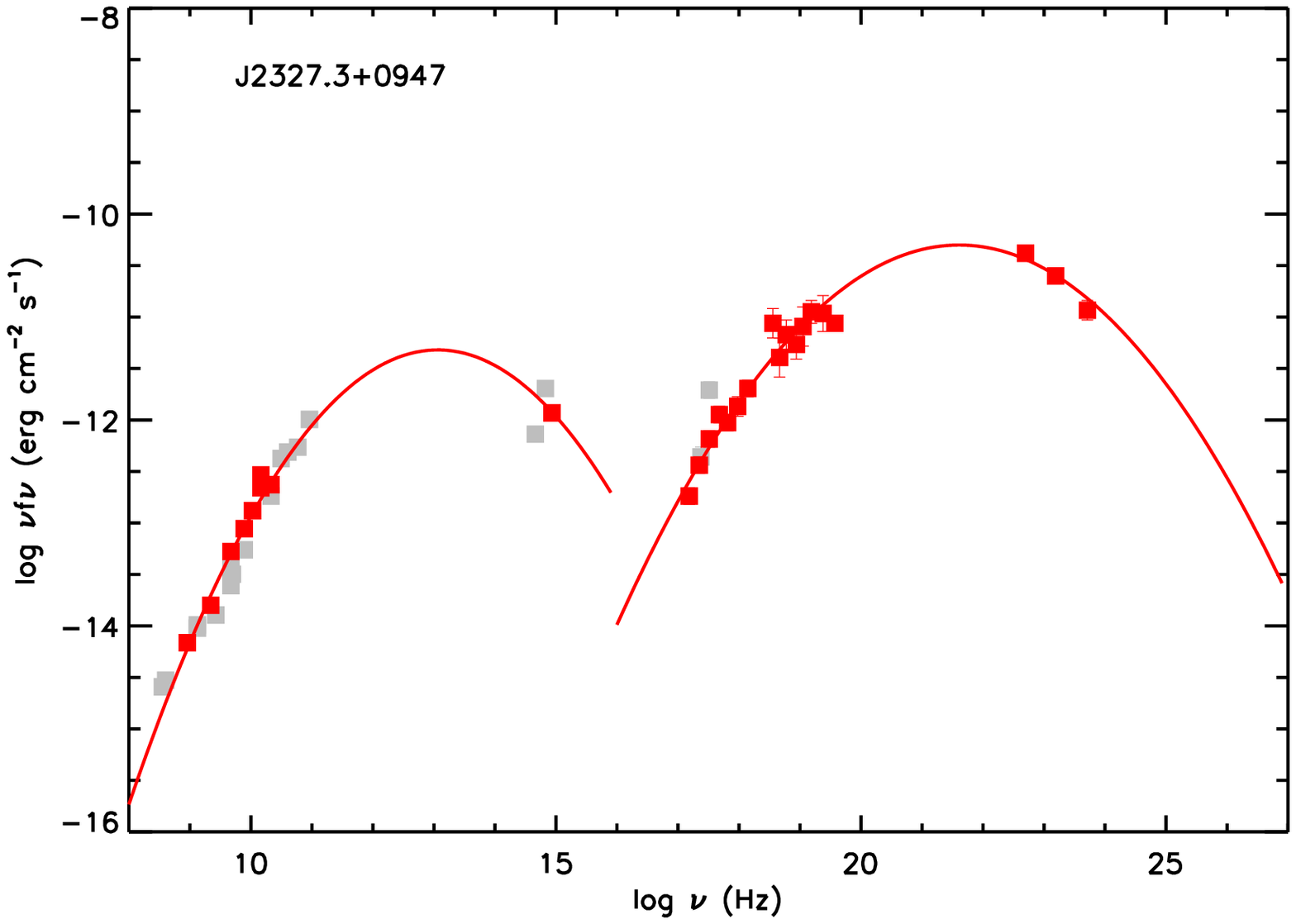}
\includegraphics[height=5cm]{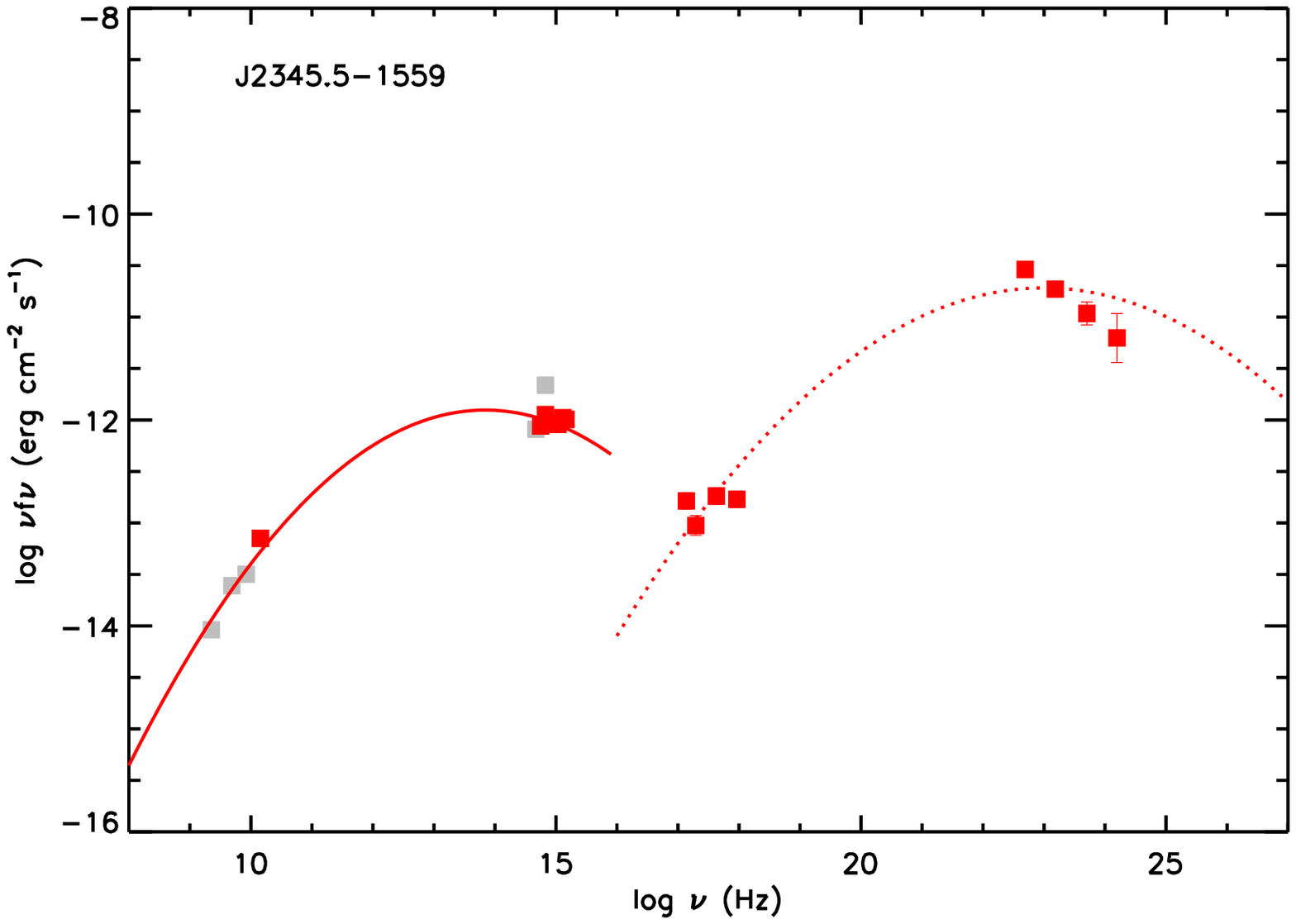}
\center{Fig. \ref{sed} --- continued.}
\end{figure*}

\section{compare with empirical relation/correlation}

\begin{figure}
\epsscale{0.8} \plotone{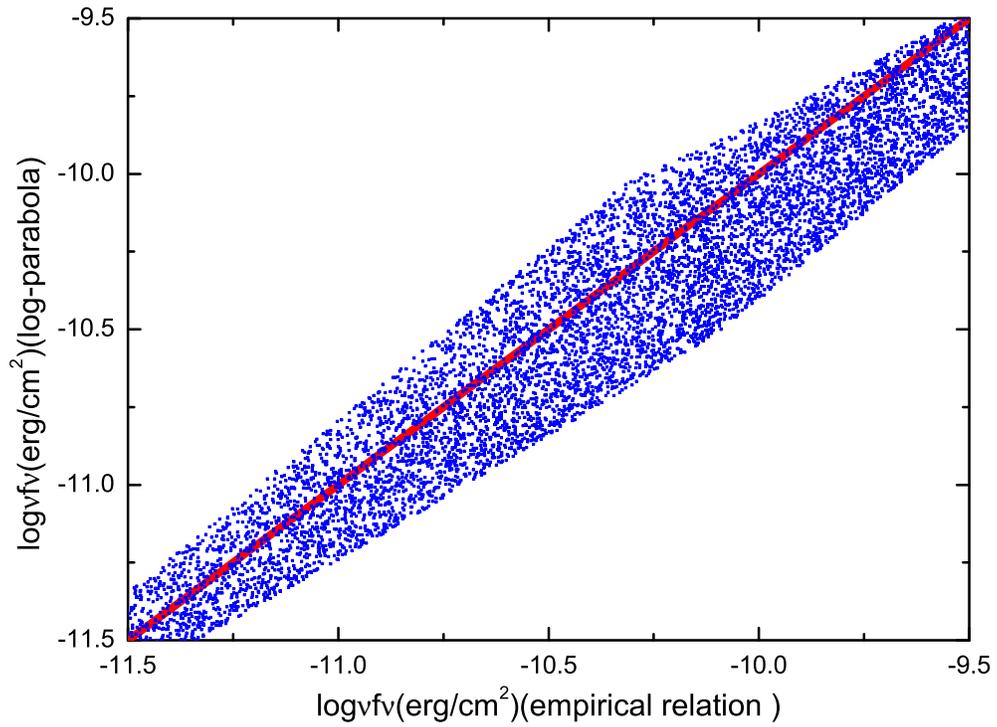} \caption{Comparison between the synchrotron peak fluxes derived from empirical relation \citep{2010ApJ...716...30A} and that derived from Equation \ref{peakfluxeq}. The blue dotes are from Mote-Carlo simulation by randomizing values of synchrotron peak frequency and 5 GHz flux density. The red line is the perfect one-to-one relation. See context for detail.}
\label{peakflux}
\end{figure}

\citet{2010ApJ...716...30A} presented an empirical relation between radio flux density, synchrotron peak frequency, and synchrotron peak flux, i.e., $\log\nu_{p}f_{\nu_{p}}=0.5\log\nu_{p}-20.4+0.9\log(R_{\rm 5GHz})$, where $R_{\rm 5GHz}$ is the radio flux density at 5 GHz in units of mJy, $\nu_{p}$ the synchrotron peak frequency in unit of Hz, and $\nu_{p}f_{\nu_{p}}$ the synchrotron peak flux in unit of erg s$^{-1}$ cm$^{-2}$ \citep[Equation 4 in][]{2010ApJ...716...30A}. A similar relation can be derived by combining a log-parabolic law of SED ($\log\nu f_{\nu}=-b\left(\log\nu-\log\nu_{p}\right)^{2}+\log\nu_{p}f_{\nu_{p}}$) and the correlation between peak frequency and curvature ($1/b=A+B\log\nu_{p}$)\footnote{Note that relation $1/b_{sy}=A+B\log\nu_{p}$ is derived in AGN frame. While this relation is assumed to be valid in the observational frame, because the uncertainty caused by transformation from AGN to observational frames (i.e., $\nu_{\rm p}^{obs}=\nu_{\rm p}/(1+z)$) is smaller than the scatter of the correlation. Similar for following analysis.},
\begin{equation}
\log\nu_{p}f_{\nu_{p}}=\log\nu f_{\nu}+\frac{\left(\log\nu-\log\nu_{p}\right)^{2}}{A+B\log\nu_{p}}.
\label{peakfluxeq}
\end{equation}
The comparison of these two relations is shown in Figure \ref{peakflux}, where the blue dotes are synchrotron peak fluxes calculated by two relations from Mote-Carlo simulation by randomizing the values of synchrotron peak frequency and 5 GHz flux density. The red line is the perfect one-to-one relation. We can see that these two relations are roughly consistent with each other.

\begin{figure}
\epsscale{0.8} \plotone{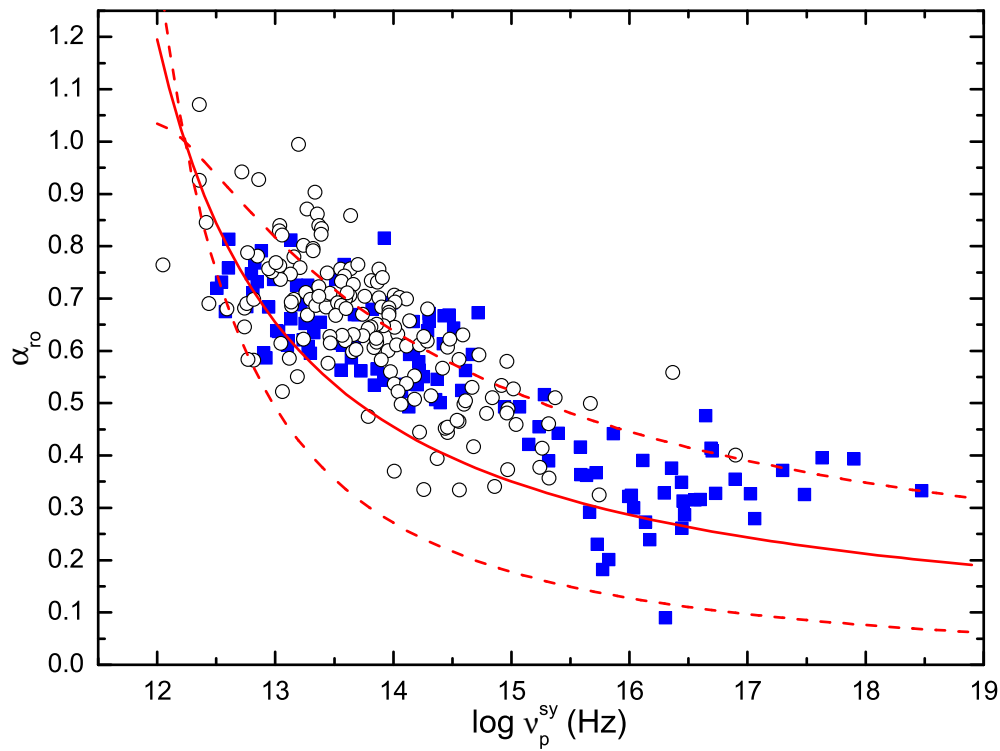} \caption{Broadband spectral indexes $\alpha_{\rm ro}$ versus synchrotron peak frequency. Blue squares and open dots are data taken from \citet{1998MNRAS.299..433F} and \citet{2003ApJ...588..128P}, respectively. The red solid line is derived from Equation \ref{nupeak}, the red dashed lines indicate the 3$\sigma$ confidence bands. The radio frequency and optical wavelength are selected at 5 GHz and 5100 {\AA}, which are same as in these samples.}
\label{alpha12}
\end{figure}

As early in 1990s people found an empirical correlation between the synchrotron peak frequency and the radio to optical broadband spectral index \citep[see e.g.,][]{1995ApJ...444..567P, 1998MNRAS.299..433F, 2003ApJ...588..128P}. A similar relation can also be derived when combining $\log\nu f_{\nu}=-b\left(\log\nu-\log\nu_{p}\right)^{2}+\log\nu_{p}f_{\nu_{p}}$ and $1/b_{sy}=A+B\log\nu_{p}$. If $f_{\nu_{1}}$ and $f_{\nu_{2}}$ are flux densities at $\nu_{1}$ and $\nu_{2}$, respectively, one has,
\begin{equation}
\begin{cases}
\log\nu_{1}f_{\nu_{1}}=-b\left(\log\nu_{1}-\log\nu_{p}\right)^{2}+\log\nu_{p}f_{\nu_{p}} \\
\log\nu_{2}f_{\nu_{2}}=-b\left(\log\nu_{2}-\log\nu_{p}\right)^{2}+\log\nu_{p}f_{\nu_{p}}.
\end{cases}
\label{nupeakmid}
\end{equation}
Substituting the relation $1/b_{sy}=A+B\log\nu_{p}$ and a broadband spectral index $\alpha_{12}=-\left(\log f_{\nu_{2}}-\log f_{\nu_{1}}\right)/\left(\log\nu_{2}-\log\nu_{1}\right)$ into the Equation  \ref{nupeakmid}, one gets,
\begin{equation}
\log\nu_{p}=\frac{A(1-\alpha_{12})+(\log\nu_{2}+\log\nu_{1})}{2-B(1-\alpha_{12})}.
\label{nupeak}
\end{equation}
The Figure \ref{alpha12} indicates the correlation between the synchrotron peak frequency and the radio to optical broadband spectral index, where the blue squares and the black open dots are data collected from \citet{1998MNRAS.299..433F} and \citet{2003ApJ...588..128P}, respectively. The red solid line is derived from Equation \ref{nupeak}, and the red dashed lines indicate the 3$\sigma$ confidence bands. The radio frequency and optical wavelength are taken at 5 GHz and 5100 {\AA}, which are same as those in \citet{1998MNRAS.299..433F} and \citet{2003ApJ...588..128P}. We can see that even though the data are systemically higher than the theory around $\nu\sim10^{14}$ Hz, the trend of the data generally agrees with theory.

\newpage



\begin{thebibliography}{99}


\bibitem[Abdo et al.(2009a)]{2009ApJ...700..597A} Abdo, A.~A., Ackermann,
M., Ajello, M., et al.\ 2009a, \apj, 700, 597


\bibitem[Abdo et al.(2009b)]{2009ApJ...699...31A} Abdo, A.~A., Ackermann,
M., Ajello, M., et al.\ 2009b, \apj, 699, 31


\bibitem[Abdo et al.(2009c)]{2009ApJ...707.1310A} Abdo, A.~A., Ackermann,
M., Ajello, M., et al.\ 2009c, \apj, 707, 1310


\bibitem[Abdo et al.(2010a)]{2010ApJ...721.1425A} Abdo, A.~A., Ackermann,
M., Agudo, I., et al.\ 2010a, \apj, 721, 1425


\bibitem[Abdo et al.(2010b)]{2010ApJ...716...30A} Abdo, A.~A., Ackermann,
M., Agudo, I., et al.\ 2010b, \apj, 716, 30


\bibitem[Abdo et al.(2010c)]{2010ApJ...715..429A} Abdo, A.~A., Ackermann,
M., Ajello, M., et al.\ 2010c, \apj, 715, 429


\bibitem[Ackermann et al.(2011)]{2011ApJ...743..171A} Ackermann, M.,
Ajello, M., Allafort, A., et al.\ 2011, \apj, 743, 171


\bibitem[Aharonian et al.(2007)]{2007ApJ...664L..71A} Aharonian, F.,
Akhperjanian, A.~G., Bazer-Bachi, A.~R., et al.\ 2007, \apjl, 664, L71


\bibitem[Aleksi{\'c} et
al.(2012)]{2012A&A...542A.100A} Aleksi{\'c}, J., Alvarez, E.~A., Antonelli, L.~A., et al.\ 2012, \aap, 542, A100


\bibitem[B{\"o}ttcher
\& Dermer(2002)]{2002ApJ...564...86B} B{\"o}ttcher, M., \& Dermer, C.~D.\ 2002, \apj, 564, 86


\bibitem[Bai(2005)]{2005ChJAS...5..207B} Bai, J.-M.\ 2005, Chinese Journal
of Astronomy and Astrophysics Supplement, 5, 207


\bibitem[Blandford
\& Rees(1978)]{1978bllo.conf..328B} Blandford, R.~D., \& Rees, M.~J.\ 1978, BL Lac Objects, 328


\bibitem[Brodie et al.(1987)]{1987ApJ...318..175B} Brodie, J., Bowyer, S.,
\& Tennant, A.\ 1987, \apj, 318, 175


\bibitem[Brown et al.(1989)]{1989ApJ...340..129B} Brown, L.~M.~J., Robson,
E.~I., Gear, W.~K., et al.\ 1989, \apj, 340, 129


\bibitem[Cerruti et al.(2013)]{2013ApJ...771L...4C} Cerruti, M., Dermer,
C.~D., Lott, B., Boisson, C., \& Zech, A.\ 2013, \apjl, 771, L4


\bibitem[Chen
\& Bai(2011)]{2011ApJ...735..108C} Chen, L., \& Bai, J.~M.\ 2011, \apj, 735, 108


\bibitem[Chen et al.(2012)]{2012ApJ...748..119C} Chen, L., Cao, X.,
\& Bai, J.~M.\ 2012, \apj, 748, 119


\bibitem[Chen et al.(2009)]{2009MNRAS.397.1713C} Chen, Z., Gu, M.,
\& Cao, X.\ 2009, \mnras, 397, 1713


\bibitem[Fossati et al.(1998)]{1998MNRAS.299..433F} Fossati, G., Maraschi,
L., Celotti, A., Comastri, A., \& Ghisellini, G.\ 1998, \mnras, 299, 433


\bibitem[Georganopoulos
\& Kazanas(2003)]{2003ApJ...594L..27G} Georganopoulos, M., \& Kazanas, D.\ 2003, \apjl, 594, L27


\bibitem[Ghisellini et al.(1998)]{1998MNRAS.301..451G} Ghisellini, G.,
Celotti, A., Fossati, G., Maraschi, L.,
\& Comastri, A.\ 1998, \mnras, 301, 451


\bibitem[Ghisellini et al.(2009)]{2009MNRAS.396L.105G} Ghisellini, G.,
Maraschi, L., \& Tavecchio, F.\ 2009, \mnras, 396, L105


\bibitem[Ghisellini et al.(1986)]{1986ApJ...310..317G} Ghisellini, G.,
Maraschi, L., Treves, A., \& Tanzi, E.~G.\ 1986, \apj, 310, 317


\bibitem[Ghisellini
\& Tavecchio(2008)]{2008MNRAS.386L..28G} Ghisellini, G., \& Tavecchio, F.\ 2008, \mnras, 386, L28


\bibitem[Ghisellini et
al.(2005)]{2005A&A...432..401G} Ghisellini, G., Tavecchio, F., \& Chiaberge, M.\ 2005, \aap, 432, 401


\bibitem[Ghisellini et al.(2010)]{2010MNRAS.402..497G} Ghisellini, G.,
Tavecchio, F., Foschini, L., et al.\ 2010, \mnras, 402, 497


\bibitem[Ghisellini et al.(1989)]{1989MNRAS.241P..43G} Ghisellini, G.,
George, I.~M., \& Done, C.\ 1989, \mnras, 241, 43P


\bibitem[Giommi et al.(2012)]{2012MNRAS.420.2899G} Giommi, P., Padovani,
P., Polenta, G., et al.\ 2012, \mnras, 420, 2899


\bibitem[Hartman et al.(1999)]{1999ApJS..123...79H} Hartman, R.~C.,
Bertsch, D.~L., Bloom, S.~D., et al.\ 1999, \apjs, 123, 79


\bibitem[Kardashev(1962)]{1962SvA.....6..317K} Kardashev, N.~S.\ 1962,
\sovast, 6, 317


\bibitem[Komatsu et al.(2011)]{2011ApJS..192...18K} Komatsu, E., Smith,
K.~M., Dunkley, J., et al.\ 2011, \apjs, 192, 18


\bibitem[Krennrich et al.(1999)]{1999ApJ...511..149K} Krennrich, F.,
Biller, S.~D., Bond, I.~H., et al.\ 1999, \apj, 511, 149


\bibitem[Landau et al.(1986)]{1986ApJ...308...78L} Landau, R., Golisch, B.,
Jones, T.~J., et al.\ 1986, \apj, 308, 78


\bibitem[Ledden
\& Odell(1985)]{1985ApJ...298..630L} Ledden, J.~E., \& Odell, S.~L.\ 1985, \apj, 298, 630


\bibitem[Ledden et al.(1981)]{1981ApJ...243...47L} Ledden, J.~E., Odell,
S.~L., Stein, W.~A., \& Wisniewski, W.~Z.\ 1981, \apj, 243, 47


\bibitem[Massaro et
al.(2004a)]{2004A&A...413..489M} Massaro, E., Perri, M., Giommi, P., \& Nesci, R.\ 2004a, \aap, 413, 489


\bibitem[Massaro et
al.(2004b)]{2004A&A...422..103M} Massaro, E., Perri, M., Giommi, P., Nesci, R., \& Verrecchia, F.\ 2004b, \aap, 422, 103


\bibitem[Massaro et
al.(2006)]{2006A&A...448..861M} Massaro, E., Tramacere, A., Perri, M., Giommi, P., \& Tosti, G.\ 2006, \aap, 448, 861


\bibitem[Massaro et al.(2011a)]{2011ApJ...739...73M} Massaro, F., Paggi, A.,
Elvis, M., \& Cavaliere, A.\ 2011a, \apj, 739, 73


\bibitem[Massaro et
al.(2008)]{2008A&A...478..395M} Massaro, F., Tramacere, A., Cavaliere, A., Perri, M., \& Giommi, P.\ 2008, \aap, 478, 395


\bibitem[Massaro et al.(2013)]{2013ApJS..207...16M} Massaro, F., Paggi, A.,
Errando, M., et al.\ 2013, \apjs, 207, 16


\bibitem[Nieppola et
al.(2006)]{2006A&A...445..441N} Nieppola, E., Tornikoski, M., \& Valtaoja, E.\ 2006, \aap, 445, 441


\bibitem[Nodes et
al.(2004)]{2004A&A...423...13N} Nodes, C., Birk, G.~T., Gritschneder, M., \& Lesch, H.\ 2004, \aap, 423, 13


\bibitem[Odell et al.(1977)]{1977ApJ...214L.105O} Odell, S.~L., Puschell,
J.~J., Stein, W.~A., \& Warner, J.~W.\ 1977, \apjl, 214, L105


\bibitem[Padovani
\& Giommi(1995)]{1995ApJ...444..567P} Padovani, P., \& Giommi, P.\ 1995, \apj, 444, 567


\bibitem[Padovani et al.(2003)]{2003ApJ...588..128P} Padovani, P., Perlman,
E.~S., Landt, H., Giommi, P., \& Perri, M.\ 2003, \apj, 588, 128


\bibitem[Paggi et
al.(2009a)]{2009A&A...504..821P} Paggi, A., Massaro, F., Vittorini, V., et al.\ 2009a, \aap, 504, 821


\bibitem[Paggi et
al.(2009b)]{2009A&A...508L..31P} Paggi, A., Cavaliere, A., Vittorini, V., \& Tavani, M.\ 2009b, \aap, 508, L31


\bibitem[Punch et al.(1992)]{1992Natur.358..477P} Punch, M., Akerlof,
C.~W., Cawley, M.~F., et al.\ 1992, \nat, 358, 477


\bibitem[Rani et al.(2011)]{2011MNRAS.417.1881R} Rani, B., Gupta, A.~C.,
Bachev, R., et al.\ 2011, \mnras, 417, 1881


\bibitem[Rieke
\& Kinman(1974)]{1974ApJ...192L.115R} Rieke, G.~H., \& Kinman, T.~D.\ 1974, \apjl, 192, L115


\bibitem[Rybicki
\& Lightman(1979)]{1979rpa..book.....R} Rybicki, G.~B., \& Lightman, A.~P.\ 1979, New York, Wiley-Interscience, 1979.~393 p.,


\bibitem[Sambruna et al.(1996)]{1996ApJ...463..444S} Sambruna, R.~M.,
Maraschi, L., \& Urry, C.~M.\ 1996, \apj, 463, 444


\bibitem[Samuelson et al.(1998)]{1998ApJ...501L..17S} Samuelson, F.~W.,
Biller, S.~D., Bond, I.~H., et al.\ 1998, \apjl, 501, L17


\bibitem[Scarpa
\& Falomo(1997)]{1997A&A...325..109S} Scarpa, R., \& Falomo, R.\ 1997, \aap, 325, 109


\bibitem[{\c S}ent{\"u}rk et al.(2013)]{2013ApJ...764..119S} {\c
S}ent{\"u}rk, G.~D., Errando, M., B{\"o}ttcher, M.,
\& Mukherjee, R.\ 2013, \apj, 764, 119


\bibitem[Sikora et al.(2002)]{2002ApJ...577...78S} Sikora, M.,
B{\l}a{\.z}ejowski, M., Moderski, R.,
\& Madejski, G.~M.\ 2002, \apj, 577, 78


\bibitem[Sitko et al.(1983)]{1983PASP...95..724S} Sitko, M.~L., Stein,
W.~A., Zhang, Y.-X., \& Wisniewski, W.~Z.\ 1983, \pasp, 95, 724


\bibitem[Tavecchio et al.(1998)]{1998ApJ...509..608T} Tavecchio, F.,
Maraschi, L., \& Ghisellini, G.\ 1998, \apj, 509, 608


\bibitem[Tramacere et
al.(2007)]{2007A&A...467..501T} Tramacere, A., Giommi, P., Massaro, E., et al.\ 2007, \aap, 467, 501


\bibitem[Tramacere et
al.(2009)]{2009A&A...501..879T} Tramacere, A., Giommi, P., Perri, M., Verrecchia, F., \& Tosti, G.\ 2009, \aap, 501, 879


\bibitem[Tramacere et al.(2011)]{2011ApJ...739...66T} Tramacere, A.,
Massaro, E., \& Taylor, A.~M.\ 2011, \apj, 739, 66


\bibitem[Tramacere et
al.(2007)]{2007A&A...466..521T} Tramacere, A., Massaro, F., \& Cavaliere, A.\ 2007, \aap, 466, 521


\bibitem[Urry
\& Padovani(1995)]{1995PASP..107..803U} Urry, C.~M., \& Padovani, P.\ 1995, \pasp, 107, 803


\bibitem[Wu et al.(2009)]{2009RAA.....9..168W} Wu, Z.-Z., Gu, M.-F.,
\& Jiang, D.-R.\ 2009, Research in Astronomy and Astrophysics, 9, 168


\bibitem[Xu et al.(2009)]{2009ApJ...694L.107X} Xu, Y.-D., Cao, X.,
\& Wu, Q.\ 2009, \apjl, 694, L107


\bibitem[Zhang et al.(2012)]{2012ApJ...752..157Z} Zhang, J., Liang, E.-W.,
Zhang, S.-N., \& Bai, J.~M.\ 2012, \apj, 752, 157








\end{thebibliography}
\end{document}